\documentclass[sigconf]{acmart}
\setcopyright{none}                      
\renewcommand\footnotetextcopyrightpermission[1]{} 
\acmConference[]{}{}{}                   

\AtBeginDocument{%
  }

\usepackage{tikz}
\usepackage{amsmath}
\usepackage{pgfplotstable}
\usepackage{booktabs}
\usepackage{pdflscape}
\usepackage{anyfontsize}
\usepackage{makecell}
\usepackage{longtable}
\usepackage{csvsimple}
\usepackage{supertabular,booktabs}
\usepackage{subcaption}
\usepackage{framed}
\usepackage{tabularx}
\usepackage{rotating}
\usepackage{xurl}
\usepackage{hyperref}
\usepackage[capitalise]{cleveref}
\usepackage{booktabs}
\usepackage{multirow}

\usepackage{amssymb}
\usepackage{tikz}
\usepackage[usenames,dvipsnames]{xcolor}
\usepackage{mdframed}
\usepackage{footnote}
\makesavenoteenv{mdframed}
\usepackage[most]{tcolorbox}

\definecolor{CustomGreen}{HTML}{00A000}
\definecolor{CustomRed}{HTML}{D62728}

\newcommand\circledtick[1][CustomGreen]{%
  \tikz[baseline=(A.south)]{
    \node[circle,fill=#1, inner sep=0.5ex] (A) {};
    \node at (A) {\textcolor{white}{\scriptsize$\checkmark$}};
  }%
}

\newcommand\circledcross[1][CustomRed]{%
  \tikz[baseline=(A.south)]{
    \node[circle,fill=#1, inner sep=0.5ex] (A) {};
    \node at (A) {\textcolor{white}{\scriptsize$\times$}};
  }%
}

\def\name{\textsf{Credential Assistant}}
\def\names{\textsf{Credential Assistants}}

\begin{document}

\title{Credential Disclosure in (EU) Digital Identity Wallets: \texorpdfstring{\\}{ } Privacy Risks and Practical Mitigations} 
\renewcommand{\shorttitle}{Credential Disclosure in (EU) Digital Identity Wallets: Privacy Risks and Practical Mitigations}

\author{Sheila Zingg}
\affiliation{%
  \institution{ETH Zurich}
  \city{Zurich}
  \country{Switzerland}
}

\author{Daniele Lain}
\affiliation{%
  \institution{ETH Zurich}
  \city{Zurich}
  \country{Switzerland}
}

\author{Yoshimichi Nakatsuka}
\affiliation{%
  \institution{ETH Zurich}
  \city{Zurich}
  \country{Switzerland}
}

\author{Kari Kostiainen}
\affiliation{%
  \institution{ETH Zurich}
  \city{Zurich}
  \country{Switzerland}
}

\author{Stefan Bechtold}
\affiliation{%
  \institution{ETH Zurich}
  \city{Zurich}
  \country{Switzerland}
}

\author{Srdjan Čapkun}
\affiliation{%
  \institution{ETH Zurich}
  \city{Zurich}
  \country{Switzerland}
}

\renewcommand{\shortauthors}{Zingg et al.}

\begin{abstract}
    The European Union will introduce the EUDI Wallet by late 2026, which allows users to hold digital credentials (i.e., representations of physical official identity documents) on their devices. This will allow users to securely and privately disclose identity attributes to websites. Although such a system has many benefits, it also introduces risks caused by poor credential disclosure decisions. In this paper, we (i) conduct a large-scale survey on credential disclosure with users and experts and (ii) evaluate the effectiveness and feasibility of our \name{} that displays expert recommendations and user opinions. Our results show that users are likely to overshare (e.g., $\sim$20\% of users disclosed their official ID to news websites). This indicates that users struggle to protect their privacy, which will impact the usability of the EUDI Wallet and lead to privacy violations, identity theft, and other abuses of leaked credentials. Finally, we show that our \name{} significantly reduces users' credential disclosure mistakes from $\sim$15\% to $\sim$7\%. However, it does not fully eliminate poor credential disclosure decisions, indicating that stronger interventions may be necessary, especially for sensitive attributes.
\end{abstract}

\maketitle

\section{Introduction}

The need for online identity verification is increasing. Services are moving online, and many of those services have legal obligations to perform stringent identity checks (e.g., KYC for banks and payment providers, identity checks for lodging providers, identity checks for e-Visas). This move towards online services is also spreading to sectors that must verify data about users beyond their identity (e.g., an online pharmacy verifying a prescription). Furthermore, governments are increasingly requiring strict enforcement of user age verification~\cite{UK_age_restriction,EU_age_restriction,France_age_restriction,US_age_restriction}. 
Currently, users can provide information about their identity online through cumbersome means, such as pictures of their identity documents or through video identity verification systems.
However, these methods force users to disclose more data than necessary (e.g., to buy alcohol online, they would need to transmit their entire identity card instead of just proving that they are over 18). Furthermore, they expose users to potential privacy violations and identity theft, while also not being secure for services, as these systems can be tricked~\cite{zheng2019survey, zheng2023spoofing}.

To address these challenges, the European Union introduced the EU Digital Identity (EUDI) Wallet~\cite{EUDI_Wallet, EUDI_Wallet_GitHub} in 2024, with the aim of having a functional system operational in late 2026. 
The EUDI Wallet will allow users to hold digital credentials (officially signed representations of physical official documents) in an app on their phone.
Verifiers (e.g., websites and apps) can request attributes contained in a credential from users through the EUDI Wallet, and users can accept those requests to disclose the attributes. This will significantly improve the user experience for online identity verification. For example, an e-shop selling alcohol can request the attribute encoding whether the user is over 18. Initially, the EUDI Wallet will focus on identity documents; however, in the future, it will store documents, such as prescriptions and educational certificates.

However, giving users full control over their data and reducing disclosure friction also brings risks. Users are likely to overshare data online (i.e., disclose data that is not actually needed for the service)~\cite{acquisti2017nudges, egelman2013choice, hann2002online, acquisti2013privacy, brandimarte2013misplaced, cvrcek2006study, norberg2007privacy, taddicken2014privacy}, and websites often request more data than they would actually need to provide their services~\cite{nouwens2020dark, mathur2019dark, tran2025dark, luguri2021shining}.
It will not be possible to reuse data harvested through the EUDI Wallet inside the EUDI ecosystem. However, there are concrete privacy risks in disclosing authenticated digital credentials, as they could be used for tracking and profiling, and even identity theft by creating fake identity documents to trick traditional identity verification systems\footnote{The eIDAS Regulation states that it must be voluntary for users to use the EUDI Wallet, and that other means of identification must always be available~\cite[Art. 5a(15)]{eu2024digitalwallet}.}.
For this reason, the EUDI framework will require mandatory registration for websites that want to request EUDI credentials, which includes the attributes they will request. The EUDI Wallet then rejects any request that exceeds what was registered, with risk of suspension of the offending website from the system if this happens~\cite[Art. 9]{eu2025walletregistration}.
However, similar to websites over-claiming what data they need in their privacy policies, there is a risk that websites will over-claim what attributes they need during registration.
While this does not respect the GDPR's data minimization principle~\cite{edps2025wallets}, having potentially hundreds of thousands of websites registering to the ecosystem means it will not be possible to verify the data requirements of each of them, and users will be exposed to over-requesting websites.

\paragraph{This Paper}

We assume that regulation alone will not be sufficient to prevent websites from over-claiming what EUDI attributes they need, and that without support, users will likely overshare EUDI credentials.
To help users, we study whether nudging at the moment of the disclosure decision can be effective. For this, we developed and evaluated a \name{} that displays user opinions or expert recommendations on whether a credential should be disclosed to a website, depending on the website's type and category.
We base our \name{} on opinions because there is only limited ground truth on whether a credential should be disclosed to a website (e.g., while identity information is necessary for KYC processes of all banks, whether it is necessary for a news website depends on the context and the website's services, and thus, is subjective).
This choice further informs the design of our \name{}. As it is unfeasible to collect opinions for each website and use case individually, we collect data at the level of website \textit{categories} instead. Furthermore, we study the most challenging case in which no context is provided to users (e.g., the website requests credentials at registration for its wide range of services, similar to current Single Sign-On), which is the most likely case in which users will make poor disclosure decisions.

As the EUDI Wallet does not exist yet, we conducted a large-scale survey and simulated user study to study the effect of the \name{}. We simplified the EUDI to better align with users' current mental models.
In the survey, conducted with 27 experts and 1,035 users, participants evaluated 14 credentials and 166 websites from 17 website categories. Their task was to decide which credentials they would disclose to each website. 
In the user study, conducted with 1,002 users, participants were shown a simulated web-browsing scenario in which a website requested a credential, and needed to decide whether they wanted to accept the request. Some participants were supported by the \name{}.

Our study provides evidence on (i) the data collection requirements for the \name{}, (ii) how effective the \name{} is in supporting users, and (iii) what the potential (over)sharing patterns are for this new type of credential, i.e., digital versions of official identity documents.
Regarding (i), we observe that for most categories, $\sim$24 expert evaluations are sufficient, although for some categories, higher disagreement may require more expert evaluations. Furthermore, the comparison of our survey and user study indicates that data for the \name{} should come from user surveys/focus groups instead of usage data, as users overshare more when seeing a direct credential request than when asked about their willingness to disclose (e.g., for disclosing medical records with LinkedIn, $\sim$31\% of users accepted the credential request, but $\sim$0\% of surveyed users said they would disclose). Regarding (ii), we find that the \name{} is able to significantly reduce the number of mistakes users make in most cases (e.g., it reduced the average number of mistakes users made from 15\% to 7\% when showing expert recommendations). However, it does not have any nudging effect when it shows low confidence numbers (i.e., 51\%-55\% recommend to disclose), which may be an issue, as there was low expert agreement in many cases where there is a good justification for a website to request a credential. 
Regarding (iii), we find that users are likely to overshare credentials (e.g., $\sim$20\% of users disclosed their official ID to news websites), and that this risk is increased in cases where users make false connections between the credential and the services a website provides (e.g., $\sim$34\% of users disclosed their bank account validation to an e-commerce website thinking it could be used for payment).

These results indicate that users might struggle to protect their privacy in the EUDI, as they might not be more cautious with their official digital documents than they are with other personal data. Furthermore, while our \name{} is able to mitigate the risk of poor credential disclosure decisions, a residual risk remains. Although 7\% may seem low, on the scale of the EUDI, this still puts the sensitive identity information of tens of thousands of users at risk. Thus, more intrusive interventions (e.g., similar to browser warnings) may be needed to ensure users do not ignore them. We note that we make several simplifications to study this topic by using a simulated environment (because the EUDI does not exist yet), omitting selective disclosure (as this does not align with users' current mental models, and thus, may lead to confusion), and studying disclosure without context (because the \name{} cannot feasibly account for context). 

To summarize, the main contributions of this paper are:
\begin{enumerate}
    \item \textit{\name{} for the EUDI Wallet.} We evaluated a scalable \name{} that displays expert recommendations and user opinions. We conducted a survey with 1,035 users and 27 experts to understand the data collection requirements to develop such a \name{}.
    \item \textit{Effectiveness of disclosure support.} We conducted a simulated user study with 1,002 participants to evaluate the effectiveness of support at decision time for EUDI disclosure decisions. We show that the \name{} reduces poor disclosure decisions; however, a residual risk remains.
    \item \textit{EUDI credential disclosure risks.} We found evidence of oversharing, which may result not only in privacy violations but also in identity theft. We also identified new patterns of oversharing resulting from users misunderstanding what a credential can be used for, even when seeing its attributes.
\end{enumerate}
\section{Background}
\label{sec:background}

The European Union will introduce the EU Digital Identity (EUDI) by the end of 2026~\cite{EUDI_Wallet, EUDI_Wallet_GitHub}. This consists of an infrastructure for online identity verification and the EUDI Wallet app, where EUDI credentials (digital representations of official documents, e.g., a digital ID) can be stored. 
The EU published a regulation~\cite{eu2024digitalwallet}, architecture reference framework (ARF)~\cite{EUDI_ARF}, and more than two dozen Implementing Acts~\cite{euwalletregulation}, based on which EU member states must revise their laws and procedures to implement their EUDI Wallets.  

\begin{figure}[tb]
    \centering
    \includegraphics[width=\linewidth]{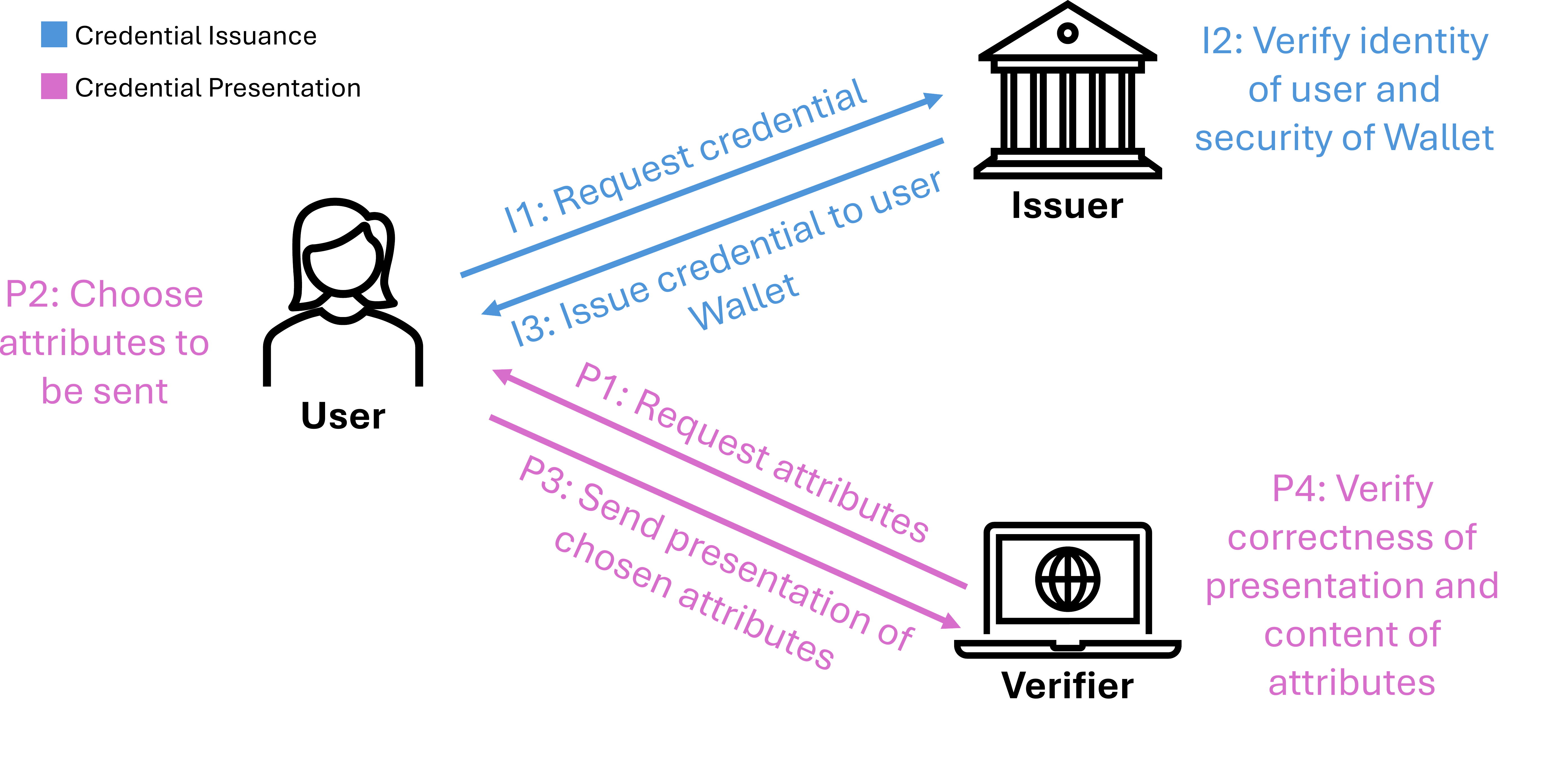}
    \caption{The parties and processes involved in the EUDI.}
    \label{fig:EUDI_overview}
\end{figure}

\cref{fig:EUDI_overview} shows an overview of the planned EUDI. There are three main parties: 1) the credential holder (which we will call the user), 2) the issuer, and 3) the verifier or relying party. The issuer signs a credential containing attributes and issues it to a user, who stores it on their own device in the EUDI Wallet app (steps I1--I3). Later, a verifier can request attributes from a user, and the user can accept or reject the request. If the user accepts the request, the EUDI Wallet generates a presentation that proves that the credential was issued to the user by a specific issuer and discloses the requested attributes contained in the credential. The verifier validates that the credential was issued by a trusted issuer and that the credential was issued to the user who presented the credential (steps P1--P4). 

The following aspects of the EUDI are important for this paper: (I)~a user can present an EUDI credential to a verifier without communicating with any other party. 
(II)~Initially, the EUDI Wallet will only store digital ID cards, but it is planned to be able to store various credentials. 
(III)~EUDI credentials are planned to consist of individually hashed attributes that can be disclosed individually. 
(IV)~The UI of the EUDI Wallet has not been decided and will be determined by each member state.
(V)~All issuers and verifiers will be registered, allowing anyone to verify who issued a credential and who is requesting attributes. A verifier registration that contains all attributes that a verifier wants to request is planned, and the EUDI Wallet would reject any request that does not follow this registration~\cite{EUDI_Law_Verifier_Credentials, EUDI_Wallet_GitHub}. 
Finally, (VI)~currently, there is no plan to allow websites to display the purpose for a request in the EUDI Wallet. 

\paragraph{Threat Model}
We assume that all parties follow the EUDI protocol correctly; however, the verifier may over-request data.
\section{Methodology}
\label{sec:mr}
Our work is motivated by the issue that websites over-request data~\cite{nouwens2020dark, mathur2019dark, tran2025dark, luguri2021shining}, and users unnecessarily disclose data~\cite{acquisti2017nudges, egelman2013choice, hann2002online, acquisti2013privacy, brandimarte2013misplaced, cvrcek2006study, norberg2007privacy, taddicken2014privacy}. 
We assume that the same risk will be present in the EUDI. Regulators plan to address this risk by forcing verifiers (i.e., websites) to register themselves and all attributes they plan to request; however, this will not stop verifiers from over-registering attributes, and it is unlikely that the registry can be checked for over-registered attributes in a scalable manner. Thus, we investigate non-intrusive nudges by designing a \name{} that can help users make better credential disclosure decisions by providing information at decision time. In the following, we describe the challenges of creating a \name{} and motivate its design. We then describe the research areas we investigate in this paper and the methods we use to investigate them.

\subsection{Design of the \name{}}

We aim to provide support to users when responding to a credential request by helping them decide whether the request is appropriate. 
However, there are several challenges when designing such an intervention.
First, there is no objective ground truth for whether a credential request is appropriate, as this is a subjective decision that depends on the specific website, use case, and user preference.
Thus, we nudge users by showing them expert recommendations and user opinions, as this was shown to be effective in similar support systems~\cite{acquisti2017nudges}.
However, collecting expert recommendations and user opinions for each website individually is not scalable: thus, we design the \name{} to work at the \textit{website category} level instead, with the assumption that websites in the same category will share use cases and require similar data.
Finally, even a single website may provide different services, and might request credentials in different contexts: as it would not be scalable to collect data for all contexts, we design the \name{} to give context-free recommendations.
While this might reduce its effectiveness (as context is important for users' disclosure decisions~\cite{nissenbaum2004privacy}), we study the most challenging setting where users have no context (e.g., when credentials will be requested at registration time, similarly to current Single Sign-On systems), as this is likely to be the case where users make the most mistakes. We show an example of the \name{} in Figure~\ref{fig:survey_image_scenario}.

\subsection{Research Areas}

\paragraph{A1: Data Collection Requirements}

We first investigate how data can be collected for the \name{}.
As discussed, we aim to display expert recommendations or user opinions and operate at a higher level, i.e., the website category.
However, if users decide very differently for websites within a category, this indicates that users make website-by-website decisions, and showing information based on only the category is unlikely to be helpful. Thus, we investigate whether there are significant differences in disclosure decisions for websites within the same category and whether these differences can be explained by other factors such as the website's size or location. Furthermore, we investigate whether user opinions can be collected through usage data or if surveys/focus groups are better suited. Lastly, we investigate if there is an expert consensus on disclosure recommendations. 

\paragraph{A2: Effectiveness of the \name{}}

Decision support systems have been studied and shown to be effective in multiple contexts~\cite{wash2018provides, acquisti2017nudges, besmer2010impact, goecks2009challenges, liu2016follow, acquisti2015privacy, acquisti2012impact, kelley2009nutrition}. However, previous approaches cannot be directly applied to the EUDI, as they either rely on displaying users' usage data (which likely contains mistakes) or are not scalable enough for the EUDI. Thus, we propose our \name{} as a scalable assistant for EUDI disclosure decisions. 
We investigate whether the \name{} is effective at reducing disclosure mistakes, and what the risks of displaying false information would be. 
Furthermore, we investigate design considerations, such as at what confidence (i.e., percentages displayed) the \name{} is effective and if different demographic groups have different preferences for the \name{}. 
Lastly, we investigate what would happen if the \name{} gets displayed besides website-provided purposes for credential requests. 

\paragraph{A3: (Over)sharing Patterns}

Oversharing online has been shown in many contexts~\cite{acquisti2017nudges, egelman2013choice, hann2002online, acquisti2013privacy, brandimarte2013misplaced, cvrcek2006study, norberg2007privacy, taddicken2014privacy}, and is likely to be a problem for the EUDI as well.
As signed digital versions of official documents are a new concept, we investigate whether users are more careful with such credentials (as they are familiar with their physical equivalents, compared to more abstract concepts such as cookies) and what misconceptions or oversharing patterns exist. 

\subsection{Study Methods}
\label{subsec:study_methods}

\paragraph{Survey}

We designed a survey of users and legal, cybersecurity, policy, and ethics experts to understand which credentials they find sensible to disclose to different websites.
Additionally, the survey was designed to understand which credentials legal experts deemed necessary for legal compliance and the core functionality of the website, and which credentials other experts deemed unreasonable to disclose to this website. 
In the survey, users and experts saw a specific website name and some information about the website (its description, category, and traffic size and the country it is based in). They were asked to select which credentials they would disclose, which credentials were necessary to disclose, and which credentials were dangerous to disclose to the website from a list of credentials.

\paragraph{User Study}

We designed a user study to understand how users respond to credential requests with and without a \name{}. In the user study, users saw a specific website name and some information about the website (its description, category, and traffic size and the country it is based in), as well as a credential name and its attributes. Participants of the control group saw no \name{}, participants of the baseline group saw a standard \name{} that only showed correct information, and participants of the test groups saw a mix of the standard \name{} and a \name{} that showed mistakes or non-standard information. 
In select cases, users additionally saw a website-provided purpose for the credential request. Users were asked to decide whether they would accept the credential request.

\paragraph{Ground Truth} 
One challenge with measuring oversharing is that there is no ground truth, since disclosure decisions are subjective. 
However, in some cases, there is no valid reason for a website to request a credential, and in other cases, there is a use case that makes disclosure sensible. 
Based on this, 3 members of the research team categorized all credential-website category pairs and discussed the pairs for which there was no unanimous decision until one could be made. 
The possible categorizations were: 1) Justified, where a credential was useful for the core functionality of a website (e.g., a job platform collecting employment histories to show relevant job postings) or necessary for legal compliance (e.g., a bank collecting official identity cards for mandatory KYC requirements); 
2) Unjustified, where there is no use case that made disclosing the credential to the website sensible (e.g., a news website requesting users' health insurance information); 3) Uncertain, where the researchers could come up with possible use cases, but they were minor or improbable (e.g., there may be a case where a government website collects users' diplomas for benefits). 
We then defined undersharing as not disclosing when justified, and oversharing as disclosing when unjustified. We validated our categorization with the expert survey results (details in \cref{sec:app_justification}).

\paragraph{How We Study Our Research Areas}

For A1, we aim to understand three things: (i) Is it sufficient to show information based on the website category?
(ii) Is it better to collect user opinions through usage data or surveys/focus groups? (iii) Is there sufficient consensus among experts to collect expert opinions with moderate effort? 
For (i), we use the results of the user study and evaluate whether users' disclosure decisions differed within a website category for factors that users care about according to prior work (i.e., the country the website is based in~\cite{riefa2017challenge, 2025_consumer_scoreboard, world_e-commerce} and the traffic size~\cite{bart2005drivers, jarvenpaa2000consumer}). 
We then analyze the standard deviation of credential disclosure rates for all credential-website category pairs. If differences are low, users likely base their decision on the website category, and not on individual websites. 
For (ii), in the unjustified cases, we compared how often users in the control group of the user study accepted a credential request for a website with how often users in the survey chose that credential for the same website.
If we see increased oversharing in the user study, this indicates that collecting data through surveys/focus groups is likely to yield more accurate results.
For (iii), we use the results of the expert survey and compare the decisions of different experts. We compute the average number of expert responses needed for high-confidence results and compute the expert disagreement to understand when this may be insufficient.

For A2, we compare the control, baseline, and test groups of the user study to assess whether the \name{} can effectively support users and how they react to different versions of the \name{}. 
We first compare the control (no \name{}) and baseline (with standard \name{} showing correct information) groups to assess whether the \name{} can reduce the number of disclosure mistakes that users make. Then we compare the control group and the test group with a \name{} that shows mistakes to understand if polluted data could mislead users into making more mistakes. 
Then we compare the control group, baseline group, and the test groups with different \name{} confidence tiers (percentages for user opinions) to understand if the confidence of the \name{} affects its effectiveness. Then we compare how the decisions of different demographic groups change for different \name{} versions to understand if there are demographic preferences. Lastly, we compare the control group, baseline group, and the test groups for different website-provided purposes to understand if it could be helpful to allow websites to display a purpose for the credential request in the EUDI Wallet.

For A3, we use the data collected to develop the \name{} to study oversharing. This means that we study oversharing in the same setup as our \name{} (i.e., through website categories and without context). This makes sense as users may not always have context, and context is less important for oversharing (i.e., if a credential is unnecessary for the website and a user still discloses, there is likely a context in which the user overshares). We use the results from the user survey and compare them with the ground truth that we established. For any credential-website category pair where the ground truth is that there is no justification for disclosure, but a high percentage of users chose the credential, oversharing is likely to occur. We also analyze if there are any oversharing patterns that indicate common misconceptions.
\section{Experimental Setup}
\label{sec:setup}
In this section, we provide a detailed description of how the survey and user study were designed and conducted. Furthermore, we describe the data we collect and the data analysis methods we use.

\subsection{Survey on EUDI Credentials}

\paragraph{Credential Choice}

We chose 14 credentials based on the W3C verifiable credential specification~\cite{W3C_VC_Use_Cases} and the use cases in the EUDI ARF~\cite{EUDI_ARF}. 
As the attributes of the credentials have not been decided, we chose a set of attributes for each credential, which generally match current physical documents, with removed redundancies. For example, most documents contain the name of the owner; however, the EUDI may be different, so we removed this redundancy.

\paragraph{Website Choice}

We chose website categories based on Cloudflare Radar's domain categories~\cite{Cloudflare_Radar_Taxonomy}. We used 17 website categories for users and 15 categories for experts that covered a wide range of services. For experts, in each category, we chose one website for Germany, one for France, and one for Italy, such that the website in each country is comparable. Thus, each expert saw one website for each category.
For users, each category had 6, 9, 10, or 18 websites for a total of 166 websites. Where possible, the websites covered different traffic sizes, and were Europe-based unless all well-known websites of a category came from outside of Europe. For 3 website categories, we used websites from Europe, the USA, and China to investigate whether the country mattered to users. Each user received 20 websites to evaluate: one for each category, and the remaining 3 were randomly chosen proportionally to the category size. There were multiple websites in each category to control for websites for which people have strong positive or negative opinions.

\paragraph{Participants}

27 experts completed our survey: 3 law, 15 cybersecurity, 5 policy, and 4 ethics. We recruited experts by reaching out to them directly, and no payment was provided. To protect the identity of the experts, we did not ask demographic questions.
Experts were defined as researchers who completed a PhD (e.g., professors, senior researchers, postdocs) or practitioners who work in a relevant field (e.g., lawyers, policy makers).
Assuming a worst-case p of 50\%, the number of experts required for $\pm$20pp and 95\% confidence interval is 24. We use a lower precision, as the experts' responses were mainly used to establish ground truth and to estimate how many experts would need to be consulted. Furthermore, this number aligns with expert studies that often use 10-30 experts\cite{alarabiat2019delphi, belton2021delphi}.

1107 users completed our survey. Participants were recruited through the Prolific platform and were paid $\sim$\$6.7 for the 25 min survey. Participants were chosen with a balanced gender and age split, and were required to reside in Germany, Italy, or France. We removed submissions that were incomplete, in which the participant revoked consent, or that were of low quality with an exceptionally fast completion time. After this, we were left with 1035 valid responses.
Even though we balance categories, websites in larger categories are sampled less frequently. Thus, in the worst case of a category having 18 websites, each website is seen on average 7.1\% of the time, whereas for 10 it is 11.6\%, and for 6 it is 18.1\%. The number of participants needed for $\pm$10pp at the 95\% confidence interval assuming a worst-case p of 50\% is 1'352 for categories with 18 websites, 827 for categories with 10 websites, and 530 for categories with 6 websites. The categories with 18 websites were important, as they measured attitudes towards different regions. Thus, we targeted 1'000 valid responses, in order to satisfy all categories except for the largest without overpowering smaller categories. With this choice, we sample $\sim$73-187 responses per website.

\paragraph{Survey Flow}

\begin{figure}[tb]
     \centering
     \begin{subfigure}[b]{\linewidth}
        \centering
        \includegraphics[width=0.9\linewidth]{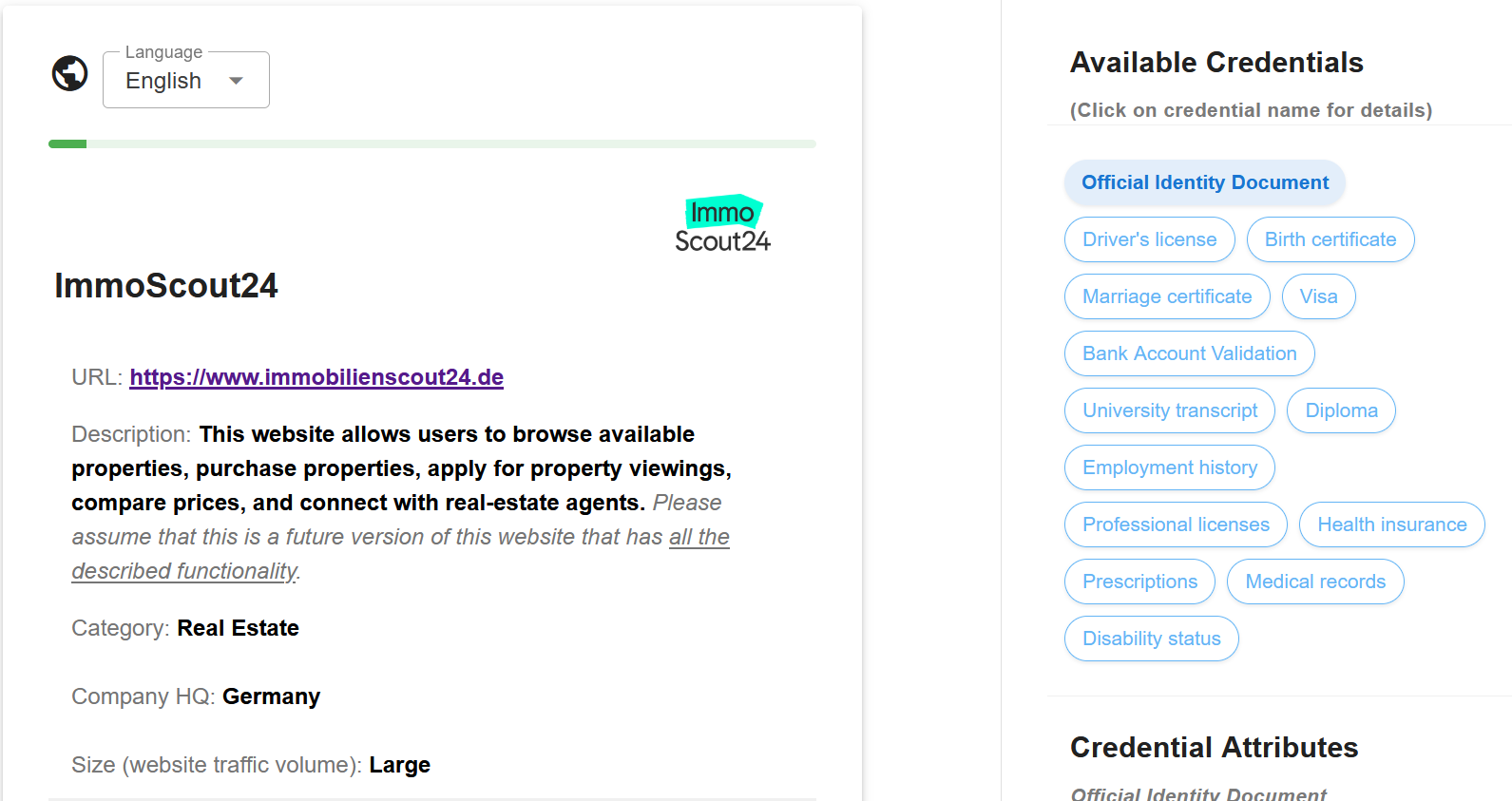}
        \caption{Website description and credentials in the sidebar.}
        \label{fig:survey_image_website_task_website_description}
     \end{subfigure}
     \hfill
          \begin{subfigure}[b]{\linewidth}
         \centering
        \includegraphics[width=0.9\linewidth]{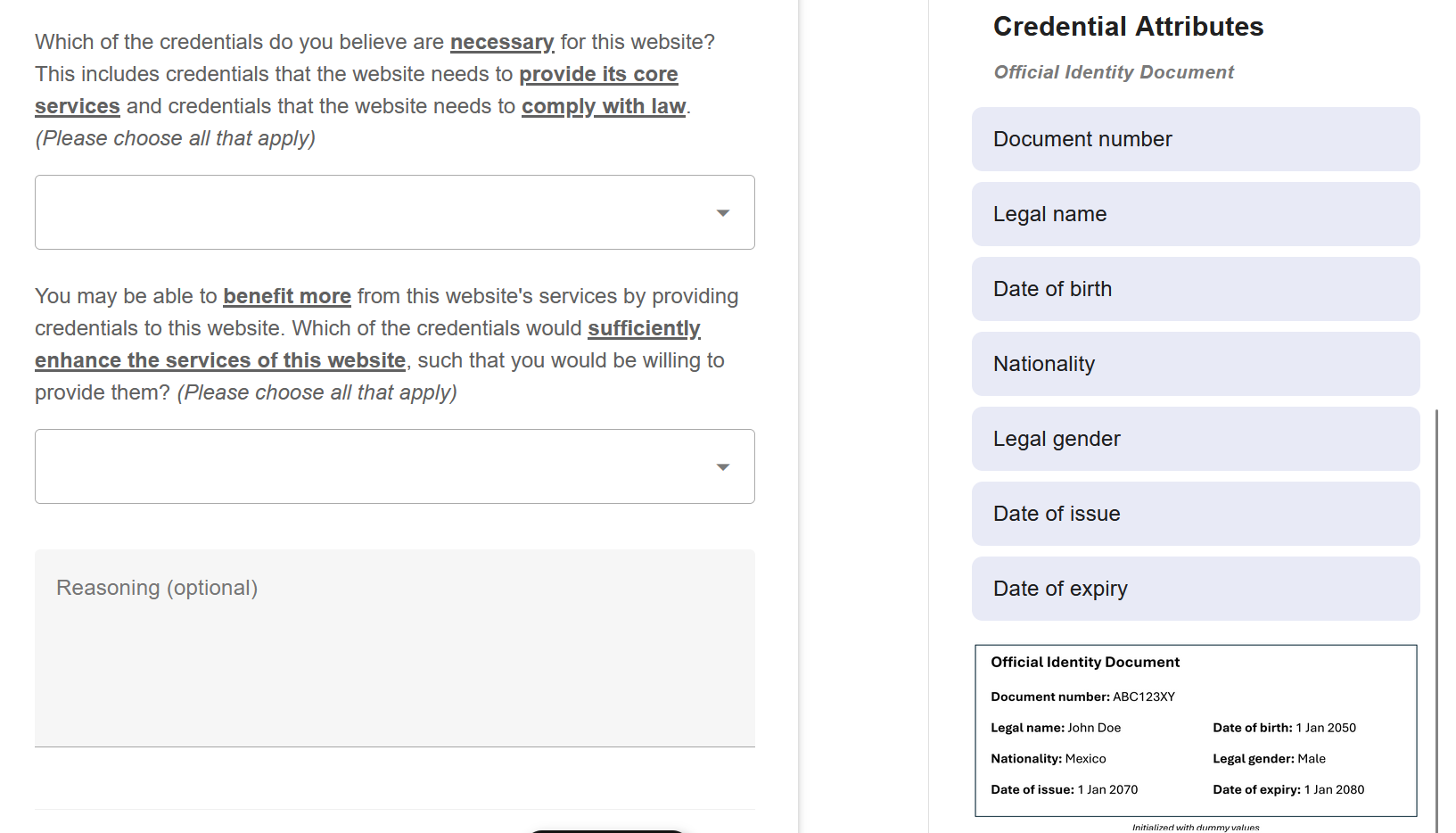}
        \caption{Survey questions and the credential attributes in the sidebar.}
        \label{fig:survey_image_website_task_law}
     \end{subfigure}
    \caption{Survey screens of the website evaluation task (more in \cref{sec:detailed_survey_images}). The sidebar shows details on the credentials.}
    \label{fig:survey_image_website}
\end{figure}

The survey was provided in English for experts and in English, German, French, and Italian for users.
After filling out a consent form and (users only) answering a brief demographic questionnaire, participants completed the following tasks.
First, (users only) performed a \textit{credentials evaluation} task: they were asked to pick which of the credentials they own(ed) as physical or electronic official documents. After that, for each document they picked, they were asked how often they use that document and how sensitive that document is to them. As the privacy paradox has shown that users often act differently than they think they should act~\cite{kokolakis2017privacy, dienlin2023longitudinal, taddicken2014privacy}, we decided to split the sensitivity question into two questions: 1) How comfortable are they with sharing this document? 2) When would they share the document even if uncomfortable?

Afterwards, experts and users performed a \textit{website evaluation} task: they were first shown a short explanation about what the EUDI Wallet is and how EUDI credentials are planned to look and function. Then they were shown different websites. For each website, participants were asked to choose: [law experts, users] the credentials that are necessary for the website, [cybersecurity, policy, and ethics expert] the credentials that should never be requested by the website, and [all] the other credentials that they would disclose to the website. For users, the credentials they chose for the first and second question were combined as credentials that they would disclose to the website. This approach is justified as users generally are willing to disclose credentials that they deem necessary, and the second question specifically asks about credentials that they would disclose. Participants could access details on all credentials in a sidebar. \cref{fig:survey_image_website} shows the survey screens.

\subsection{User Study on the \name{}}
\label{subsec:setup_study}

\paragraph{Scenario Choice}
A scenario contained a website, a requested credential, and, in select cases, a website-provided purpose. The scenarios were grouped into 6 sets, S1-S6, each of which was designed for a different test group. We ranked each credential's sensitivity based on the user survey responses and made a balanced choice of credential sensitivities for the scenarios in each set. S2/S4 contained a website-provided purpose. In S5/S6, the \name{} displayed both expert recommendations and user opinions, while only showing one of the two otherwise. S2/S3/S5 used unjustified, S4/S6 used justified, and S1 used a mix of justified, unjustified, and uncertain website-credential pairs. Thus, we say that choosing yes in S2/S3/S5 and choosing no in S4/S6 is a mistake. S1 tested different \name{} confidence tiers (i.e., displayed percentages).

\paragraph{Test Groups}
Participants were assigned to the control, the baseline, or the test group and saw 20 scenarios in total. Participants in the control group did not see the \name{}. Participants in the baseline group saw a \name{} that always had standard confidence, made no mistakes, and did not contradict a website-provided purpose. Half of the baseline group saw specific expert types in the \name{}, while the other half saw an unspecified expert. The test group was divided into 16 classes that each tested one condition in which the \name{} showed unexpected behavior, i.e., low confidence, very high confidence, mistakes in the displayed information, or displayed information at odds with the website-provided purpose. Since it would not make sense for the \name{} to always show unexpected information, participants in the test group saw 2 or 4 test scenarios, and otherwise baseline scenarios. A detailed overview of the scenarios and test groups is shown in \cref{sec:app_deteils_groups}.

\paragraph{Participants}

1024 participants completed our user study. Participants were recruited through the Prolific platform and were paid $\sim$\$6.7 for the 25 min study. Participants were chosen with a balanced gender and age split and were required to reside in the EU. We removed submissions that were incomplete or in which the participant revoked consent. After this, we were left with 1002 valid responses. In most cases, the probability of a user being assigned to a test group was 5\%, and each participant provided two responses for the group, giving an effective response rate of 10\%. Thus, assuming a worst-case p of 50\%, for $\pm$10pp with 95\% confidence interval, the number of responses needed is 970, which we aimed for.

\paragraph{User Study Flow}

\begin{figure}[tb]
    \includegraphics[width=0.9\linewidth]{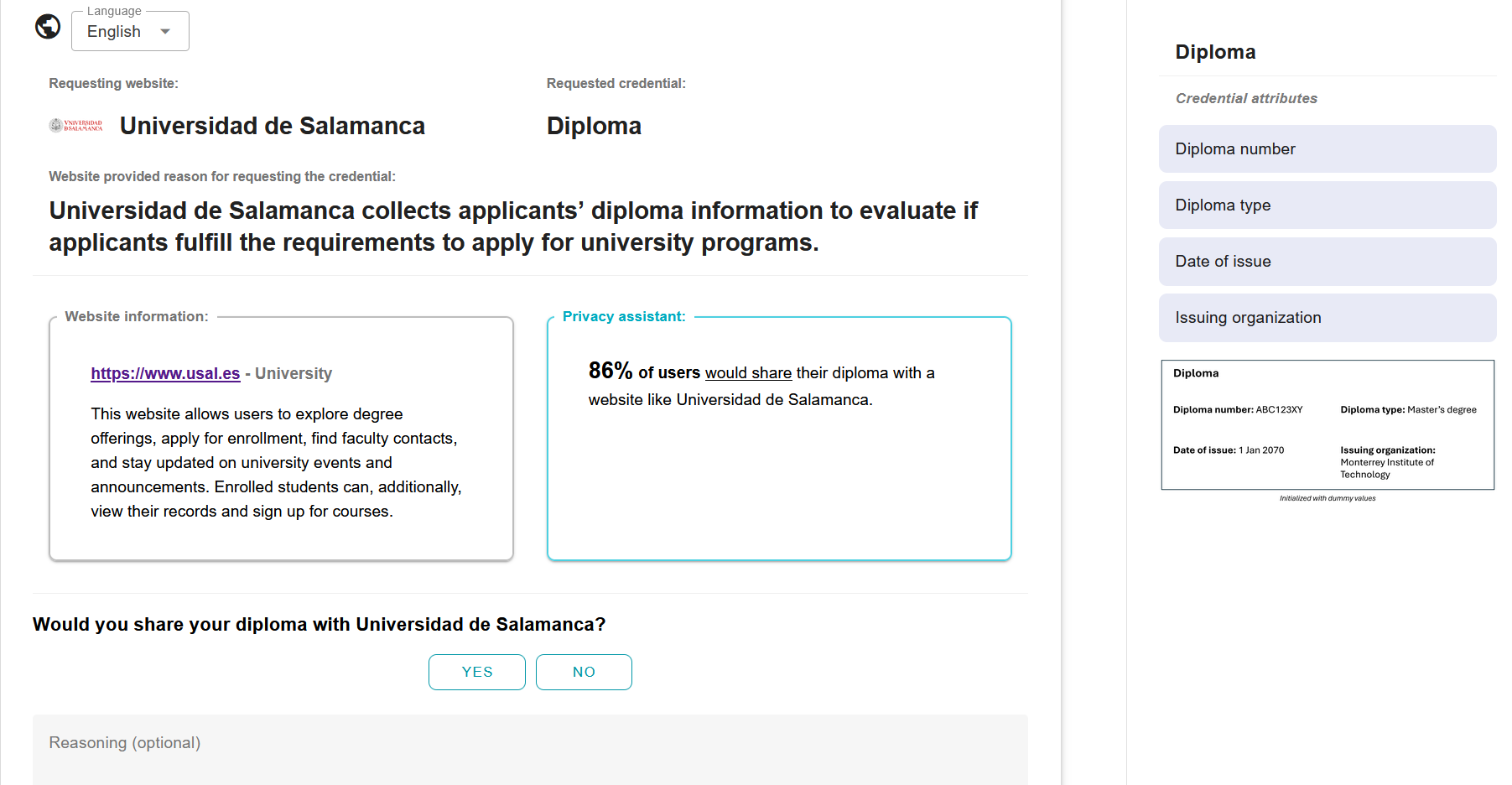}
    \caption{Study screen for the credential disclosure task with the \name{} and a purpose (more in \cref{sec:detailed_survey_images}). The sidebar shows the attributes of the requested credential.}
    \label{fig:survey_image_scenario}
\end{figure}

The user study was provided in English, German, French, Italian, and Spanish.
We used a commitment check at the beginning of the study, which has been shown to increase response quality~\cite{Qualtrics_Commitment}.
After filling out a consent form and answering a brief demographic questionnaire, participants completed a \textit{credential disclosure task}: they were first shown a short explanation about what the EUDI Wallet is and how EUDI credentials are planned to look and function. 
Then, participants (except for the control group) were introduced to the \name{}. After that, participants were shown multiple simulated scenarios of a website requesting a credential. 
Users did not have to navigate a real or simulated website. The task was to decide whether they would accept or reject the request. 
As the EUDI Wallet does not exist yet, we used a UI based on similar credential requests (e.g., OAuth). \cref{fig:survey_image_scenario} shows the study screens. 

\subsection{Data Analysis}

\paragraph{Collected Data} 
In the expert survey, for 15 website categories and 14 credentials, we collect whether the expert would recommend disclosing the credential with the website. Furthermore, for legal experts, we collect whether they believe the credential is necessary for the core functionality of the website or for legal compliance, and from the other experts, whether the credential should never be disclosed to the website.
In the user survey, for 166 websites divided into 17 categories and 14 credentials, we collect whether the user would disclose the credential to the website. Furthermore, we collect for each credential whether the user owns the credential, how frequently they use the credential, and how sensitive they feel the credential is. Lastly, we collect demographic data.

In the user study, for simulated scenarios of a website requesting a credential, we collect whether the user accepts the request. The scenarios are split into groups testing: low- and high-confidence recommendations, purpose statements, correct and wrong \name{} information, single and multi-statement \name{}. Furthermore, we collect demographic data.

\paragraph{Analysis Methods}
For the user survey, we computed the percentage of users who chose a credential for each website, website category, country, and traffic size. For the expert survey, we computed the percentage of non-law experts who would disclose a credential to each website category and the percentage who would recommend never to disclose that credential to that website. We also compute the percentage of law experts who believe a credential is necessary for a website category. For the user study, we computed the percentage of users who chose to yes for a credential request in each test, control, and baseline scenario, and aggregated the numbers over the set of equivalent scenarios (i.e., S1-S6).

Significance was computed using the Chi-squared test if a sufficient number of responses were collected and the groups to compare were similar in size. Otherwise, Fisher's exact test was used. The number of experts needed was computed using the observed expert-to-expert inter-class correlation and used a $\pm$5 percentage-point precision at 95\% confidence. Disagreement among experts was computed as the normalized variance of the number of credentials experts would disclose to websites in each category.
\section{Results}
\label{sec:results}

In this section, we discuss the results of our survey and user study described in \cref{sec:setup}.
Additional results are shown in \cref{sec:app_additional_results}. 
As mentioned in \cref{sec:mr}, we categorized website-credential pairs into justified, unjustified, and uncertain depending on whether there is a valid reason for the website to request the credential. We validated our categorization with the results of the expert survey (details in \cref{sec:app_justification}), and find that they match in most cases.

\subsection{A1: Data Collection Requirements}\label{results:A1}
We first evaluate whether collecting information at the website category level is sufficient, and then evaluate requirements and challenges when collecting user and expert opinions.

\paragraph{Impact of Country and Traffic Size}

We evaluated how often user survey participants would disclose a credential to a Europe-based compared to a USA- or China-based website, and how often they would disclose a credential to a high-traffic (large) compared to a low-traffic (small) website (details in \cref{subsec:app_additional_figures}).
The results show limited differences between users' disclosure decisions for different countries and traffic sizes, with only a few entries showing a significant difference when using the Chi-squared test. 
Thus, even though some users mentioned considering the country (e.g., "Due to its connections to China, I would not entrust any data to this website", "I do not want to give any data to the USA, nor to its subsidiaries"), on average, users are unlikely to base their disclosure decisions on the country or traffic size. This is surprising, as surveys show that EU consumers feel more confident purchasing from their own country or the EU~\cite{riefa2017challenge, 2025_consumer_scoreboard, world_e-commerce}, and prior research has shown that reputation and perceived size matter for website trust~\cite{bart2005drivers, jarvenpaa2000consumer}.

\paragraph{In-Category Differences Between Websites}

\begin{figure}[tb]
    \centering
    \includegraphics[width=0.9\linewidth]{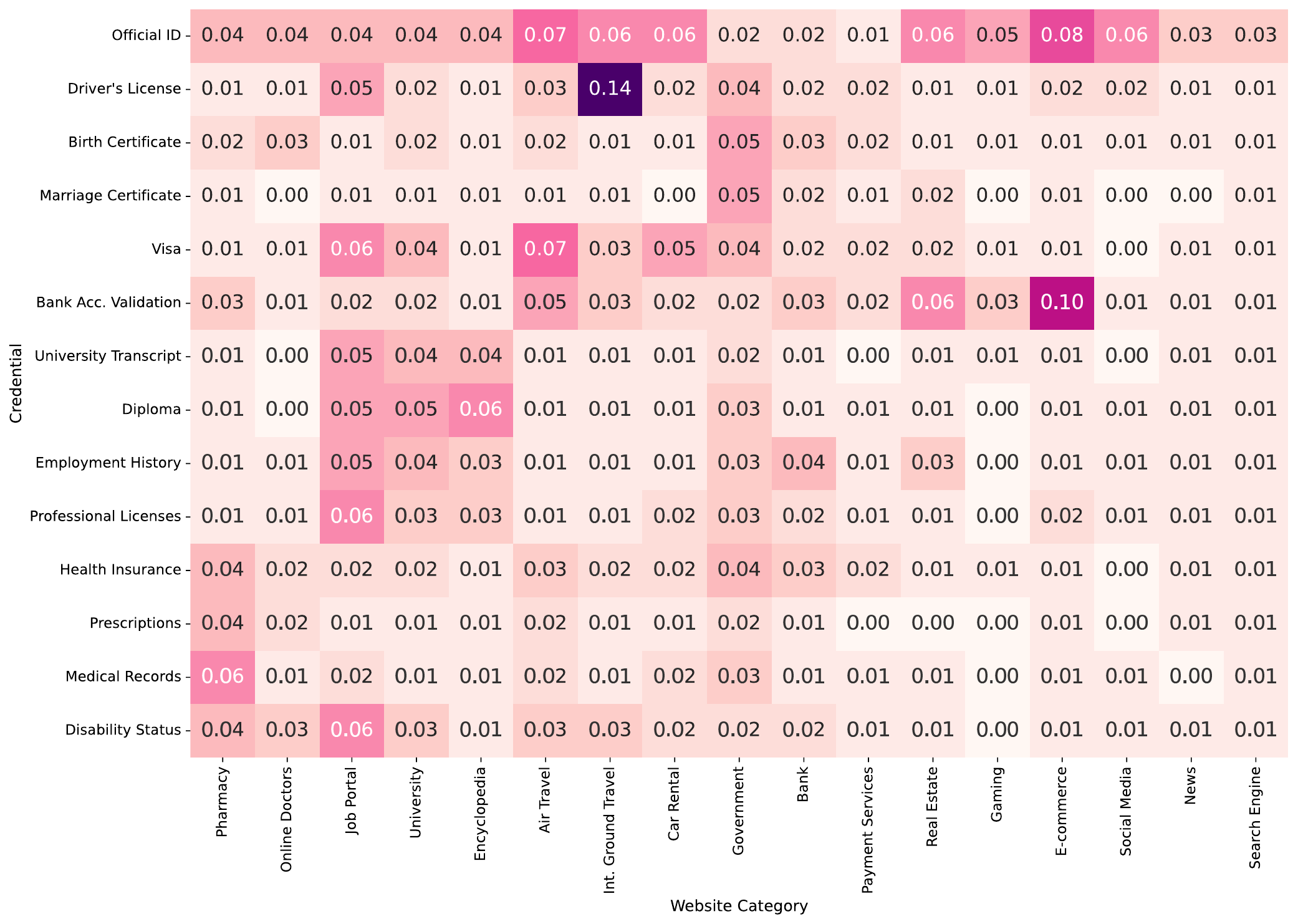}
    \caption{User survey results showing the standard deviation of the percentage of participants who would disclose a credential between different websites within a category.}
    \label{fig:standard_deviation_heatmap}
\end{figure}

\cref{fig:standard_deviation_heatmap} presents the results of the user survey showing the standard deviation of the percentage of users who would disclose a credential between websites in a category. 
A high standard deviation indicates that users may require support on an individual website level. 
The results show that the standard deviation is low for most credential-website category pairs, indicating that most disclosure decisions are based on the website category. However, there are two notable exceptions. 1) The standard deviation for driver's license-international ground travel is high (0.14). This comes from BlaBlaCar, which has a bus and a ride-sharing service. Although we asked participants about the bus services (clearly stated in the survey), many participants may have thought about the ride-sharing service, for which a driver's license is needed to sign up as a driver. 2) The standard deviation for official ID-e-commerce is high (0.8). This is likely due to some e-commerce websites selling age-restricted items (for which an official ID may be needed for age verification) while others do not. Thus, a \name{} that displays information based on the website category is likely sufficient; however, multi-service websites may need to be handled separately, and a tailored website categorization for the EUDI may be needed.

\paragraph{Collecting User Opinions}

\begin{figure}[tb]
    \centering
    \includegraphics[width=0.9\linewidth]{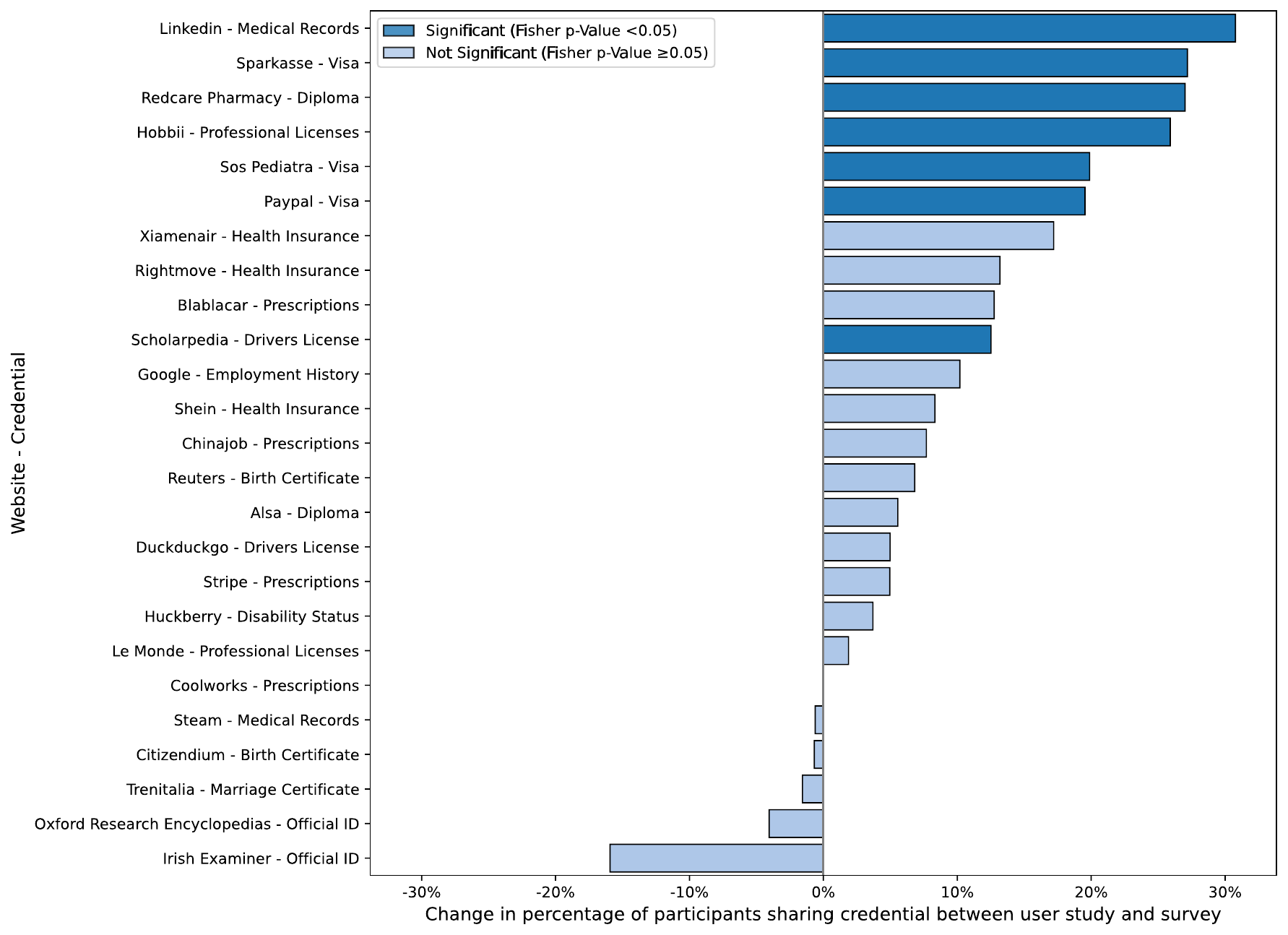}
    \caption{Comparison of how many participants said they would disclose a credential for a website in the user survey, as compared to how many control group participants in the user study disclosed the credential to the website. Positive percentages indicate increased disclosure in the user study. For all website-credential pairs, disclosure is unjustified.}
    \label{fig:chart_overshare_no}
\end{figure}

For all user study scenarios where disclosure is unjustified, we compared the responses of the control group with the number of users who said they would disclose the same credential to the same website in the user survey.  
The results of this comparison are presented in \cref{fig:chart_overshare_no}, and show that in most cases, more users disclosed a credential when responding to a request, in many cases significantly so. Examples include: $\sim$31\% of users disclosed their medical records to LinkedIn as compared to $\sim$0\% of users saying they would do so, $\sim$29\% of users disclosed their diploma with Redcare Pharmacy as compared to $\sim$2\% of users saying they would do so, and $\sim$24\% of users disclosed their visa with Sos Pediatra as compared to $\sim$4\% of users saying they would do so. 
This indicates that users are more likely to overshare when responding to a request. Thus, user opinions should be collected through surveys and user focus groups instead of usage patterns to ensure the \name{} does not propagate oversharing.
We note that, in practice, users may be even less likely to reject credential requests, as websites may use dark patterns~\cite{mathur2021makes, mathur2019dark, nouwens2020dark, luguri2021shining} or refuse to provide services. Prior work has shown similar effects, such as that users act less privacy conscious than they state they are (the privacy paradox)~\cite{kokolakis2017privacy, dienlin2023longitudinal, taddicken2014privacy}, and that users usually fill out all fields, including optional fields, in online forms~\cite{krol2016control, preibusch2013privacy}. However, to the best of our knowledge, increased oversharing when responding to requests has not yet been explored. 

\paragraph{Number of Experts Needed}

\begin{figure}[tb]
    \centering
    \includegraphics[width=0.9\linewidth]{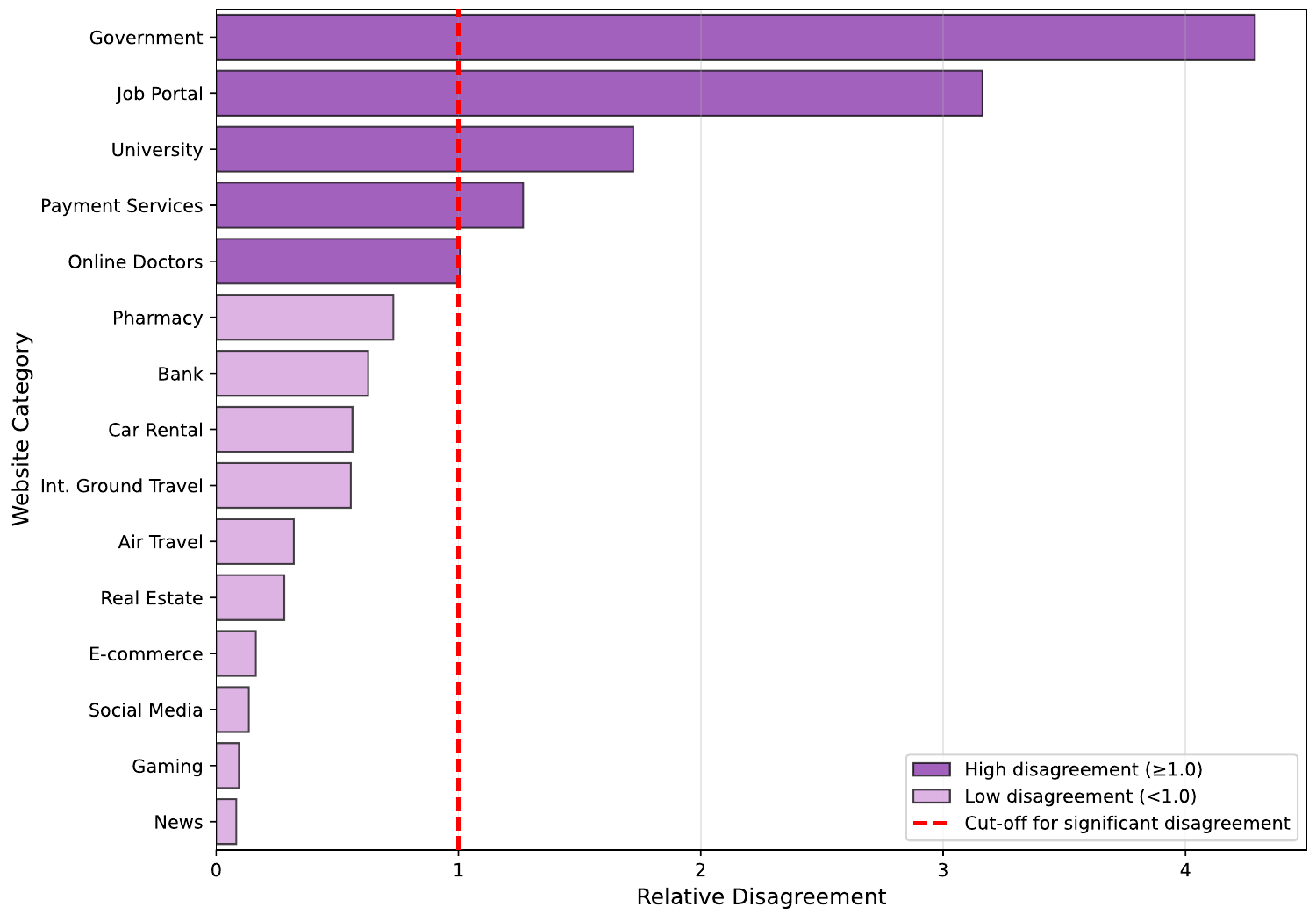}
    \caption{Expert survey results showing the disagreement among experts for each website category. Disagreement was computed by normalizing the variance of the number of credentials experts would disclose to websites in each category.}
    \label{fig:expert_disagreement}
\end{figure}

Using the observed expert-to-expert inter-class correlation of $\sim$0.057, the number of experts needed on average across all website categories and credentials for $\pm$5 percentage-point precision at 95\% confidence is 24. 
However, for some categories, expert disagreement is high, as shown in \cref{fig:expert_disagreement}. For websites with high disagreement, it is unlikely that experts will converge on a majority opinion. For such categories, a larger panel of experts (100+, depending on the desired accuracy) may be needed to get an accurate representation of the expert recommendation. 

\paragraph{Expert Consensus}

\begin{figure}[tb]
     \centering
    \includegraphics[width=0.9\linewidth]{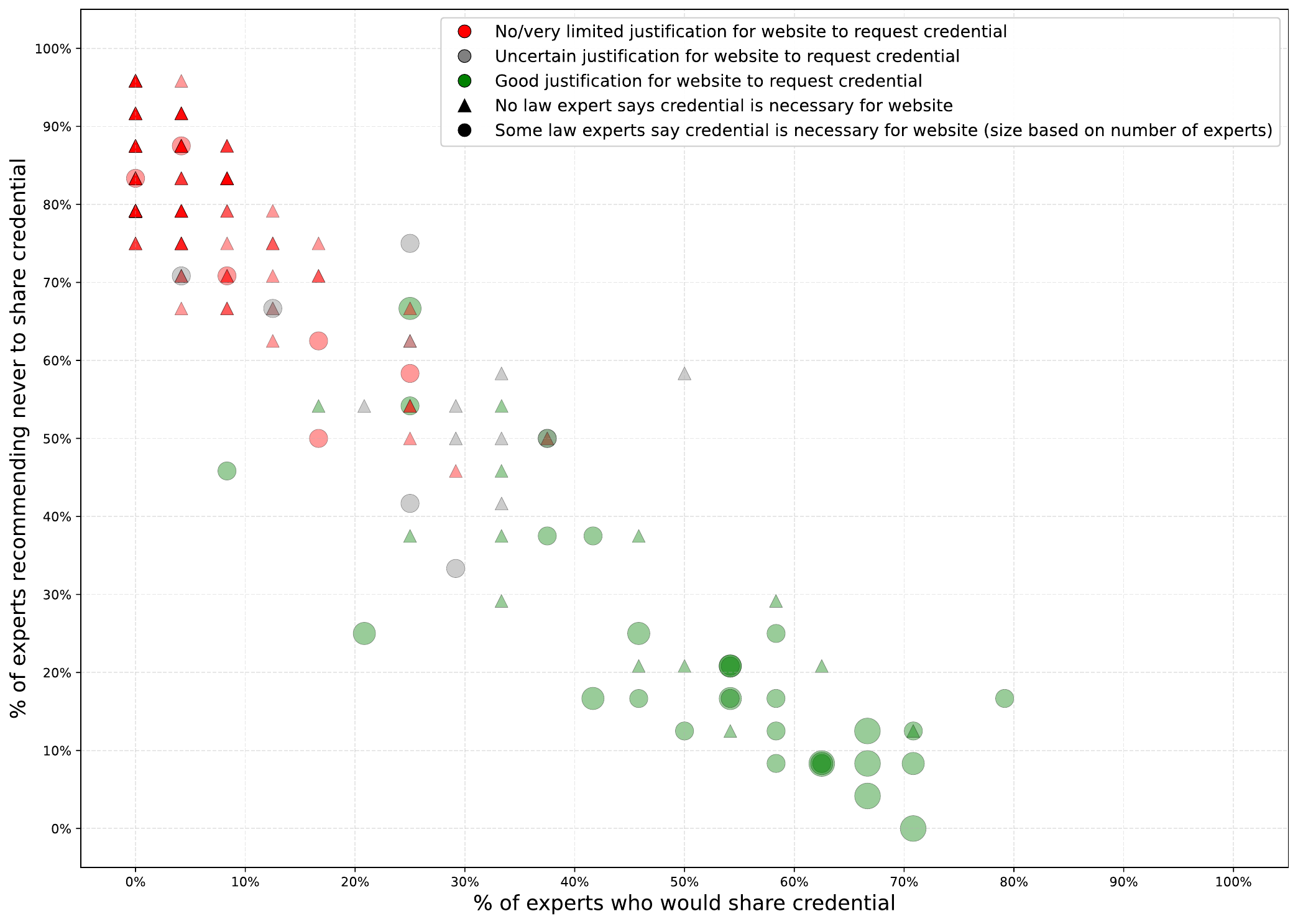}
    \caption{Expert survey results showing how many experts would disclose a credential to a website vs. how many experts recommend never to disclose that credential to that website.}
    \label{fig:alignment_scatter}
\end{figure}

To better understand expert consensus, we analyze expert responses for justified, unjustified, and uncertain credential-website category pairs.
\cref{fig:alignment_scatter} presents the results from the expert survey showing the percentage of experts who would disclose compared to the percentage who would recommend never to disclose a credential to a website. Additionally, it shows the number of law experts who state it is necessary to disclose the credential to the website. 
The figure shows that for unjustified pairs, expert opinion is largely aligned, with a high percentage of experts recommending never to disclose, and a low percentage of experts willing to disclose a credential. However, for justified pairs, responses are more mixed, with in many cases, $\sim$40-60\% of experts stating that they would disclose the credential, and mixed results on whether the credential is necessary. This indicates that expert recommendations will be effective at warning users against dangerous disclosure decisions, but less effective at helping users decide when disclosure may be beneficial. This is unfortunate, as users may be uncertain in those cases, and prior research has shown that users lean on expert advice when uncertain~\cite{emami2018influence, meyers2020inducing, harvey1997taking, bailey2023meta, qi2018collaborator}. However, we note that more research is needed to understand if the low percentages for justified pairs come from the lack of context in our survey or from a general hesitancy of experts to recommend disclosure.
(Additional figures are in \cref{subsec:app_additional_figures})

\subsection{A2: Effectiveness of the \name{}}
In this section, we discuss whether the \name{} is effective at supporting users by looking at different \name{} versions and ways to display information.

\paragraph{Reduction of Mistakes}

\begin{figure}[tb]
     \centering
     \begin{subfigure}[b]{\linewidth}
         \centering
        \includegraphics[width=0.9\linewidth]{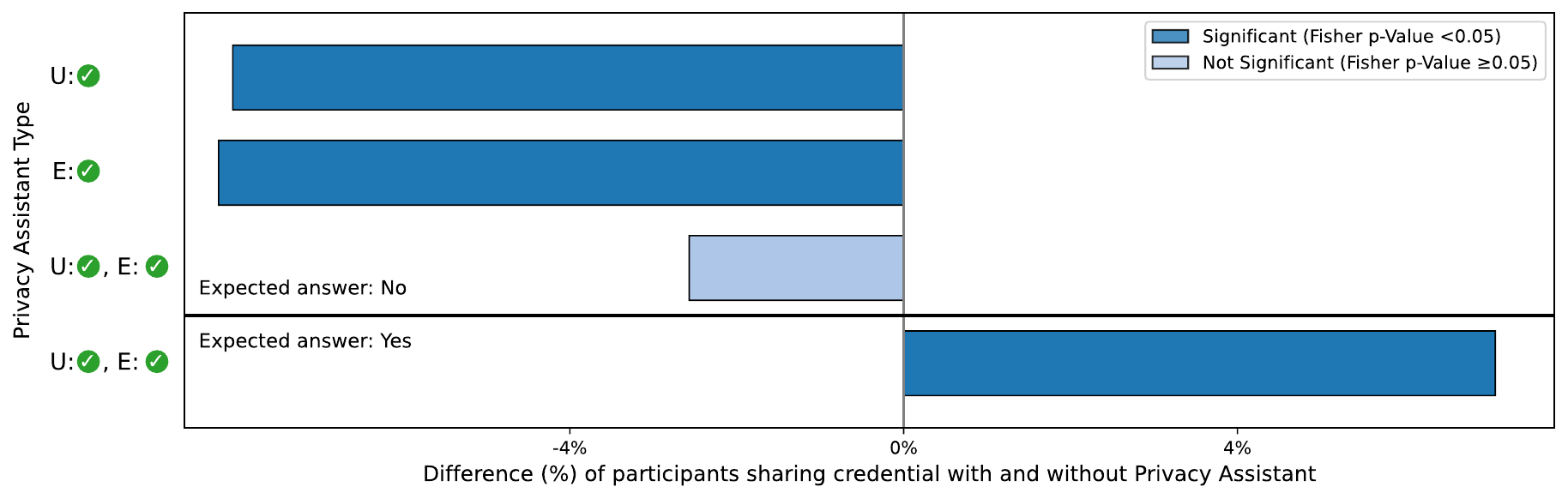}
        \caption{\name{} only shows correct information.}
        \label{fig:chart_mistake_reduce}
     \end{subfigure}
     \hfill
     \begin{subfigure}[b]{\linewidth}
         \centering
        \includegraphics[width=0.9\linewidth]{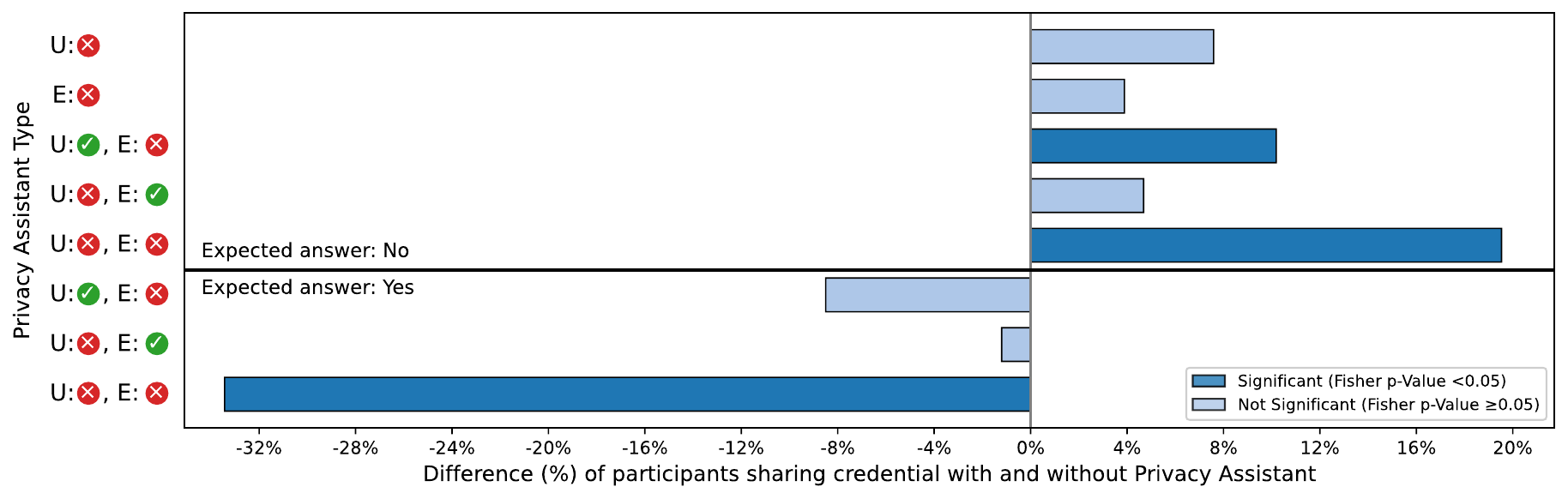}
        \caption{\name{} shows mistakes in the information.}
        \label{fig:chart_mistake_increase}
     \end{subfigure}
    \caption{User study results comparing the percentage of users disclosing a credential in the control group with the users with a \name{}. A positive number indicates more users with a \name{} disclosed the credential. The \name{} displays: U = user opinion, E = expert recommendation, \circledtick = correct data, \circledcross = mistake.}
    \label{fig:mistake_change}
\end{figure}

\cref{fig:chart_mistake_reduce} presents the results of the user study comparing the average percentage of users who disclose a credential with a website in the control group (i.e., no \name{}) with the users who saw the correct information in the \name{}. 
For 3 out of 4 designs, the results show a significant decrease (i.e., $p\leq0.05$ in Fisher's exact test) in the number of mistakes users made when receiving support from the \name{}.
In those cases, the average mistake rate dropped by $\sim$8\% (e.g., from $\sim$15\% to $\sim$7\% when seeing expert recommendations). We note that this occurred even though the \name{} was fully non-intrusive and showed the same display for all credentials regardless of sensitivity. Thus, even though a residual risk of $\sim$7\% may be unacceptably high, it is likely that more pinpointed and intrusive interventions will lead to a further reduction in the number of mistakes. These results are in line with prior research in other contexts, which has shown that users change their behavior based on nudges~\cite{wash2018provides, acquisti2017nudges, besmer2010impact, goecks2009challenges, liu2016follow, acquisti2015privacy, acquisti2012impact}. However, we add to this a specific focus on helping users avoid clear mistakes.

\paragraph{Potential for Misleading}
We expect that it is highly unlikely that mistakes get displayed to users if a tool like the \name{} gets implemented in the EUDI Wallet, since there would be a strong interest in quality. 
However, it cannot be fully ruled out that mistakes could occur, and thus, we tested whether mistakes in the \name{} could lead users to make more mistakes (i.e., herding effects). \cref{fig:chart_mistake_increase} presents the results of the user study that compares the average percentage of users who disclose a credential with a website in the control group (i.e., no \name{}) with the users who saw the \name{} displaying mistakes. While the users in the control group always made fewer mistakes, this effect was only significant (i.e., $p\leq0.05$ in Fisher's exact test) for 3 out of the 8 \name{} versions, 2 of which showed mistakes in both the user and expert information, which is highly unlikely. Removing those cases, the average increase in mistakes is around $\sim$5\% (e.g., from $\sim$9\% to $\sim$14\% when seeing wrong user information and the correct expert recommendation).

\paragraph{Impact of Confidence}

\begin{figure}[tb]
    \centering
    \includegraphics[width=0.9\linewidth]{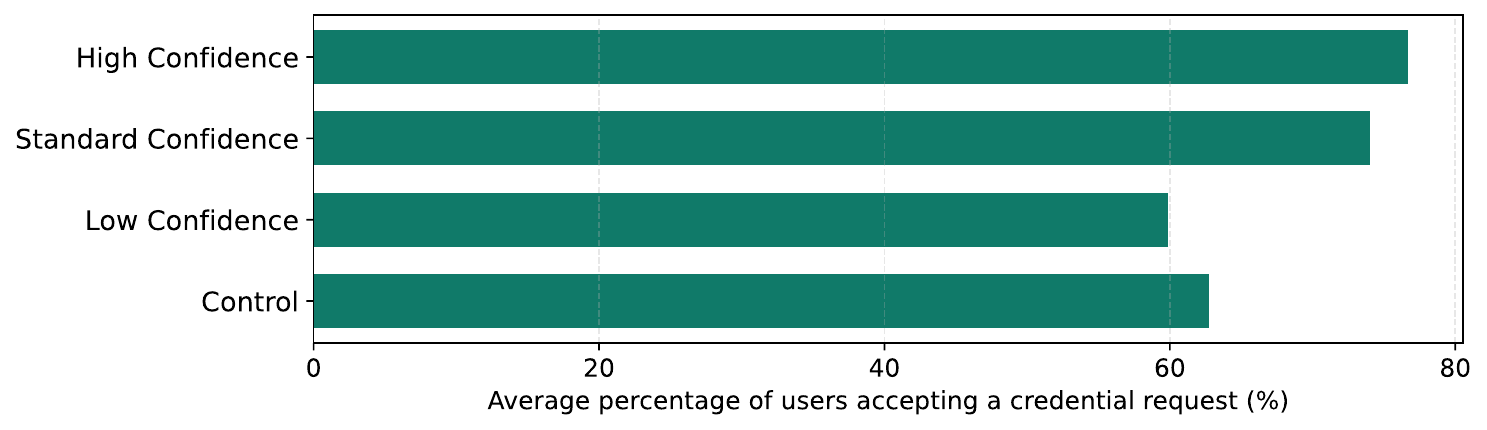}
    \caption{User study result for the average percentage of users who disclosed a credential for different \name{} confidence tiers. The control group had no \name{}. Low confidence = 51-55\% of users would disclose, standard confidence 81-85\% of users would disclose, and high confidence = 91-95\% of users would disclose.}
    \label{fig:confidence}
\end{figure}

\cref{fig:confidence} presents the user study results showing how many users, on average, disclosed a credential for different \name{} confidence tiers. Significance was computed with Fisher's exact test using the cut-off of $p\leq0.05$. The results show that the average percentage of users disclosing a credential is significantly higher ($\sim$74\% vs. $\sim$63\%) for the users who saw a \name{} with standard confidence (i.e., showing 81-85\%) than for those without a \name{}. Furthermore, the average percentage of users disclosing a credential is similar ($\sim$63\% vs. $\sim$60\%) for the users without a \name{} and those who saw a \name{} with low confidence (i.e., showing 51-55\%). Lastly, the average percentage of users disclosing a credential is similar ($\sim$74\% vs. $\sim$77\%) for the users who saw a \name{} with standard confidence and those who saw a \name{} with high confidence (i.e., showing 91-95\%). This indicates that the \name{} only has a nudging effect if the confidence is sufficiently high; however, the nudging effect does not increase beyond some confidence tier. This is in line with prior research in other contexts that has shown that higher confidence nudges have a greater impact~\cite{overton2025people, martin2002levels}

\paragraph{Demographic Differences}

\begin{figure}[tb]
     \centering
     \begin{subfigure}[b]{\linewidth}
         \centering
         \includegraphics[width=0.9\linewidth]{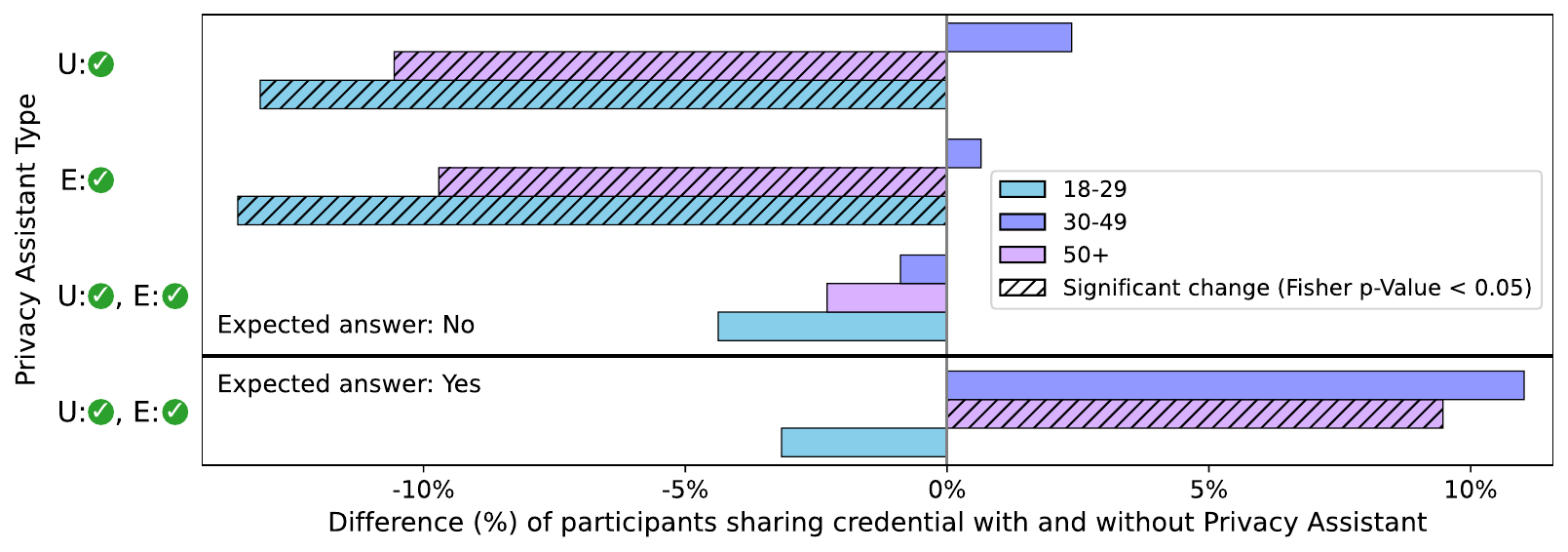}
         \caption{\name{} only shows correct information}
         \label{fig:mistake_reduce_by_age}
     \end{subfigure}
     \hfill
     \begin{subfigure}[b]{\linewidth}
         \centering
         \includegraphics[width=0.9\linewidth]{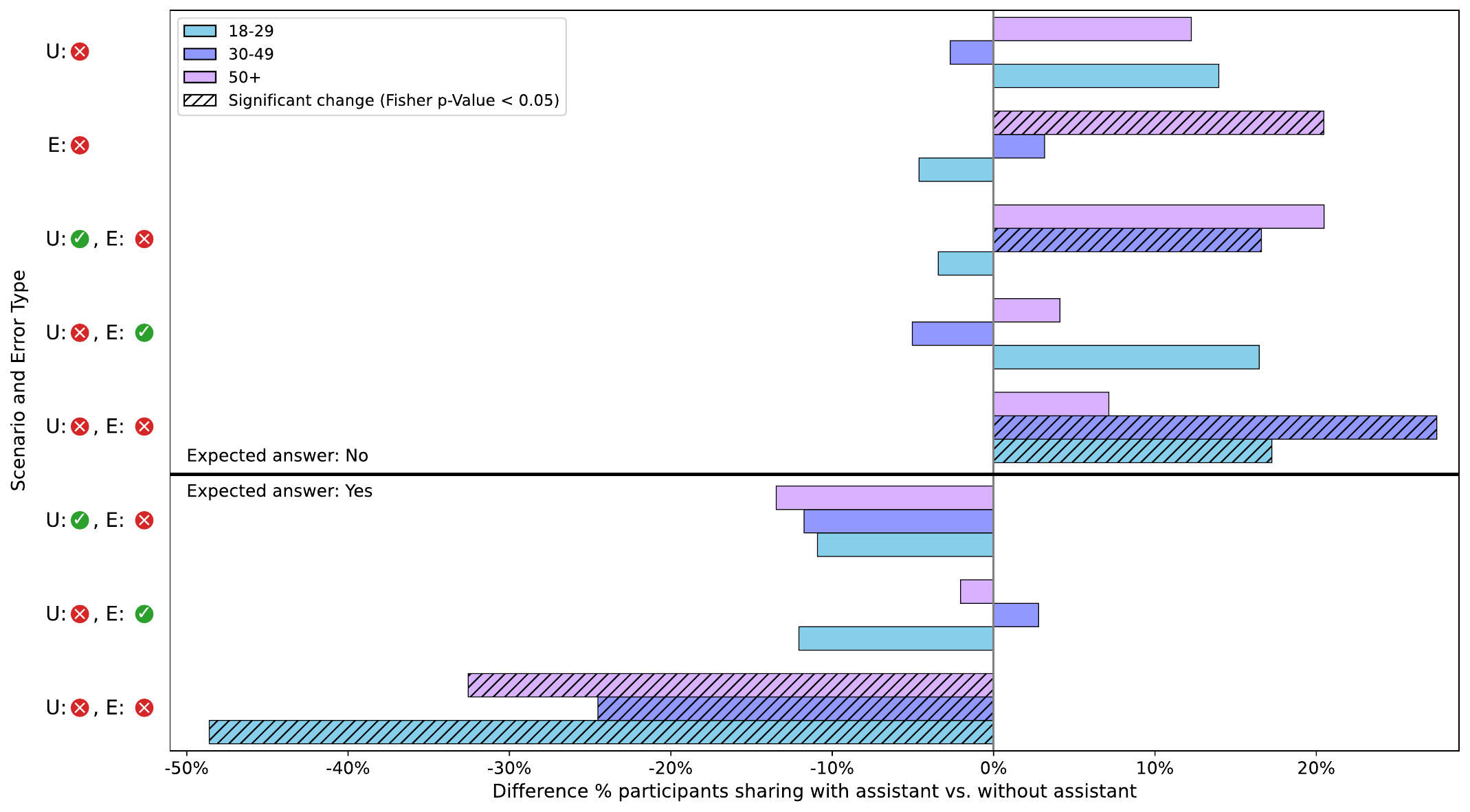}
         \caption{\name{} shows mistakes in the information}
         \label{fig:mistake_increase_by_age}
     \end{subfigure}
    \caption{User study results comparing the percentage of users disclosing a credential in the control group with the users with a \name{} for the age groups 18-29, 30-49, and 50+. A positive number indicates more users who saw a \name{} disclosed the credential. The \name{} displays: U = user opinion, E = expert recommendation, \circledtick = correct data, \circledcross = mistake.}
    \label{fig:mistake_change_by_age}
\end{figure}

\cref{fig:mistake_change_by_age} presents the user study results showing how the disclosure decisions of users in the 18-29, 30-49, and 50+ age groups changed based on the \name{}. Our results indicate that users in different age groups prefer different information. For instance, the \name{} showing user opinions had a stronger nudging effect on the 18-29 age group than the \name{} showing expert recommendations. However, for the 30-49 and 50+ age groups, the effect was opposite. This aligns with prior research showing that dependence on peers decreases and trust in experts increases with age~\cite{steinberg2007age, knoll2017age, wang2020expert}. Thus, tailored nudges based on demographics may be more effective, as has been proposed in some prior research~\cite{mills2022personalized, calboli2025ai}.

\paragraph{Impact of Website-Provided Purposes}

\begin{figure}[tb]
     \centering
     \begin{subfigure}[b]{\linewidth}
         \centering
        \includegraphics[width=0.9\linewidth]{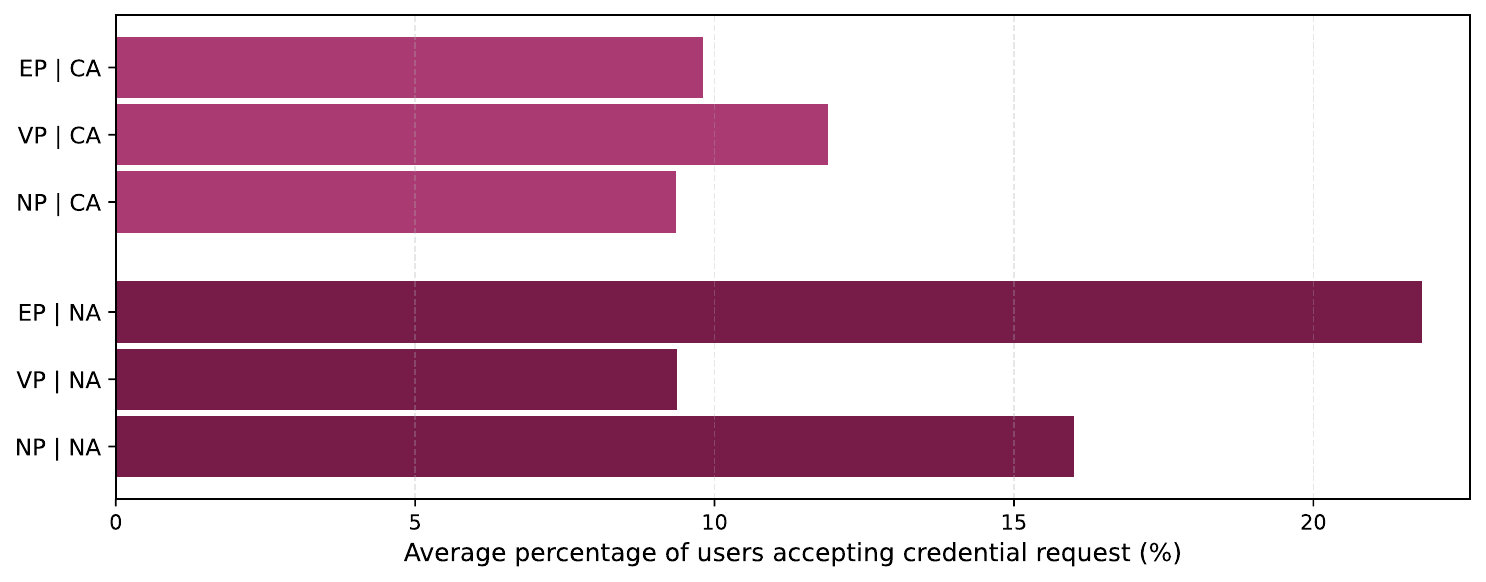}
        \caption{Credential request not supported by the website provided purpose.}
        \label{fig:purpose_bogus}
     \end{subfigure}
     \hfill
     \begin{subfigure}[b]{\linewidth}
         \centering
        \includegraphics[width=0.9\linewidth]{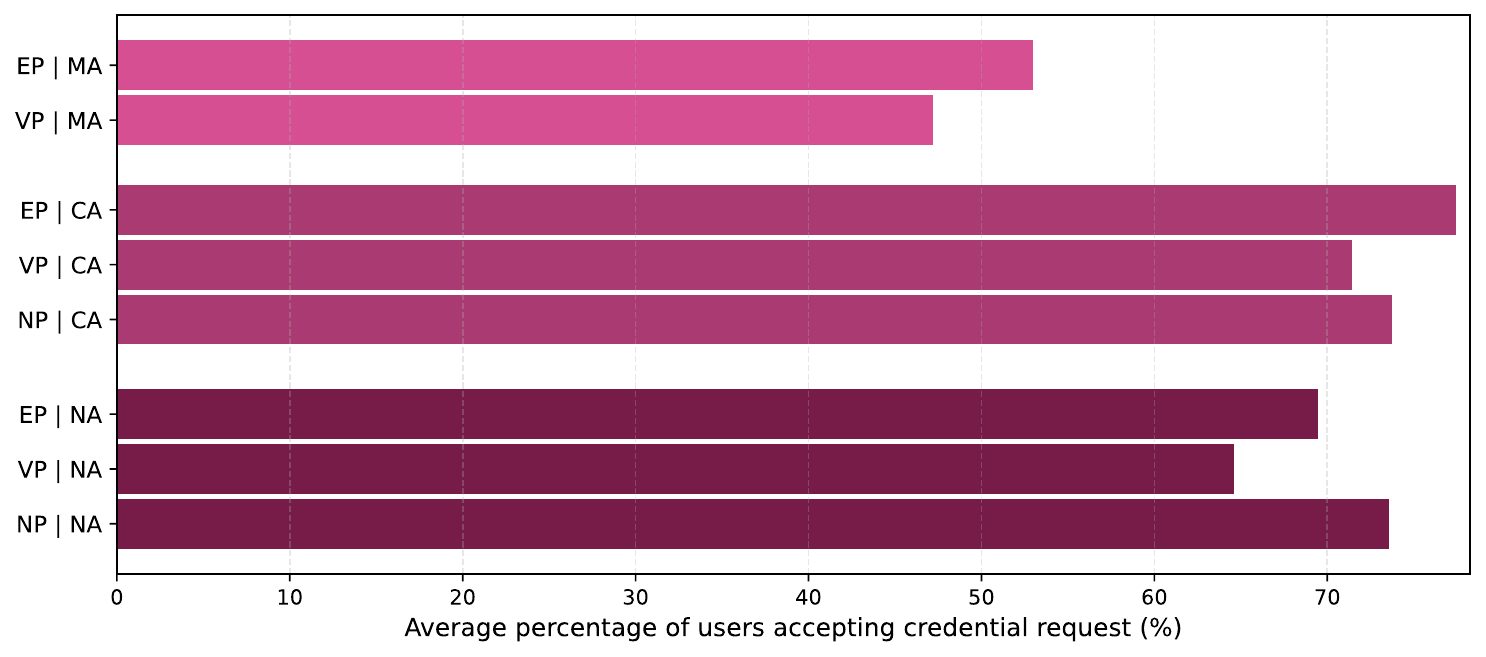}
        \caption{Credential request justified by the website provided purpose.}
        \label{fig:purpose_proper}
     \end{subfigure}
    \caption{User study results showing the average percentage of users disclosing credentials to websites for different website-provided purposes and \names{}. NP = no purpose, VP = vague purpose, EP = extensive purpose. NA = no \name{}, CA = \name{} with correct information, MA = \name{} with mistakes.}
    \label{fig:response_purpose}
\end{figure}

\cref{fig:response_purpose} presents the results of the user study showing how often users disclosed a credential to a website for different website-provided purposes, and different \name{} types. Our results show that a vague purpose almost always leads to less disclosure (e.g., $\sim$71\% with purpose vs. $\sim$74\% without purpose for a justified request with a correct \name{}). On the other hand, showing an extensive purpose almost always increases the amount of disclosure, regardless of whether the request was justified (e.g., $\sim$22\% with purpose vs. $\sim$16\% without purpose for an unjustified request without a \name{}). However, if the \name{} incorrectly recommends not to disclose a credential, then the number of users who disclose decreases significantly (i.e., $p\leq0.05$ in Fischer's exact test), even with an extensive website-provided purpose ($\sim$69\% without vs. $\sim$77\% with correct vs. $\sim$53\% with mistaken \name{}). This indicates that even an extensive website-provided purpose cannot counteract mistakes in the \name{}, and that allowing websites to add purpose statements in the EUDI Wallet could be risky if those statements are not verified. Our research adds a further indication that an extensive purpose can lead to more disclosure, which has some prior work showing more~\cite{tan2014effect, liu2016follow} and other prior work showing less~\cite{shih2015privacy, knijnenburg2013making} disclosure. We add to this initial results on whether a purpose can counteract bad nudges, which has, to the best of our knowledge, not yet been explored.

\subsection{A3: (Over)sharing Patterns}
\label{subsec:res_a1}
As discussed in \cref{results:A1}, users overshare more when seeing a request as compared to what they believe they should disclose, mirroring the privacy paradox~\cite{kokolakis2017privacy, dienlin2023longitudinal, taddicken2014privacy}. Thus, any oversharing we identify is likely worse in practice, where users react to requests.

\begin{figure}[tb]
    \centering
    \includegraphics[width=0.9\linewidth]{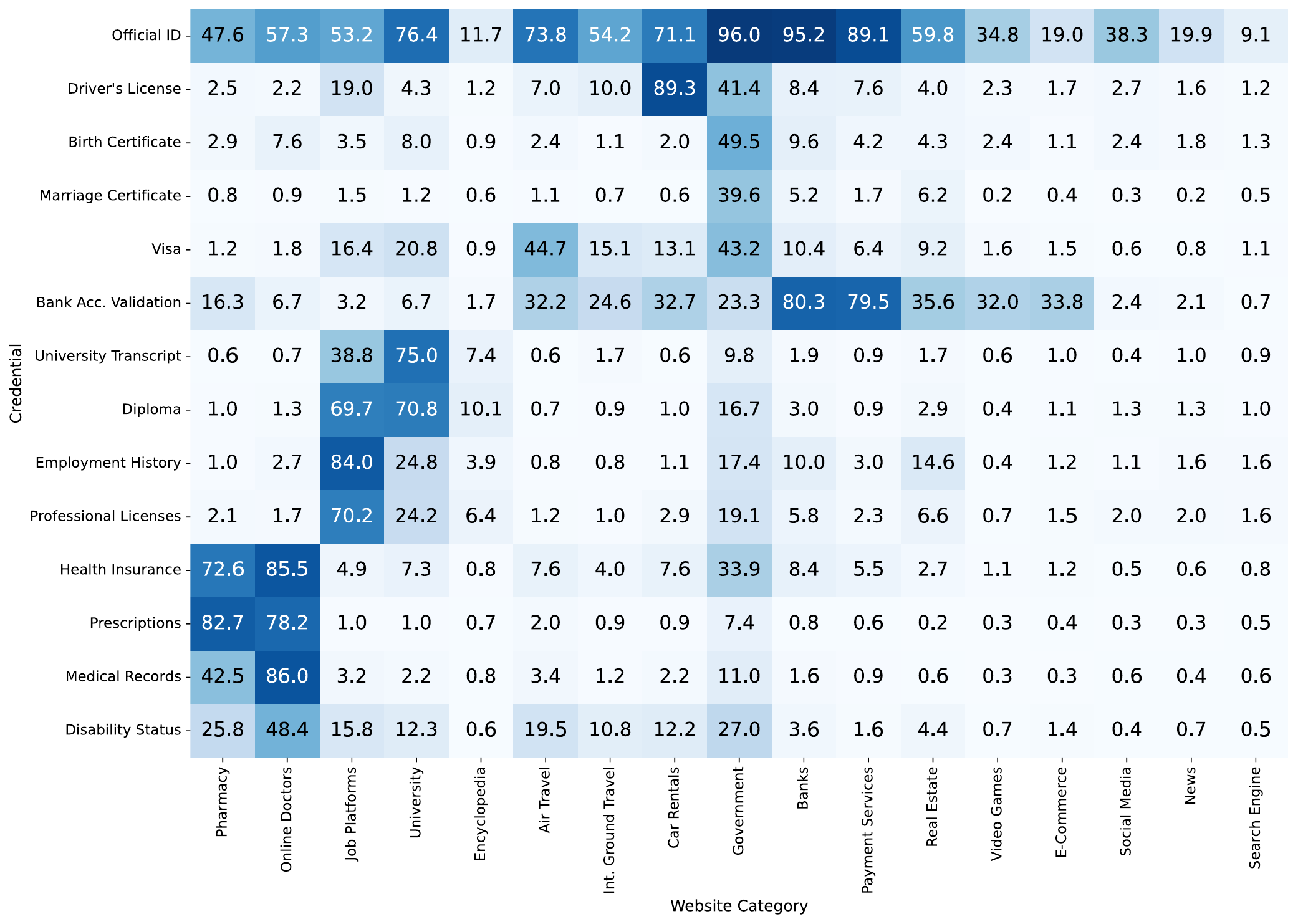}
    \caption{User survey results showing the percentage of users who would disclose a credential per website category.}
    \label{fig:percentage_heatmap}
\end{figure}

\paragraph{General Oversharing}
\cref{fig:percentage_heatmap} presents the results of the user survey showing what percentage of participants would disclose a credential for each website category. While the numbers often follow expectations, there are multiple instances of oversharing. 
Examples include the following cases, where there is no justification for disclosure. $\sim$20\% of users would disclose their official ID to news websites, $\sim$9\% to search engine websites, and $\sim$12\% to encyclopedia websites. Furthermore, $\sim$6\% of users would disclose their health insurance to payment service websites, and $\sim$7\% would disclose their prescriptions to government websites. Although these percentages may seem low, the absolute number of credentials a verifier can collect can be high for high-traffic websites, enabling scalable collection of credentials without much effort. For instance, it is not unrealistic to assume that a search engine reaches a traffic volume of a million unique users. If $\sim$9\% are willing to disclose their official ID, then a verifier can collect 90,000 official IDs. This is in line with prior research in other contexts that has shown that users were willing to disclose data even for small benefits~\cite{acquisti2017nudges, egelman2013choice, hann2002online, acquisti2013privacy, brandimarte2013misplaced, cvrcek2006study, norberg2007privacy, taddicken2014privacy}.

\paragraph{Naming Confusion} 
The results of the user survey presented in \cref{fig:percentage_heatmap} also show oversharing when users connect the website or credential name to a use case that is not present. For example, encyclopedia websites (e.g., Wikipedia) would likely require no credentials at all, as they simply compile and allow the search for information. 
However, users seemed to connect these websites to education and profession due to their content.
This is likely why a higher number of participants said they would disclose university-related ($\sim$10\% for diploma, $\sim$7\% for university transcript) or profession-related credentials ($\sim$6\% for professional licenses, $\sim$4\% for employment history). Another example is the bank account validation credential, which did not contain the data needed to execute a payment. Still, many participants believed that they could use it to pay ("... bank account verification is sometimes required when purchasing a premium subscription ..."), and thus, frequently said they would disclose it ($\sim$32\% for video game, $\sim$34\% for e-commerce, $\sim$32\% for air travel websites). These examples are especially concerning, as they indicate that a verifier could name their website in a specific way or leverage misconceptions to collect more data. To the best of our knowledge, this effect has not been investigated before.
\section{Discussion}

In this section, we discuss our main results and why these results are important. Then we summarize some recommendations for the EUDI that can be concluded from our results.

\subsection{Summary of Main Results}

Displaying information in the \name{} based on the website category is sufficient. However, websites providing multiple services or website categories with varying credential requirements (e.g., age verification for e-commerce) need to be handled specially. Collecting user opinions through surveys leads to better data quality as compared to usage data. On average, 24 experts need to be surveyed to receive sufficiently high-confidence results; however, in cases with low expert consensus, more experts may be needed.

The \name{} is effective at reducing the number of mistakes that users make. In most cases, the average mistake rate dropped by $\sim$8\% (e.g., from $\sim$15\% to $\sim$7\% when seeing only expert recommendations in the assistant). Furthermore, the confidence (i.e., displayed percentages) of the \name{} is important, as the \name{} showed no nudging effect with low percentages. Lastly, there are demographic differences regarding the preferences for the \name{}. For example, users in the 50+ age group reacted more to seeing expert recommendations while users in the 18-29 age group reacted more to user opinions.

Users overshared credentials, even those that are generally considered sensitive in the physical world (e.g., $\sim$20\% of users would disclose their official ID to a news website). 
Oversharing increased when users misunderstood what a credential could be used for (e.g., thinking the bank account validation can be used for payment) or what credentials a website needs for its services (e.g., thinking encyclopedia websites need education or profession credentials).

\paragraph{Why This Matters}

Data shown in the \name{} needs to be collected with care. If users overshare more when seeing requests, then using such data would inflate the numbers in the \name{}, which may increase oversharing rather than decreasing it. Furthermore, even though the \name{} can reduce the oversharing risk, it does not fully remove it. Even a residual risk of $\sim$7\% can be detrimental at the planned scale of the EUDI. For example, if the EUDI is used by 10 million users, this would still put the data of 70'000 users at risk, which is problematic considering the sensitive nature of the planned EUDI credentials.

Oversharing EUDI credentials is not only a privacy risk but can also increase the risk of identity theft for users. The EUDI is not planned to be mandatory; therefore, there will be value for malicious parties to collect identity information that they can later use to impersonate their victims using traditional identity verification methods.  
Examples of how harvested EUDI data can be used include \textit{SIM swapping attacks}~\cite{ENSI_SIM_Swapping_Beware, ENSI_SIM_Swapping_Report, lee2020empirical} (harvested data is used to receive a replacement SIM with the victims phone number allowing them to take over accounts with SMS 2FA), \textit{call-in identity verification} (harvested data is used to trick the weak identity verification of health insurers and banks\footnote{Data usually asked: name, date of birth, address, residency number, last large payment received. Tested with banks and health insurers in 3 European countries.}), and \textit{opening a bank account in someone else's name} (harvested data is used to create deepfakes that can trick liveness checks in the onboarding process of banks~\cite{li2022seeing, KYC_attack}).

\subsection{Recommendations for the EUDI}
\label{sec:recommednations}

\paragraph{Add a \name{} to the EUDI Wallet}
Our results show that users are likely to make disclosure mistakes with the EUDI Wallet, which could be a risk to users and may reduce trust in the system. Furthermore, our results show that the \name{} can reduce the number of mistakes users make. Thus, we recommend adding a \name{} to the EUDI Wallet, allowing them to receive support at disclosure time, which is most effective. 

\paragraph{Deploy Nudges Selectively and Prominently}

Prior research has shown that the effect of warnings decreases over time if users see them too often~\cite{vance2019fog, anderson2016your, vance2018tuning}. This is likely to happen for the \name{} if users see nudges for every credential request.
Thus, we recommend only showing nudges in important cases (i.e., when the credential is required, or when disclosure is risky). However, as these are the cases where a user is likely to face negative consequences from a poor disclosure decision, the nudge should be intrusive (e.g., similar to browser warnings).
Furthermore, it may make sense to time the warnings such that users see them when they are making a poor disclosure decision and to change the warning design based on the sensitivity of the data.

\paragraph{Adapt Nudges for Different User Groups}
Our research indicates that different demographic groups have different preferences for the \name{} (e.g., older users prefer expert recommendations while younger users prefer user opinions). Based on this, we recommend adapting nudges based on user group preferences.

\paragraph{Check Verifier Registry Using the \name{}}

As described in \cref{sec:background}, the EUDI system will include a mandatory registry for verifiers, in which they must register the EUDI attributes they plan to request. Although the EUDI Wallet and the registrars ensure that websites can only request the attributes that they have registered, this safeguard does not protect against over-claiming: the EUDI framework lacks any systematic, scalable expert review of whether a website genuinely needs the attributes it registers.
We believe that a tool like the \name{} could help to partially automate a review of the verifier registry, by comparing the registered attributes with the \name{} information, and flagging registrations that ask for credentials for which the \name{} would recommend not to disclose. Experts could then focus on reviewing the flagged registrations.

\paragraph{Use Surveys Instead of Usage Data to Assess User Opinions}
Our results show that users are more likely to overshare when they see a credential request, and usage data would only contain cases where a user received a credential request from a website. Furthermore, it is possible for websites to use dark patterns and paid users to increase the number of users who disclose a credential, skewing the data in their favor. Thus, it is better to survey a balanced set of users to identify which credentials users would disclose to websites.

\paragraph{Prioritize Transparency for Increased Public Trust}
Prior research has shown that trust is important for technology adoption~\cite{bahmanziari2003trust}. However, open letters~\cite{openletterEUDIJune2023, openletterEUDIFeb2023} and comments in our survey mention concerns about the EUDI (e.g., "I don't like the idea of a digital id system", "I worry about government overreach in this area", "There is no way i will get a digital id so you can be tracked and monitored by the EU"). 
The same is true for the \name{}. 
Some participants noted that they did not trust the expert recommendations ("It's impossible to trust those so-called experts because I don't know who they are." "I would not trust the opinions of other users or whatever panel of experts were selected, unless it was some kind of reputable institute."). Thus, it is important that the technologies used in the EUDI and the source of information shown in the \name{} is clearly communicated.

\section{Limitations}
\label{sec:limitations}

\paragraph{Simulated Environment}

Since the EUDI Wallet and ecosystem do not yet exist, it was not possible to test disclosure in the real world. Instead, we opted for a simulated environment with fixed scenarios. We see two main ways in which the simulated environment may have affected participants. First, users may have disclosed more freely as they were not providing any real data, and thus, there was no risk of their data being misused. Second, participants may have been able to judge the necessity of disclosure more thoroughly, as they did not see dark patterns and would not be locked out of services. A measurement study after the EUDI Wallet is deployed is required to verify to what extent these factors play a role.

\paragraph{Lack of Context}

While we did provide some information about what services the website provides, we did not provide an exhaustive list of services or information about why the user is using this website. This choice was made because the \name{} cannot be built scalably if it needs to take the context into consideration, and we wanted to assess what credentials participants would disclose across all website services. However, participants may not have thought about some credential use cases, which can explain some of the lower numbers in the results. For example, only a few participants were willing to disclose their official ID to video game websites; however, if we had given users the context that they are buying an age-restricted game, this number would likely be higher. Our main objective was to study oversharing, and thus, this choice is less of an issue. If there is no reason to disclose a credential to a website and a user still chooses to do so, then additional context is unlikely to change this tendency to overshare. Even though studying undersharing in the context of the EUDI would be interesting, we acknowledge that context is very important to measure this, and thus, our experiments are unsuitable. 

\paragraph{Selective Disclosure}

The EUDI will allow selective disclosure of attributes (i.e., one can only disclose the name attribute of the identity card credential).
Including this into our study would have made it too complex, as we already measured a large number of website categories, credentials, and \name{} options. However, this means that more research is needed to understand what attributes users would be willing to disclose to websites, and if those attributes match what the website needs to operate. 
\section{Related Work}
\label{sec:related}

\paragraph{Oversharing}

Studies have found that users are willing to use options providing less privacy even for low compensation for app~\cite{egelman2013choice}, website~\cite{hann2002online, acquisti2013privacy, hurwitz2012user}, and location~\cite{cvrcek2006study} data. Additionally, users often click accept on dialogs without understanding them~\cite{felt2012android, machuletz2019multiple}. Furthermore, users often make worse privacy decisions due to misconceptions. For example, users often underestimate negative consequences, do not think about privacy unless a violation occurred, believe their privacy is protected by the presence of privacy policies, or believe that more control over data equals more privacy~\cite{acquisti2017nudges, acquisti2015privacy, brandimarte2013misplaced}. Lastly, the privacy paradox shows that users often believe that they are privacy conscious, but act counter to that belief~\cite{kokolakis2017privacy, dienlin2023longitudinal, taddicken2014privacy}. In our work, we study oversharing in the new area of digital official identity documents. We find that oversharing occurs and identify new misconceptions that can lead to oversharing in this area.

\paragraph{Nudging}

Nudging systems give users additional information without forcing them to make a specific decision. Studies have shown that nudging helps combat privacy misconceptions and biases~\cite{acquisti2017nudges, herzog2007user}. Nudging systems have been evaluated and shown to be effective in contexts, such as cookies~\cite{goecks2005supporting}, app permissions~\cite{liu2016follow}, file sharing~\cite{digioia2005social}, social media posts~\cite{wang2014field}, configurations (e.g., privacy settings)~\cite{knijnenburg2013helping, patil2011little}, and location sharing~\cite{toch2014crowdsourcing}. However, nudging also has risks and challenges. For example, herding effects can lead to worse decisions as users decide solely based on the majority opinion~\cite{goecks2009challenges}. Lastly, the timing of nudges matters significantly for their effectiveness~\cite{egelman2009timing}. In our work, we study nudging in a new area: disclosure of digital versions of official identity documents, and find that nudging can be helpful in this area.

\paragraph{Identity Wallets and Single Sign-On (SSO)}

The EUDI Wallet is similar to SSI (self-sovereign identity) Wallets. Prior studies have investigated how well users understand the privacy protections in SSI Wallets and what usability challenges they face when using SSI Wallets~\cite{korir2022empirical, teuschel2023don}. These studies have found that users misunderstand SSI Wallets (e.g., some users thought identities were controlled by the identity provider) and put high importance on usability and convenience. Additionally, the EUDI Wallet takes inspiration from SSO services, where users can allow websites to access their information through an identity provider (usually for logins). Prior work on SSOs has shown that data is often over-requested and that there is a lack of transparency, making it hard for users to choose the most privacy-preserving option~\cite{morkonda2025sign, dimova2023everybody}.
\section{Conclusions}
\label{sec:conclusion}

In this paper, we investigated whether a support system at disclosure time can help users make better EUDI credential disclosure decisions.
We evaluate a scalable \name{} that displays expert recommendations or user opinions and show that it significantly reduces the number of mistakes users make from $\sim$15\% to $\sim$7\%. This result is encouraging, especially as our \name{} was fully non-intrusive. However, at the scale of the EUDI, a residual risk of $\sim$7\% still puts thousands or even millions of users at risk, which is unacceptable. More intrusive interventions (e.g., similar to browser warnings) for sensitive disclosure decisions may be needed. 
We also identify common misconceptions that may make users more likely to overshare EUDI credentials (e.g., falsely connecting a credential to a use case or a website to a service based on their name). 
Future research should investigate whether stronger interventions can reduce the disclosure risk to acceptable levels and measure oversharing in the EUDI after it has been deployed.

\bibliographystyle{ACM-Reference-Format}
\bibliography{references}

@misc{W3C_VC_Use_Cases,
    title = {Verifiable Credentials Use Cases},
    author = {W3C},
    urldate = {2026-04-03},
    url = {https://www.w3.org/TR/vc-use-cases/},
    year = {2026}
}

@misc{EUDI_ARF,
    title = {EUDI Architecture and Reference Framework},
    author = {European Commission},
    urldate = {2025-12-31},
    url = {https://eudi.dev/1.1.0/arf/},
    year = {2023}
}

@misc{EUDI_Wallet,
    title = {A digital ID and personal digital wallet for EU citizens, residents and businesses},
    author = {European Commission},
    urldate = {2026-04-03},
    url = {https://ec.europa.eu/digital-building-blocks/sites/spaces/EUDIGITALIDENTITYWALLET/pages/694487738/EU+Digital+Identity+Wallet+Home},
    year = {2026}
}

@misc{EUDI_Wallet_GitHub,
    title = {European Digital Identity Wallet},
    author = {European Commission},
    urldate = {2026-04-03},
    url = {https://eudi.dev/latest/},
    year = {2026}
}

@article{EUDI_Law_Verifier_Credentials,
    author={European Commission},
    title={Commission Implementing Regulation (EU) 2025/848 of 6 May 2025 Laying Down Rules for the Application of Regulation (EU) No 910/2014 of the European Parliament and of the Council as Regards the Registration of Wallet-Relying Parties},
    journal = {Official Journal of the European Union},
    year    = {2025},
    volume  = {L 2025/848},
    pages   = {1--15},
    url     = {https://eur-lex.europa.eu/legal-content/EN/TXT/PDF/?uri=OJ:L_202500848}
}

@misc{Cloudflare_Radar_Taxonomy,
    title = {Domain categories},
    author = {Cloudflare},
    urldate = {2026-04-03},
    url = {https://developers.cloudflare.com/cloudflare-one/traffic-policies/domain-categories/},
    year = {2026}
}

@misc{Qualtrics_Commitment,
    title = {Improve data quality by using a commitment request instead of attention checks},
    author = {Qualtrics},
    urldate = {2026-04-03},
    url = {https://www.qualtrics.com/articles/strategy-research/attention-checks-and-data-quality/},
    year = {2022}
}

@article{acquisti2017nudges,
  title={Nudges for privacy and security: Understanding and assisting users’ choices online},
  author={Acquisti, Alessandro and Adjerid, Idris and Balebako, Rebecca and Brandimarte, Laura and Cranor, Lorrie Faith and Komanduri, Saranga and Leon, Pedro Giovanni and Sadeh, Norman and Schaub, Florian and Sleeper, Manya and others},
  journal={ACM Computing Surveys (CSUR)},
  volume={50},
  number={3},
  pages={1--41},
  year={2017},
  publisher={ACM New York, NY, USA}
}

@incollection{egelman2013choice,
  title={Choice architecture and smartphone privacy: There’s a price for that},
  author={Egelman, Serge and Felt, Adrienne Porter and Wagner, David},
  booktitle={The economics of information security and privacy},
  pages={211--236},
  year={2013},
  publisher={Springer}
}

@article{hann2002online,
  title={Online information privacy: Measuring the cost-benefit trade-off},
  author={Hann, Il-Horn and Hui, Kai-Lung and Lee, Tom and Png, Ivan},
  journal={ICIS 2002 proceedings},
  pages={1},
  year={2002}
}

@article{acquisti2013privacy,
  title={What is privacy worth?},
  author={Acquisti, Alessandro and John, Leslie K and Loewenstein, George},
  journal={The Journal of Legal Studies},
  volume={42},
  number={2},
  pages={249--274},
  year={2013},
  publisher={University of Chicago Press Chicago, IL}
}

@article{brandimarte2013misplaced,
  title={Misplaced confidences: Privacy and the control paradox},
  author={Brandimarte, Laura and Acquisti, Alessandro and Loewenstein, George},
  journal={Social psychological and personality science},
  volume={4},
  number={3},
  pages={340--347},
  year={2013},
  publisher={Sage Publications Sage CA: Los Angeles, CA}
}

@inproceedings{cvrcek2006study,
  title={A study on the value of location privacy},
  author={Cvrcek, Dan and Kumpost, Marek and Matyas, Vashek and Danezis, George},
  booktitle={Proceedings of the 5th ACM workshop on Privacy in electronic society},
  pages={109--118},
  year={2006}
}

@article{norberg2007privacy,
  title={The privacy paradox: Personal information disclosure intentions versus behaviors},
  author={Norberg, Patricia A and Horne, Daniel R and Horne, David A},
  journal={Journal of consumer affairs},
  volume={41},
  number={1},
  pages={100--126},
  year={2007},
  publisher={Wiley Online Library}
}

@article{taddicken2014privacy,
  title={The ‘privacy paradox’in the social web: The impact of privacy concerns, individual characteristics, and the perceived social relevance on different forms of self-disclosure},
  author={Taddicken, Monika},
  journal={Journal of computer-mediated communication},
  volume={19},
  number={2},
  pages={248--273},
  year={2014},
  publisher={Oxford University Press Oxford, UK}
}

@inproceedings{wash2018provides,
  title={Who provides phishing training? facts, stories, and people like me},
  author={Wash, Rick and Cooper, Molly M},
  booktitle={Proceedings of the 2018 chi conference on human factors in computing systems},
  pages={1--12},
  year={2018}
}

@inproceedings{besmer2010impact,
  title={The impact of social navigation on privacy policy configuration},
  author={Besmer, Andrew and Watson, Jason and Lipford, Heather Richter},
  booktitle={Proceedings of the Sixth Symposium on Usable Privacy and Security},
  pages={1--10},
  year={2010}
}

@inproceedings{goecks2009challenges,
  title={Challenges in supporting end-user privacy and security management with social navigation},
  author={Goecks, Jeremy and Edwards, W Keith and Mynatt, Elizabeth D},
  booktitle={Proceedings of the 5th Symposium on Usable Privacy and Security},
  pages={1--12},
  year={2009}
}

@inproceedings{liu2016follow,
  title={Follow my recommendations: A personalized privacy assistant for mobile app permissions},
  author={Liu, Bin and Andersen, Mads Schaarup and Schaub, Florian and Almuhimedi, Hazim and Zhang, Shikun Aerin and Sadeh, Norman and Agarwal, Yuvraj and Acquisti, Alessandro},
  booktitle={Twelfth symposium on usable privacy and security (SOUPS 2016)},
  pages={27--41},
  year={2016}
}

@article{acquisti2015privacy,
  title={Privacy and human behavior in the age of information},
  author={Acquisti, Alessandro and Brandimarte, Laura and Loewenstein, George},
  journal={Science},
  volume={347},
  number={6221},
  pages={509--514},
  year={2015},
  publisher={American Association for the Advancement of Science}
}

@article{acquisti2012impact,
  title={The impact of relative standards on the propensity to disclose},
  author={Acquisti, Alessandro and John, Leslie K and Loewenstein, George},
  journal={Journal of Marketing Research},
  volume={49},
  number={2},
  pages={160--174},
  year={2012},
  publisher={SAGE Publications Sage CA: Los Angeles, CA}
}

@article{overton2025people,
  title={People Believe If 90\% Prefer A over B, A Must Be Much Better than B. Are They Wrong?},
  author={Overton, Graham and Evangelidis, Ioannis and Vosgerau, Joachim},
  journal={Journal of Consumer Research},
  volume={52},
  number={1},
  pages={135--156},
  year={2025},
  publisher={Oxford University Press}
}

@article{martin2002levels,
  title={Levels of consensus and majority and minority influence},
  author={Martin, Robin and Gardikiotis, Antonis and Hewstone, Miles},
  journal={European Journal of Social Psychology},
  volume={32},
  number={5},
  pages={645--665},
  year={2002},
  publisher={Wiley Online Library}
}

@article{riefa2017challenge,
  title={The challenge of protecting EU consumers in global online markets},
  author={Riefa, Christine},
  year={2017},
  publisher={The European Consumer Organisation}
}

@misc{2025_consumer_scoreboard,
    title = {2025 Consumer Conditions Scoreboard},
    author = {European Comission},
    urldate = {2026-04-03},
    url = {https://commission.europa.eu/strategy-and-policy/policies/consumers/consumer-protection-policy/key-consumer-data_en},
    year = {2025}
}

@misc{world_e-commerce,
    title = {World Consumer Rights Day 2018 Briefing: e-commerce backgrounder},
    author = {Consumers International},
    urldate = {2026-04-03},
    url = {https://www.consumersinternational.org/media/154916/e-commerce-overview-report.pdf},
    year = {2018}
}

@article{bart2005drivers,
  title={Are the drivers and role of online trust the same for all web sites and consumers? A large-scale exploratory empirical study},
  author={Bart, Yakov and Shankar, Venkatesh and Sultan, Fareena and Urban, Glen L},
  journal={Journal of marketing},
  volume={69},
  number={4},
  pages={133--152},
  year={2005},
  publisher={SAGE Publications Sage CA: Los Angeles, CA}
}

@article{jarvenpaa2000consumer,
  title={Consumer trust in an Internet store},
  author={Jarvenpaa, Sirkka L and Tractinsky, Noam and Vitale, Michael},
  journal={Information technology and management},
  volume={1},
  number={1},
  pages={45--71},
  year={2000},
  publisher={Springer}
}

@article{alarabiat2019delphi,
  title={The Delphi method in information systems research (2004-2017)},
  author={Alarabiat, Ayman and Ramos, Isabel},
  journal={Electronic Journal of Business Research Methods},
  volume={17},
  number={2},
  pages={pp86--99},
  year={2019}
}

@article{belton2021delphi,
  title={Delphi with feedback of rationales: How large can a Delphi group be such that participants are not overloaded, de-motivated, or disengaged?},
  author={Belton, Ian and Wright, George and Sissons, Aileen and Bolger, Fergus and Crawford, Megan M and Hamlin, Iain and L{\=u}ka, Courtney Taylor Browne and Vasilichi, Alexandrina},
  journal={Technological Forecasting and Social Change},
  volume={170},
  pages={120897},
  year={2021},
  publisher={Elsevier}
}

@article{emami2018influence,
  title={The influence of friends and experts on privacy decision making in IoT scenarios},
  author={Emami Naeini, Pardis and Degeling, Martin and Bauer, Lujo and Chow, Richard and Cranor, Lorrie Faith and Haghighat, Mohammad Reza and Patterson, Heather},
  journal={Proceedings of the ACM on human-computer interaction},
  volume={2},
  number={CSCW},
  pages={1--26},
  year={2018},
  publisher={ACM New York, NY, USA}
}

@article{meyers2020inducing,
  title={Inducing feelings of ignorance makes people more receptive to expert (economist) opinion},
  author={Meyers, Ethan A and Turpin, Martin H and Bia{\l}ek, Micha{\l} and Fugelsang, Jonathan A and Koehler, Derek J},
  journal={Judgment and Decision Making},
  volume={15},
  number={6},
  pages={909--925},
  year={2020},
  publisher={Cambridge University Press}
}

@article{harvey1997taking,
  title={Taking advice: Accepting help, improving judgment, and sharing responsibility},
  author={Harvey, Nigel and Fischer, Ilan},
  journal={Organizational behavior and human decision processes},
  volume={70},
  number={2},
  pages={117--133},
  year={1997},
  publisher={Elsevier}
}

@article{bailey2023meta,
  title={A meta-analysis of the weight of advice in decision-making},
  author={Bailey, Phoebe E and Leon, Tarren and Ebner, Natalie C and Moustafa, Ahmed A and Weidemann, Gabrielle},
  journal={Current Psychology},
  volume={42},
  number={28},
  pages={24516--24541},
  year={2023},
  publisher={Springer}
}

@article{qi2018collaborator,
  title={A collaborator's reputation can bias decisions and anxiety under uncertainty},
  author={Qi, Song and Footer, Owen and Camerer, Colin F and Mobbs, Dean},
  journal={Journal of Neuroscience},
  volume={38},
  number={9},
  pages={2262--2269},
  year={2018},
  publisher={Society for Neuroscience}
}

@inproceedings{shih2015privacy,
  title={Privacy tipping points in smartphones privacy preferences},
  author={Shih, Fuming and Liccardi, Ilaria and Weitzner, Daniel},
  booktitle={Proceedings of the 33rd Annual ACM Conference on Human Factors in Computing Systems},
  pages={807--816},
  year={2015}
}

@article{knijnenburg2013making,
  title={Making decisions about privacy: information disclosure in context-aware recommender systems},
  author={Knijnenburg, Bart P and Kobsa, Alfred},
  journal={ACM Transactions on Interactive Intelligent Systems (TiiS)},
  volume={3},
  number={3},
  pages={1--23},
  year={2013},
  publisher={ACM New York, NY, USA}
}

@inproceedings{tan2014effect,
  title={The effect of developer-specified explanations for permission requests on smartphone user behavior},
  author={Tan, Joshua and Nguyen, Khanh and Theodorides, Michael and Negr{\'o}n-Arroyo, Heidi and Thompson, Christopher and Egelman, Serge and Wagner, David},
  booktitle={Proceedings of the SIGCHI Conference on Human Factors in Computing Systems},
  pages={91--100},
  year={2014}
}

@article{steinberg2007age,
  title={Age differences in resistance to peer influence.},
  author={Steinberg, Laurence and Monahan, Kathryn C},
  journal={Developmental psychology},
  volume={43},
  number={6},
  pages={1531},
  year={2007},
  publisher={American Psychological Association}
}

@article{knoll2017age,
  title={Age-related differences in social influence on risk perception depend on the direction of influence},
  author={Knoll, Lisa J and Leung, Jovita T and Foulkes, Lucy and Blakemore, Sarah-Jayne},
  journal={Journal of adolescence},
  volume={60},
  pages={53--63},
  year={2017},
  publisher={Elsevier}
}

@article{wang2020expert,
  title={When expert recommendation contradicts peer opinion: Relative social influence of valence, group identity and artificial intelligence},
  author={Wang, Jinping and Molina, Maria D and Sundar, S Shyam},
  journal={Computers in Human Behavior},
  volume={107},
  pages={106278},
  year={2020},
  publisher={Elsevier}
}

@article{mills2022personalized,
  title={Personalized nudging},
  author={Mills, Stuart},
  journal={Behavioural Public Policy},
  volume={6},
  number={1},
  pages={150--159},
  year={2022},
  publisher={Cambridge University Press}
}

@article{calboli2025ai,
  title={AI-enhanced nudging in public policy: why to worry and how to respond},
  author={Calboli, Stefano and Engelen, Bart},
  journal={Mind \& Society},
  pages={1--19},
  year={2025},
  publisher={Springer}
}

@misc{ENSI_SIM_Swapping_Beware,
    title = {Beware of the Sim Swapping Fraud!},
    author = {ENISA},
    urldate = {2026-04-03},
    url = {https://www.enisa.europa.eu/news/enisa-news/beware-of-the-sim-swapping-fraud},
    year = {2021}
}

@misc{ENSI_SIM_Swapping_Report,
    title = {Countering SIM-Swapping},
    author = {ENISA},
    urldate = {2026-04-03},
    url = {https://www.enisa.europa.eu/publications/countering-sim-swapping},
    year = {2021}
}

@article{zheng2019survey,
  title={A survey on image tampering and its detection in real-world photos},
  author={Zheng, Lilei and Zhang, Ying and Thing, Vrizlynn LL},
  journal={Journal of Visual Communication and Image Representation},
  volume={58},
  pages={380--399},
  year={2019},
  publisher={Elsevier}
}

@article{zheng2023spoofing,
  title={Spoofing attacks and anti-spoofing methods for face authentication over smartphones},
  author={Zheng, Zheng and Wang, Qian and Wang, Cong},
  journal={IEEE Communications Magazine},
  volume={61},
  number={12},
  pages={213--219},
  year={2023},
  publisher={IEEE}
}

@misc{UK_age_restriction,
    title = {Online Safety Act: explainer},
    author = {Gov.UK},
    urldate = {2026-04-03},
    url = {https://www.gov.uk/government/publications/online-safety-act-explainer/online-safety-act-explainer},
    year = {2025}
}

@misc{EU_age_restriction,
    title = {Commission releases enhanced second version of the age-verification blueprint},
    author = {European Commission},
    urldate = {2026-04-03},
    url = {https://digital-strategy.ec.europa.eu/en/news/commission-releases-enhanced-second-version-age-verification-blueprint},
    year = {2025}
}

@misc{France_age_restriction,
    title = {Technical guidelines on age verification for the protection of persons under 18 from online pornography},
    author = {Arcom},
    urldate = {2026-04-03},
    url = {https://www.arcom.fr/en/find-out-more/legal-area/legal-resources/technical-guidelines-age-verification-protection-persons-under-18-online-pornography},
    year = {2024}
}

@misc{US_age_restriction,
    title = {Supreme Court upholds Texas law aimed at blocking kids from seeing pornography online},
    author = {AP News},
    urldate = {2026-04-03},
    url = {https://apnews.com/article/supreme-court-porn-age-verification-texas-12a73197796fe8c4bef0d888259543cf},
    year = {2025}
}

@inproceedings{nouwens2020dark,
  title={Dark patterns after the GDPR: Scraping consent pop-ups and demonstrating their influence},
  author={Nouwens, Midas and Liccardi, Ilaria and Veale, Michael and Karger, David and Kagal, Lalana},
  booktitle={Proceedings of the 2020 CHI conference on human factors in computing systems},
  pages={1--13},
  year={2020}
}

@article{mathur2019dark,
  title={Dark patterns at scale: Findings from a crawl of 11K shopping websites},
  author={Mathur, Arunesh and Acar, Gunes and Friedman, Michael J and Lucherini, Eli and Mayer, Jonathan and Chetty, Marshini and Narayanan, Arvind},
  journal={Proceedings of the ACM on human-computer interaction},
  volume={3},
  number={CSCW},
  pages={1--32},
  year={2019},
  publisher={ACM New York, NY, USA}
}

@inproceedings{mathur2021makes,
  title={What makes a dark pattern... dark? Design attributes, normative considerations, and measurement methods},
  author={Mathur, Arunesh and Kshirsagar, Mihir and Mayer, Jonathan},
  booktitle={Proceedings of the 2021 CHI conference on human factors in computing systems},
  pages={1--18},
  year={2021}
}

@inproceedings{tran2025dark,
  title={Dark Patterns in the Opt-Out Process and Compliance with the California Consumer Privacy Act (CCPA)},
  author={Tran, Van Hong and Mehrotra, Aarushi and Sharma, Ranya and Chetty, Marshini and Feamster, Nick and Frankenreiter, Jens and Strahilevitz, Lior},
  booktitle={Proceedings of the 2025 CHI Conference on Human Factors in Computing Systems},
  pages={1--25},
  year={2025}
}

@article{luguri2021shining,
  title={Shining a light on dark patterns},
  author={Luguri, Jamie and Strahilevitz, Lior Jacob},
  journal={Journal of Legal Analysis},
  volume={13},
  number={1},
  pages={43--109},
  year={2021},
  publisher={Oxford University Press}
}

@incollection{hurwitz2012user,
  title={User choice, privacy sensitivity, and acceptance of personal information collection},
  author={Hurwitz, Joshua B},
  booktitle={European data protection: Coming of age},
  pages={295--312},
  year={2012},
  publisher={Springer}
}

@inproceedings{herzog2007user,
  title={User help techniques for usable security},
  author={Herzog, Almut and Shahmehri, Nahid},
  booktitle={Proceedings of the 2007 symposium on Computer human interaction for the management of information technology},
  pages={11--es},
  year={2007}
}

@inproceedings{goecks2005supporting,
  title={Supporting privacy management via community experience and expertise},
  author={Goecks, Jeremy and Mynatt, Elizabeth D},
  booktitle={Communities and Technologies 2005: Proceedings of the Second Communities and Technologies Conference, Milano 2005},
  pages={397--417},
  year={2005},
  organization={Springer}
}

@inproceedings{digioia2005social,
  title={Social navigation as a model for usable security},
  author={DiGioia, Paul and Dourish, Paul},
  booktitle={Proceedings of the 2005 symposium on Usable privacy and security},
  pages={101--108},
  year={2005}
}

@inproceedings{wang2014field,
  title={A field trial of privacy nudges for facebook},
  author={Wang, Yang and Leon, Pedro Giovanni and Acquisti, Alessandro and Cranor, Lorrie Faith and Forget, Alain and Sadeh, Norman},
  booktitle={Proceedings of the SIGCHI conference on human factors in computing systems},
  pages={2367--2376},
  year={2014}
}

@inproceedings{knijnenburg2013helping,
  title={Helping users with information disclosure decisions: potential for adaptation},
  author={Knijnenburg, Bart P and Kobsa, Alfred},
  booktitle={Proceedings of the 2013 international conference on Intelligent user interfaces},
  pages={407--416},
  year={2013}
}

@inproceedings{patil2011little,
  title={With a little help from my friends: can social navigation inform interpersonal privacy preferences?},
  author={Patil, Sameer and Page, Xinru and Kobsa, Alfred},
  booktitle={Proceedings of the ACM 2011 conference on Computer supported cooperative work},
  pages={391--394},
  year={2011}
}

@article{toch2014crowdsourcing,
  title={Crowdsourcing privacy preferences in context-aware applications},
  author={Toch, Eran},
  journal={Personal and ubiquitous computing},
  volume={18},
  number={1},
  pages={129--141},
  year={2014},
  publisher={Springer}
}

@inproceedings{egelman2009timing,
  title={Timing is everything? The effects of timing and placement of online privacy indicators},
  author={Egelman, Serge and Tsai, Janice and Cranor, Lorrie Faith and Acquisti, Alessandro},
  booktitle={Proceedings of the SIGCHI Conference on Human Factors in Computing Systems},
  pages={319--328},
  year={2009}
}

@article{kokolakis2017privacy,
  title={Privacy attitudes and privacy behaviour: A review of current research on the privacy paradox phenomenon},
  author={Kokolakis, Spyros},
  journal={Computers \& security},
  volume={64},
  pages={122--134},
  year={2017},
  publisher={Elsevier}
}

@article{dienlin2023longitudinal,
  title={A longitudinal analysis of the privacy paradox},
  author={Dienlin, Tobias and Masur, Philipp K and Trepte, Sabine},
  journal={New Media \& Society},
  volume={25},
  number={5},
  pages={1043--1064},
  year={2023},
  publisher={SAGE Publications Sage UK: London, England}
}

@inproceedings{krol2016control,
  title={Control versus effort in privacy warnings for webforms},
  author={Krol, Kat and Preibusch, S{\"o}ren},
  booktitle={Proceedings of the 2016 ACM on Workshop on Privacy in the Electronic Society},
  pages={13--23},
  year={2016}
}

@incollection{preibusch2013privacy,
  title={The privacy economics of voluntary over-disclosure in web forms},
  author={Preibusch, S{\"o}ren and Krol, Kat and Beresford, Alastair R},
  booktitle={The Economics of Information Security and Privacy},
  pages={183--209},
  year={2013},
  publisher={Springer}
}

@inproceedings{vance2019fog,
  title={The fog of warnings: how non-essential notifications blur with security warnings},
  author={Vance, Anthony and Eargle, David and Jenkins, Jeffrey L and Kirwan, C Brock and Anderson, Bonnie Brinton},
  booktitle={Fifteenth Symposium on Usable Privacy and Security (SOUPS 2019)},
  pages={407--420},
  year={2019}
}

@article{anderson2016your,
  title={Your memory is working against you: How eye tracking and memory explain habituation to security warnings},
  author={Anderson, Bonnie Brinton and Jenkins, Jeffrey L and Vance, Anthony and Kirwan, C Brock and Eargle, David},
  journal={Decision Support Systems},
  volume={92},
  pages={3--13},
  year={2016},
  publisher={Elsevier}
}

@article{vance2018tuning,
  title={Tuning out security warnings},
  author={Vance, Anthony and Jenkins, Jeffrey L and Anderson, Bonnie Brinton and Bjornn, Daniel K and Kirwan, C Brock},
  journal={MIS Quarterly},
  volume={42},
  number={2},
  pages={355--380},
  year={2018},
  publisher={JSTOR}
}

@inproceedings{lee2020empirical,
  title={An empirical study of wireless carrier authentication for $\{$SIM$\}$ swaps},
  author={Lee, Kevin and Kaiser, Benjamin and Mayer, Jonathan and Narayanan, Arvind},
  booktitle={Sixteenth symposium on usable privacy and security (soups 2020)},
  pages={61--79},
  year={2020}
}

@inproceedings{kelley2009nutrition,
  title={A" nutrition label" for privacy},
  author={Kelley, Patrick Gage and Bresee, Joanna and Cranor, Lorrie Faith and Reeder, Robert W},
  booktitle={Proceedings of the 5th Symposium on Usable Privacy and Security},
  pages={1--12},
  year={2009}
}

@inproceedings{li2022seeing,
  title={Seeing is living? rethinking the security of facial liveness verification in the deepfake era},
  author={Li, Changjiang and Wang, Li and Ji, Shouling and Zhang, Xuhong and Xi, Zhaohan and Guo, Shanqing and Wang, Ting},
  booktitle={31st USENIX Security Symposium (USENIX Security 22)},
  pages={2673--2690},
  year={2022}
}

@misc{KYC_attack,
    title = {Unmasking Cybercrime: Strengthening Digital Identity Verification against Deepfakes},
    author = {Cybercrime Atlas},
    urldate = {2026-01-29},
    url = {https://reports.weforum.org/docs/WEF\_Unmasking\_Cybercrime\_Strengthening\_Digital\_Identity\_Verification\_against\_Deepfakes\_2026.pdf}
}

@article{bahmanziari2003trust,
  title={Is trust important in technology adoption? A policy capturing approach},
  author={Bahmanziari, Tammy and Pearson, J Michael and Crosby, Leon},
  journal={Journal of Computer Information Systems},
  volume={43},
  number={4},
  pages={46--54},
  year={2003},
  publisher={Taylor \& Francis}
}

@article{eu2025walletregistration,
    author={European Commission},
    title={Commission Implementing Regulation (EU) 2025/848 of 6 May 2025 Laying Down Rules of the Application of Regulation (EU) No 910/2014 of the European Parliament and of the Council as Regards the Registration of Wallet-Relying Parties},
    journal = {Official Journal of the European Union},
    year    = {2025},
    volume  = {L 2025/848},
    pages   = {1--15},
    url     = {https://eur-lex.europa.eu/legal-content/EN/TXT/PDF/?uri=OJ:L_202500848}
}

@misc{edps2025wallets,
    author  =   {European Data Protection Supervisor},
    title   =   {TechDispatch: Digital Identity Wallets},
    year    =   {2025},
    url     =   {https://www.edps.europa.eu/system/files/2025-12/25-12-16_techdispatch-digital-identity-wallet_en.pdf}
}

@article{eu2024digitalwallet,
    author={European Union},
    title={Regulation (EU) 2024/1183 of the European Parliament and of the Council of 11 April 2024 Amending Regulation (EU) No 910/2014 as Regards Establishing the European Digital Identity Framework},
    journal = {Official Journal of the European Union},
    year    = {2024},
    volume  = {L 2024/1183},
    pages   = {1--56},
    url     = {https://eur-lex.europa.eu/legal-content/EN/TXT/PDF/?uri=OJ:L_202401183}
}

@misc{euwalletregulation,
    author  =   {European Commission},
    title   =   {The European DigitalIdentity Regulation},
    urldate = {2026-04-03},
    url     =   {https://ec.europa.eu/digital-building-blocks/sites/spaces/EUDIGITALIDENTITYWALLET/pages/915931811/The+European+Digital+Identity+Regulation},
    year = {2026}
}

@misc{openletterEUDIJune2023,
    author  =   {{Epicenter.works}},
    title   =   {Open Letter to Swedish Presidency and Permanent Representations of EU Member States},
    year    =   {2023},  
    urldate =   {2026-02-04},
    url     =   {https://epicenter.works/fileadmin/import/cso-eidas-open_letter_2023.pdf},
    note    =   {Signed by 24 organizations}
}

@misc{openletterEUDIFeb2023,
    author  =   {{Epicenter.works}},
    title   =   {Open Letter to President and Vice Presidents of EU Member States},
    year    =   {2023},    
    urldate =   {2026-02-04},
    url     =   {https://epicenter.works/fileadmin/import/open_letter_eidas_2023-01_0.pdf},
    note    =   {Signed by 39 organizations}
}

@inproceedings{felt2012android,
  title={Android permissions: User attention, comprehension, and behavior},
  author={Felt, Adrienne Porter and Ha, Elizabeth and Egelman, Serge and Haney, Ariel and Chin, Erika and Wagner, David},
  booktitle={Proceedings of the eighth symposium on usable privacy and security},
  pages={1--14},
  year={2012}
}

@article{machuletz2019multiple,
  title={Multiple purposes, multiple problems: A user study of consent dialogs after GDPR},
  author={Machuletz, Dominique and B{\"o}hme, Rainer},
  journal={arXiv preprint arXiv:1908.10048},
  year={2019}
}

@inproceedings{korir2022empirical,
  title={An empirical study of a decentralized identity wallet: Usability, security, and perspectives on user control},
  author={Korir, Maina and Parkin, Simon and Dunphy, Paul},
  booktitle={Eighteenth symposium on usable privacy and security (SOUPS 2022)},
  pages={195--211},
  year={2022}
}

@article{teuschel2023don,
  title={’Don’t Annoy Me With Privacy Decisions!’—Designing Privacy-Preserving User Interfaces for SSI Wallets on Smartphones},
  author={Teuschel, Moritz and P{\"o}hn, Daniela and Grabatin, Michael and Dietz, Felix and Hommel, Wolfgang and Alt, Florian},
  journal={IEEE Access},
  volume={11},
  pages={131814--131835},
  year={2023},
  publisher={IEEE}
}

@article{morkonda2025sign,
  title={“Sign in with... Privacy”: Timely Disclosure of Privacy Differences among Web SSO Login Options},
  author={Morkonda, Srivathsan G and Chiasson, Sonia and van Oorschot, Paul C},
  journal={ACM Transactions on Privacy and Security},
  volume={28},
  number={2},
  pages={1--28},
  year={2025},
  publisher={ACM New York, NY}
}

@article{dimova2023everybody,
  title={Everybody's Looking for SSOmething: A large-scale evaluation on the privacy of OAuth authentication on the web},
  author={Dimova, Yana and Van Goethem, Tom and Joosen, Wouter},
  journal={Proceedings on Privacy Enhancing Technologies},
  year={2023}
}

@article{nissenbaum2004privacy,
  title={Privacy as contextual integrity},
  author={Nissenbaum, Helen},
  journal={Wash. L. Rev.},
  volume={79},
  pages={119},
  year={2004},
  publisher={HeinOnline}
}
\clearpage

\appendix 

\section{Ethical Considerations}

We received IRB approval for our survey and user study. When writing that application at the start of our research and throughout the project, we thoroughly considered the ethical implications of our work and believe that this work did not cause harm or encourage negative outcomes. We identify the following stakeholders of our research: 1) The eventual users of the EUDI, 2) The lawmakers developing the EUDI, 3) The Prolific participants in our survey and user study, 4) The expert participants in our survey, and 5) The research team. 

1) The EUDI is already under development, and it will be deployed in 2026. Users of the EUDI will need to make complex decisions about when to disclose their credentials and when not to. Prior research has shown that making constant decisions about privacy is challenging for users, leading them to make decisions that they regret. The goal of our work is to develop a system that can protect users, which we believe to be inherently positive. A second goal of our system is to inform users when disclosure may be necessary, which can increase the benefit that users gain from the EUDI. Furthermore, the system we evaluate only nudges users and does not force them to make a specific decision, which respects the user's right to make their own decisions. Of course, a nudging system may be used to influence users maliciously by manipulating data. However, we address this concern by analyzing to what extent users can be nudged in a wrong direction and by discussing the importance of trust for the EUDI.  

2) A risk from a paper like this would be that it reduces trust in the EUDI, which could harm lawmakers developing the system. We mitigate this risk by providing a fair assessment of a specific aspect of the EUDI and also discussing its benefits. Furthermore, we evaluate a potential solution to mitigate the identified risks. Lastly, we give a set of recommendations for the EUDI that can mitigate the identified risks. As the EUDI has not yet been deployed, lawmakers can address these risks, which could strengthen trust in the system.

3) Prolific participants were informed about the goals and content of the survey/user study through a consent form. No deception was deployed, and thus, all participants gave informed consent. Participants were compensated for their work at a rate of $\sim$\$6.7 for a 25 min survey, which exceeds the minimum wage in any country from which we recruited. Compensation was not based on participants' performance, and participants who chose to abandon the study and withdraw consent were still compensated. All responses we received from Prolific participants were pseudonomized, and we were not able to link responses to specific individuals. Additionally, we will not publish the demographics and free-text responses to ensure that the dataset does not incidentally leak the information of any participant. As such, we believe the identity of all study participants is fully protected, and thus, there is no risk to them from the responses they provided. Lastly, we made sure to tell participants that the EUDI does not exist yet, and thus, all questions are hypothetical. Similarly, we made sure to tell participants that all data requests and all \name{} recommendations were made up. Thus, we believe the risk that participants misunderstood our study as actual recommendations instead of fictional examples is low. 

4) Expert participants were informed about the goals and content of the survey through a consent form. No deception was deployed, and thus, all experts gave informed consent. We did not compensate experts; however, we offered to keep them updated on the findings of our research. As we recruited experts from fields that are likely to be interested in the findings, we believe this to be sufficient. The responses received from experts were pseudonomized; however, as we reached out to a limited number of experts directly, the research team may be able to make links to individual experts. We did not record any links we may have made and did not report them in any way. Lastly, we deleted the list of experts that we reached out to and never disclosed this information to anyone.

5) Our research did not deal with any topics that could harm the research team. As all user studies were based on solid ethical foundations with IRB approval, we believe there is no risk to the research team from this work. 

\section{Ground Truth Details}
\label[appendix]{sec:app_justification}

To test whether users make credential disclosure mistakes, we had to define what a mistake is. However, since disclosure decisions are inherently subjective, there was no absolute ground truth that we could use as a basis. Thus, three members of the research team categorized all credential-website category pairs. For pairs where there was no unanimous decision, the members discussed the pair until a unanimous decision could be made. We categorized each website-credential pair into justified, unjustified, and uncertain based on whether the website has a valid reason (i.e., a valid use case where having access to the credential is necessary for the website or improves its services) to request the credential. Justified was used when the website had a good reason to request the credential, either because it was necessary or improved the core functionality of the website. Unjustified was used when there was no use case that would make collecting the credential sensible for the website. Uncertain was used for cases where the research team could come up with a use case, but it was minor or improbable (e.g., it may be the case that governments would collect users' diplomas for government benefit programs). After the categorization was completed, there were 150 credential-website category pairs categorized as unjustified, 44 credential-website category pairs categorized as justified, and 16 credential-website category pairs categorized as uncertain. The full categorization can be found in the submitted artifacts under files>justifications.csv.

To validate our categorization, we compared all justified and unjustified website-credential pairs that we used in the user study with the results of the expert survey. \cref{tab:good_justification_eval} shows the justified pairs, and \cref{tab:no_justification_eval} shows the unjustified pairs with the results of the expert survey. 
Except for three pairs, our categorization matches the expert survey responses. The pairs where the categorization does not match can be explained as follows: 1) International Ground Travel-Official ID: Currently, the official ID is not collected; however, it is likely that this will change in the future to comply with border control requirements, especially for international travel across the Schengen borders. 2) Social Media-Official ID: Governments are tightening requirements for social media sites to verify users' age and put in place age restrictions, which would make providing an official ID mandatory. 3) Bank-Visa: It is likely that the visa was confused with a residency permit. Banks usually collect a residency permit; however, a visa is not necessary.

\begin{table*}[tbp]
\centering
\setlength{\tabcolsep}{6pt}
\renewcommand{\arraystretch}{1.1}
\caption{Overview of website-credential pairs and expert evaluations for scenarios categorized as having a good justification for the credential request.}
\label{tab:good_justification_eval}
\begin{tabular}{llccc}
\toprule
\multicolumn{2}{c}{} & \multicolumn{3}{c}{\textbf{Expert}} \\
\cmidrule(lr){3-5}
\textbf{Website Category} & \textbf{Credential} & \textbf{Would Disclose (\%)} & \textbf{Recommend not Disclose (\%)} & \textbf{Is Necessary (\%)} \\
\midrule
Pharmacy            & Health Insurance       & 54\%  & 17\% & 33\% \\
Pharmacy            & Prescriptions          & 63\%  & 8\%  & 67\% \\
Online Doctors      & Prescriptions          & 67\%  & 13\% & 100\% \\
Online Doctors      & Medical Records        & 71\%  & 8\%  & 67\% \\
Government          & Marriage Certificate   & 50\%  & 13\% & 33\% \\
Job Portal          & Employment History     & 71\%  & 13\% & 33\% \\
University          & University Transcript  & 54\%  & 21\% & 67\% \\
University          & Diploma                & 54\%  & 17\% & 33\% \\
Air Travel          & Visa                   & 58\%  & 25\% & 33\% \\
Int.\ Ground Travel & Official ID            & 38\%  & 38\% & 33\% \\
Car Rental          & Driver's License       & 67\%  & 8\%  & 100\% \\
Social Media        & Official ID            & 17\%  & 54\% & 0\% \\
\bottomrule
\end{tabular}
\end{table*}

\begin{table*}[tb]
\centering
\setlength{\tabcolsep}{6pt}
\renewcommand{\arraystretch}{1.1}
\caption{Overview of website-credential pairs and expert evaluations for scenarios categorized as having no or a very limited justification for the credential request.}
\label{tab:no_justification_eval}
\begin{tabular}{llccc}
\toprule
\multicolumn{2}{c}{} & \multicolumn{3}{c}{\textbf{Expert}} \\
\cmidrule(lr){3-5}
\textbf{Website} & \textbf{Credential} & \textbf{Would Disclose (\%)} & \textbf{Recommend not Disclose (\%)} & \textbf{Is Necessary (\%)} \\
\midrule
Pharmacy              & Diploma                & 0\%  & 96\% & 0\%  \\
Online Doctors        & Visa                   & 0\%  & 92\% & 0\%  \\
Job Portal            & Prescriptions          & 4\%  & 92\% & 0\%  \\
Job Portal            & Medical Records        & 4\%  & 92\% & 0\%  \\
Air Travel            & Health Insurance       & 8\%  & 92\% & 0\%  \\
Int.\ Ground Travel   & Diploma                & 0\%  & 88\% & 0\%  \\
Int.\ Ground Travel   & Prescriptions          & 4\%  & 88\% & 0\%  \\
Bank                  & Visa                   & 25\% & 58\% & 0\%  \\
Real Estate           & Health Insurance       & 0\%  & 79\% & 0\%  \\
Gaming                & Medical Records        & 0\%  & 83\% & 0\%  \\
E-commerce            & Professional Licenses  & 0\%  & 83\% & 0\%  \\
E-commerce            & Health Insurance       & 0\%  & 83\% & 0\%  \\
E-commerce            & Disability Status      & 4\%  & 83\% & 0\%  \\
News                  & Official ID            & 17\% & 63\% & 33\% \\
News                  & Birth Certificate      & 4\%  & 75\% & 0\%  \\
News                  & Professional Licenses  & 4\%  & 71\% & 0\%  \\
Payment Services      & Visa                   & 17\% & 71\% & 0\%  \\
Payment Services      & Prescriptions          & 4\%  & 88\% & 0\%  \\
\bottomrule
\end{tabular}
\end{table*}
\section{Additional Results}
\label[appendix]{sec:app_additional_results}

In this section, we provide complementary figures for the results discussed in \cref{sec:results} and discuss additional results from our survey and user study.

\subsection{Additional Figures}
\label[appendix]{subsec:app_additional_figures}

\begin{figure}[tb]
     \centering
     \begin{subfigure}[b]{0.45\linewidth}
         \centering
        \includegraphics[width=\linewidth]{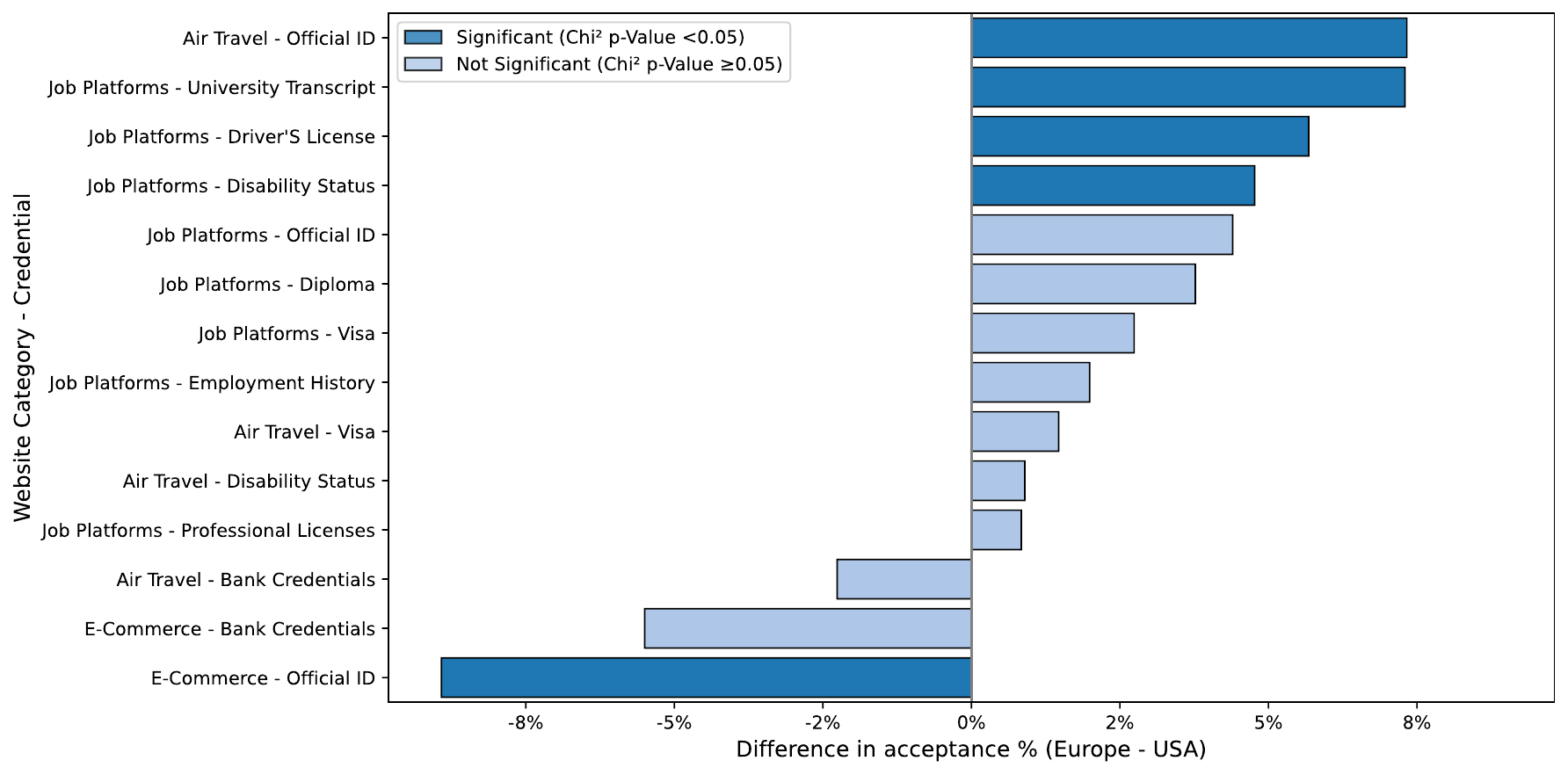}
        \caption{Europe-based websites vs. USA-based websites.}
        \label{fig:europe_usa_comparison}
     \end{subfigure}
     \hfill
     \begin{subfigure}[b]{0.45\linewidth}
         \centering
        \includegraphics[width=\linewidth]{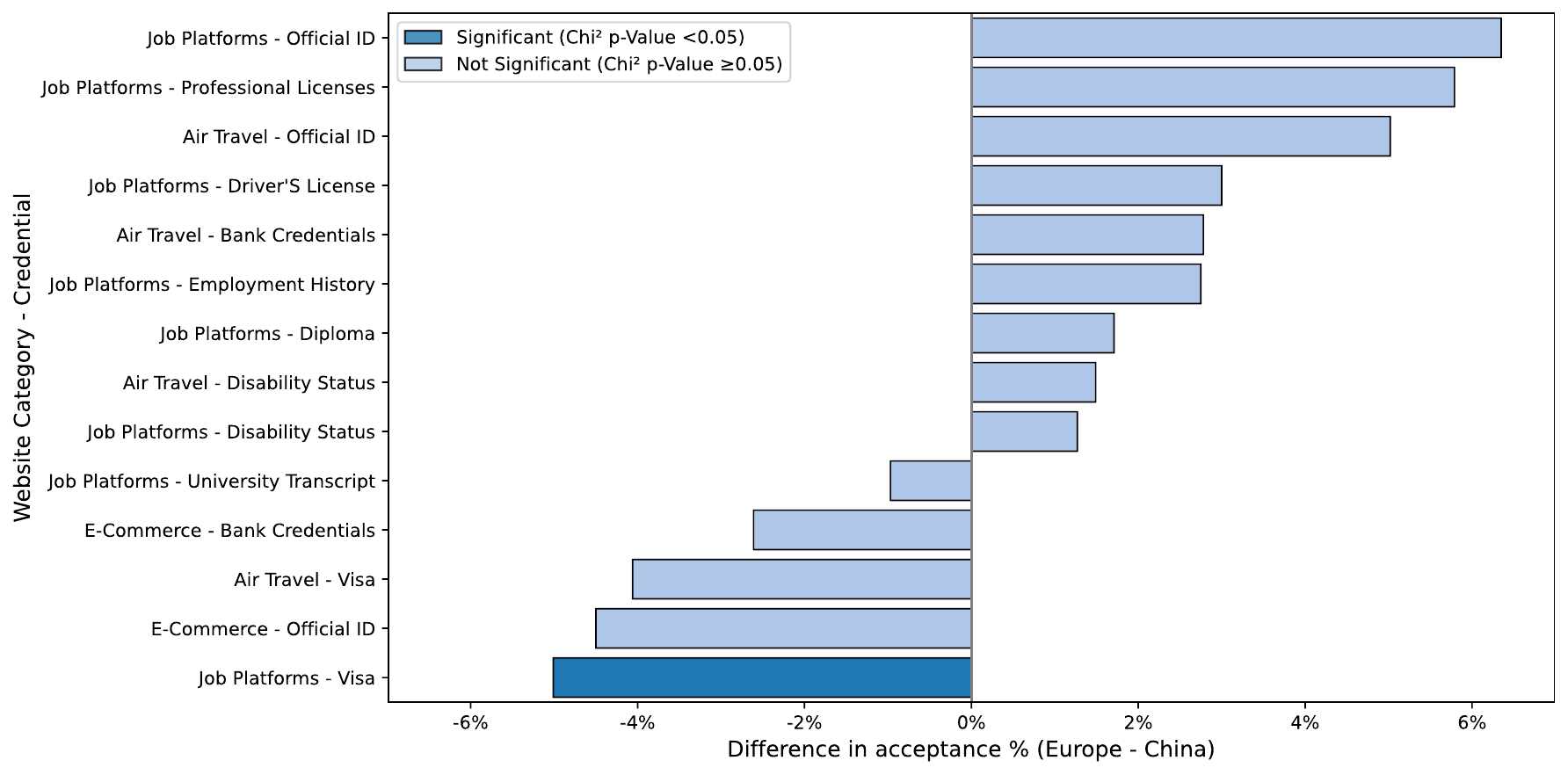}
        \caption{Europe-based websites vs. China-based websites.}
        \label{fig:europe_china_comparison}
     \end{subfigure}
    \caption{Results of the user survey showing how many participants would disclose a credential for Europe-based websites vs. China-/USA-based websites. Limited to the website categories where different countries were considered and to the website-credential pairs where at least 10\% of participants would disclose the credential for either country. Positive percentages indicate increased disclosure for Europa-based websites.}
    \label{fig:country_comparison}
\end{figure}

\begin{figure}[tb]
    \centering
    \includegraphics[width=\linewidth]{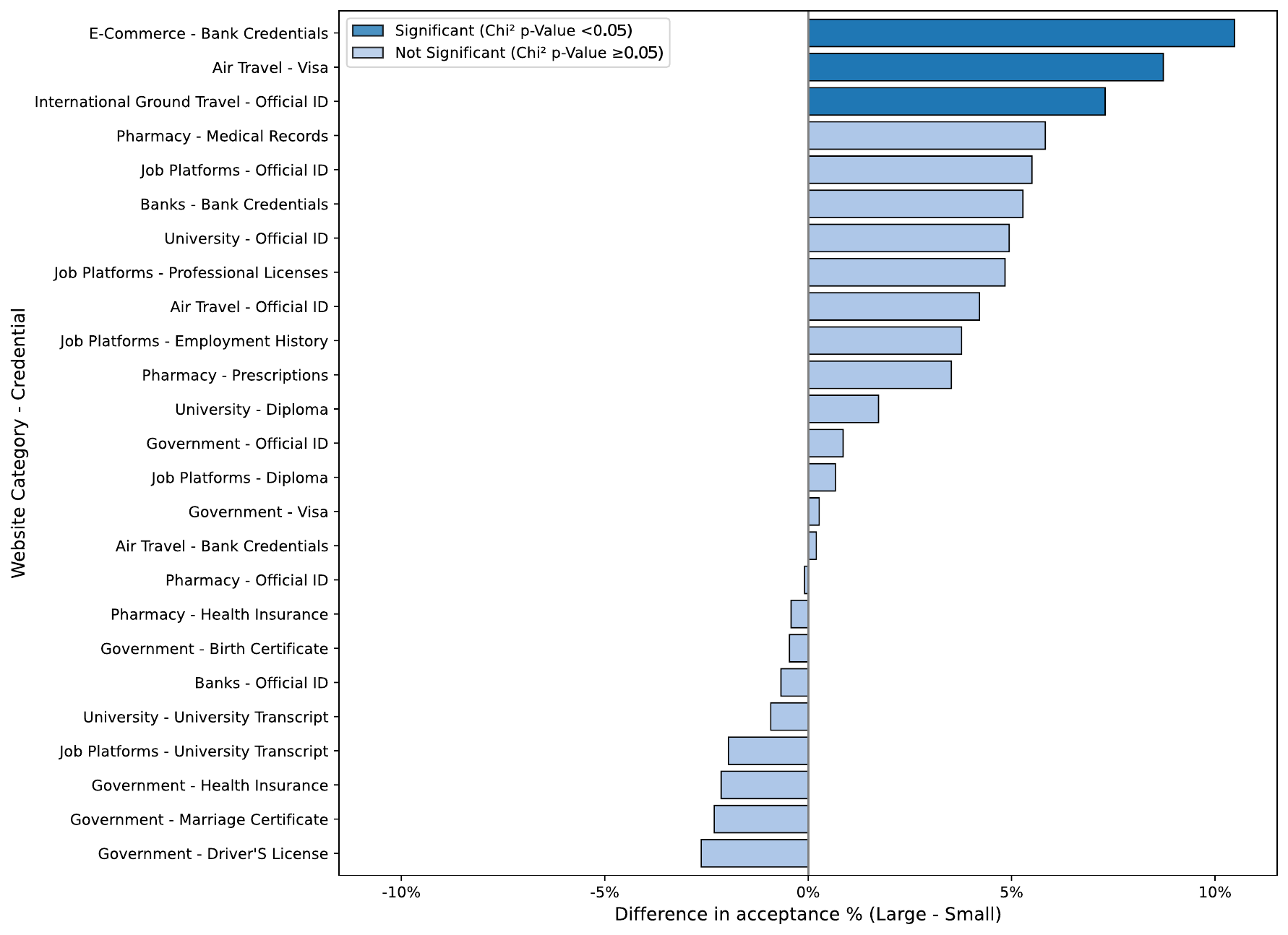}
    \caption{Results of the user survey showing how many participants would disclose a credential for high-traffic (large) websites vs. low-traffic (small) websites. Limited to the website categories where different website sizes were considered and to the website-credential pairs where at least 30\% of participants would disclose the credential for either country. Positive percentages indicate increased disclosure for large websites.}
    \label{fig:large_small_comparison}
\end{figure}

\cref{fig:country_comparison} presents the user survey results showing how often participants stated that they would disclose a credential to a Europe-based compared to a USA- or China-based website. \cref{fig:large_small_comparison} presents the user study results showing how often participants would disclose a credential to high-traffic (large) websites compared to low-traffic (small) websites. The figures show some significant differences for Europe vs. the USA and only one for Europe vs. China. The significant differences are also mixed between cases where there is more disclosure for Europe-based websites and some that show less disclosure. Thus, it seems like users do not heavily base their decisions on the country in which the website is based.

\begin{figure}[tb]
     \centering
     \begin{subfigure}[b]{0.45\linewidth}
         \centering
         \includegraphics[width=\linewidth]{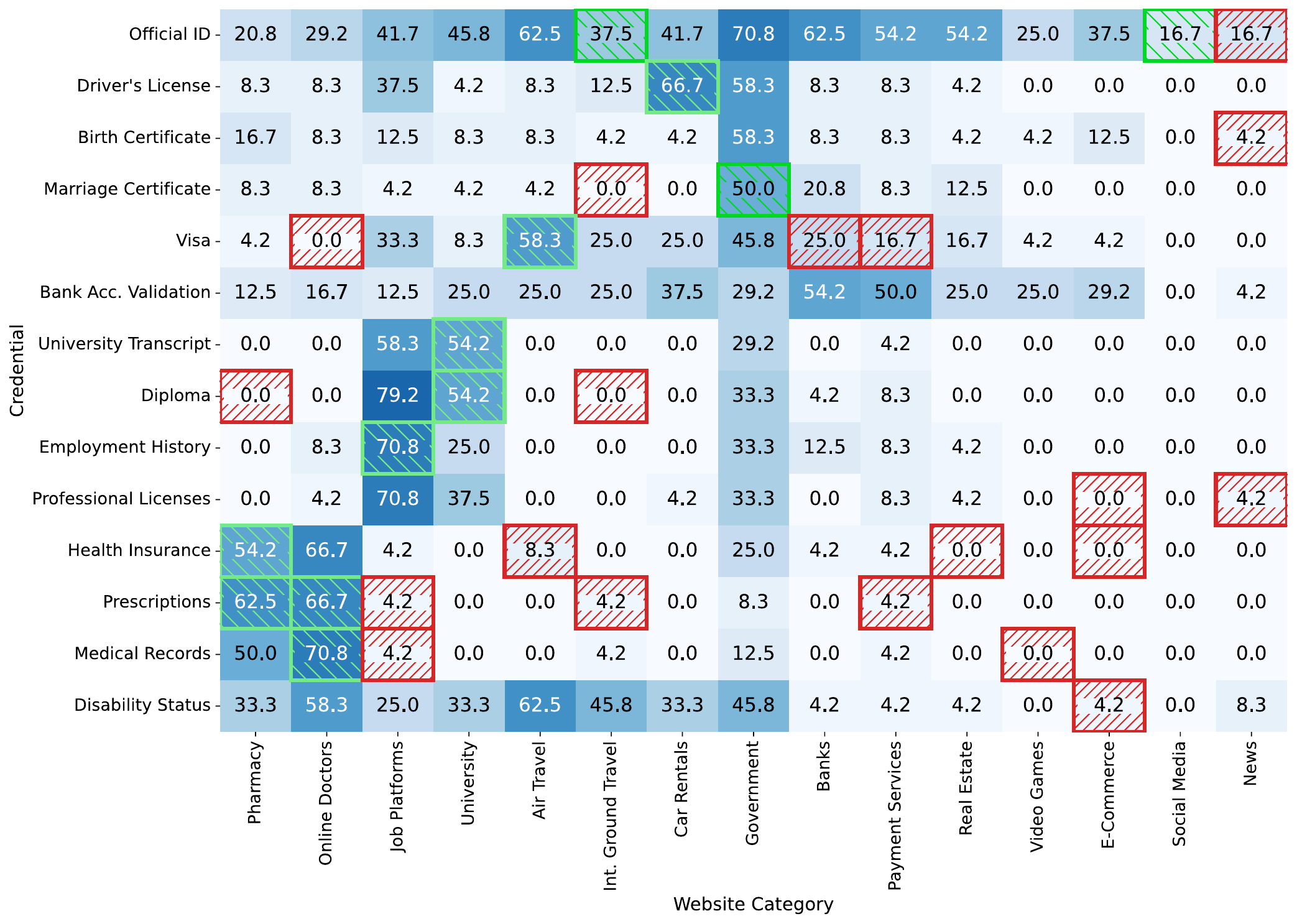}
         \caption{Expert would disclose.}
         \label{fig:expert_would_heatmap}
     \end{subfigure}
     \hfill
     \begin{subfigure}[b]{0.45\linewidth}
         \centering
         \includegraphics[width=\linewidth]{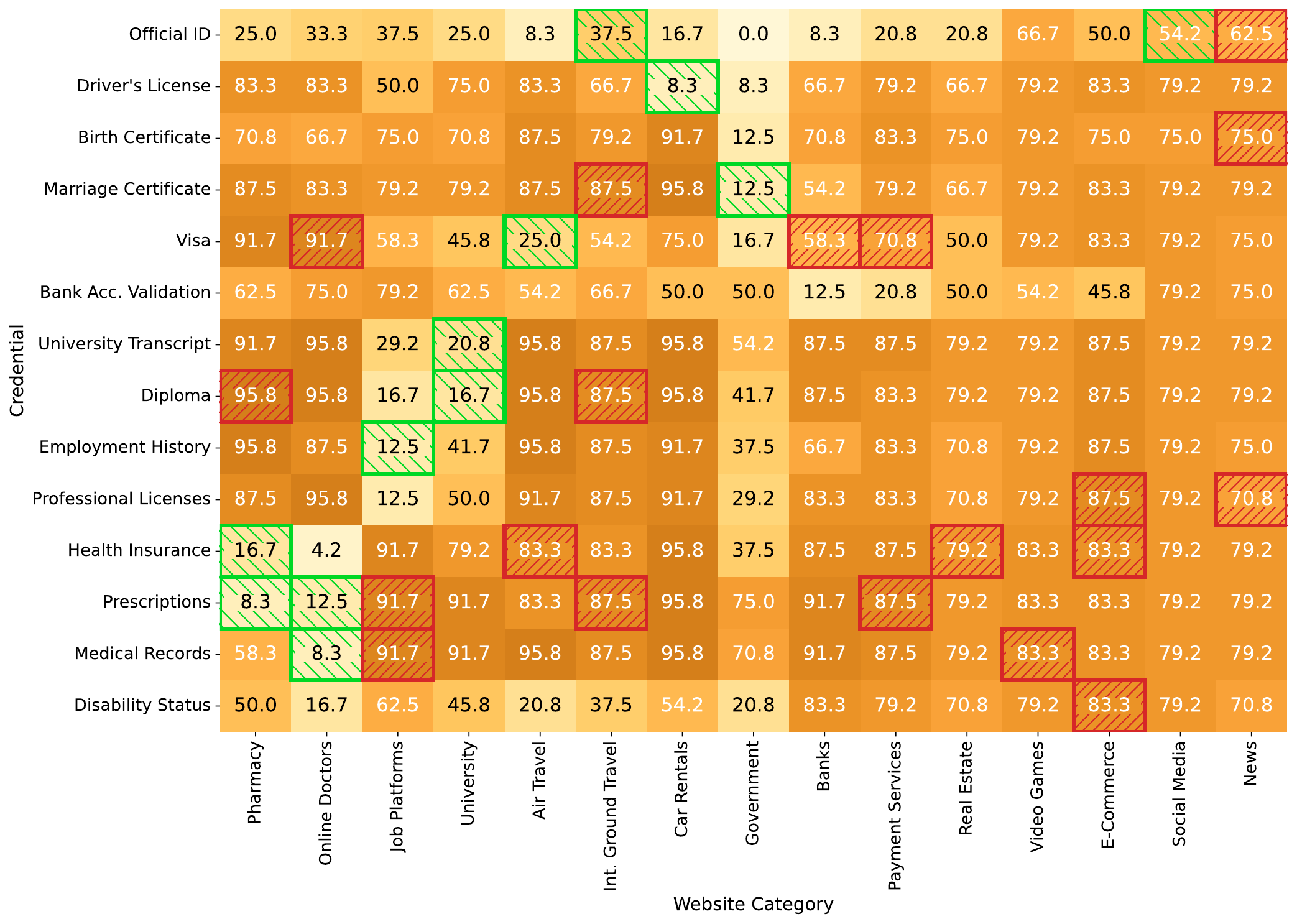}
         \caption{Experts recommend never to disclose.}
         \label{fig:expert_never_heatmap}
     \end{subfigure}
     \hfill
     \begin{subfigure}[b]{0.45\linewidth}
         \centering
         \includegraphics[width=\linewidth]{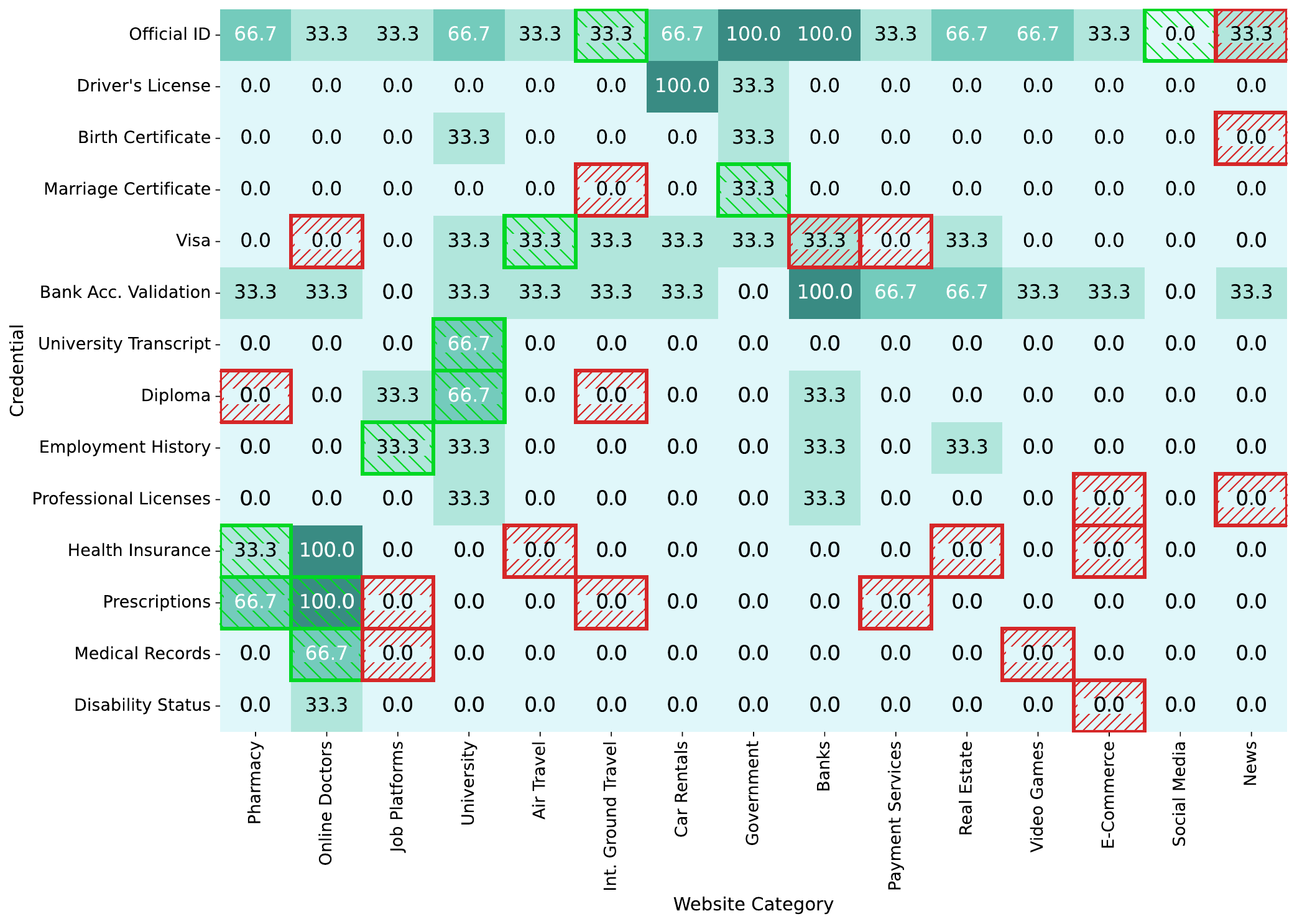}
         \caption{Law experts deem necessary.}
         \label{fig:law_nec_heatmap}
     \end{subfigure}
        \caption{Results of the expert survey showing how many cybersecurity, policy, and ethics experts recommend never to disclose a credential to a website and how many would disclose the credential to the website. Also shown is how many law experts state that the credential is necessary for the websites. Fields shaded in green are the phase 2 scenarios with a justification and fields shaded in red are the phase 2 scenarios with no justification for the credential request.}
        \label{fig:expert_heatmaps}
\end{figure}

\begin{figure}[tb]
     \centering
    \includegraphics[width=\linewidth]{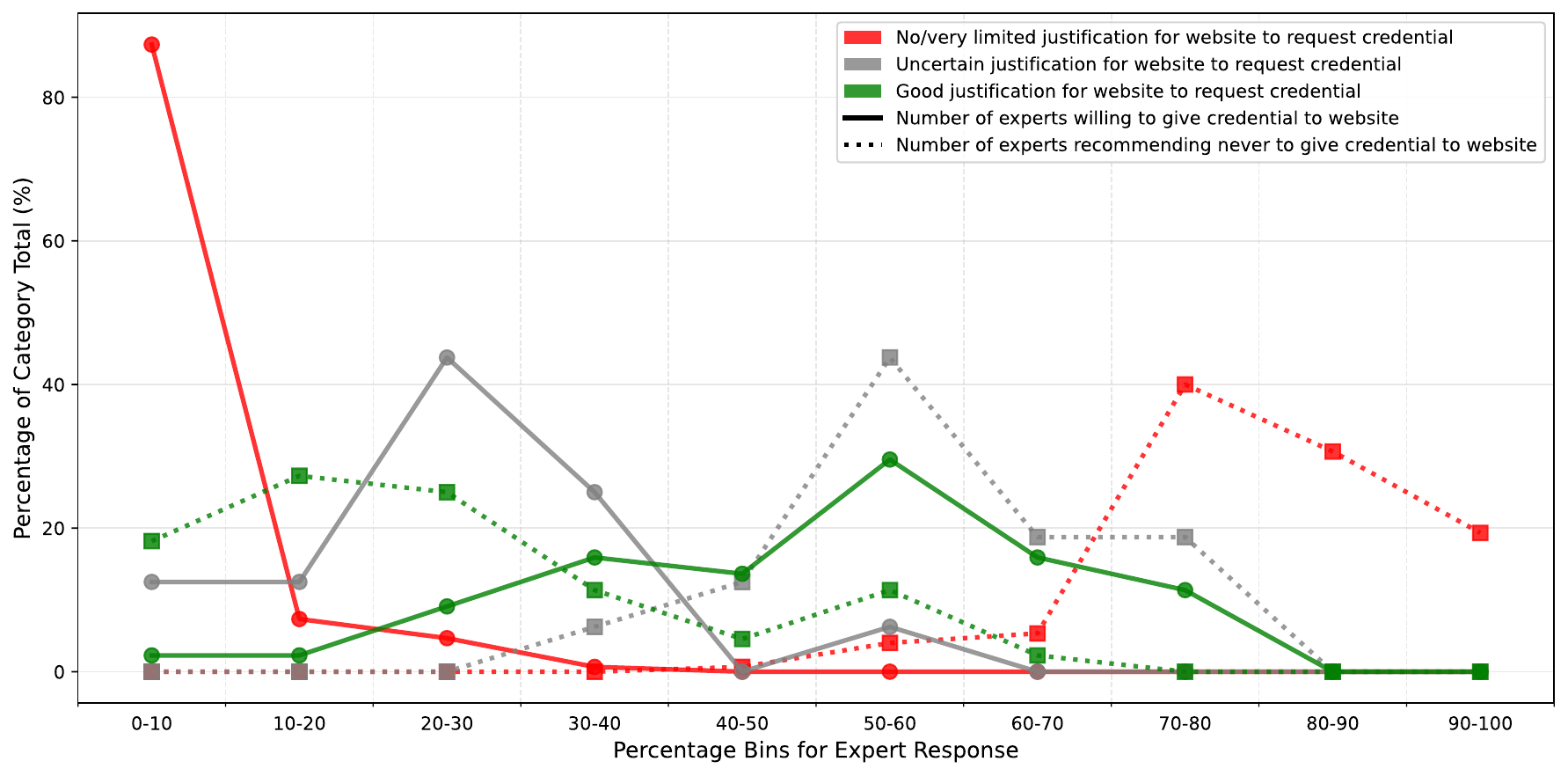}
    \caption{Results of the expert survey showing how many cybersecurity, policy, and ethics experts recommend never to disclose a credential to a website and how many would disclose the credential to the website. Red = unjustified request, green = justified request, gray = uncertain requests, solid line = experts would disclose, dotted line = expert recommend not to disclose. x-axis = percentage bins, y-axis = percentage of total number of website-credential pairs in a category for which the expert result falls in that category.}
    \label{fig:alignment_line}
\end{figure}

\cref{fig:expert_would_heatmap} presents the expert survey results showing the percentage of cybersecurity, policy, and ethics experts who would disclose a credential to a website. \cref{fig:expert_never_heatmap} presents the expert survey results showing the percentage of cybersecurity, policy, and ethics experts who recommend never to disclose a credential to a website. \cref{fig:law_nec_heatmap} presents the expert survey results showing the percentage of law experts who state that a credential is necessary for a website. In all figures, the justified scenarios from the user study are shaded in green, and the unjustified scenarios from the user study are shaded in red. \cref{fig:alignment_line} visualizes the percentage of experts who recommend never to disclose a credential and the percentage of experts who would disclose a credential. The x-axis has fixed percentage bins (each bin contains 10 percentage points). All website-credential pairs were categorized as unjustified (red), justified (green), and uncertain (gray), as described in \cref{subsec:study_methods}. For each category of the total number of website-credential pairs in the category, the percentage of experts who would disclose a credential in each percentage bin was plotted as a solid line. For each category, of the total number of website-credential pairs in the category, the percentage of experts who would recommend never to disclose a credential in each percentage bin was plotted as a solid line. For example, if the justified category has 44 website-credential pairs, and for 22 of them, the percentage of experts willing to disclose the credential is between 40\% and 50\%, then there would be a dot on the 40-50 x-axis at the 50\% y-axis for the solid green line.

\subsection{Credential Sensitivity \& Frequency of Use}

\cref{tab:app_comfort} presents the results of the user study showing for each credential, how many participants own or owned a physical or electronic document that corresponds to this credential. Furthermore, it shows, on average, for the participants who own the document, how frequently they use that document and how comfortable they feel using that document. Lastly, it shows how many participants did not choose the highest comfort response for a document, and for those participants, the average score of when they would still share the document. All scores were instantiated with a text to explain what the score meant.

\begin{table*}[tb]
\centering
\caption{User survey results showing how many participants own a credential (physically or electronically), how frequently they use it, how comfortable they are with sharing, and how often they would share if uncomfortable. Under "Not Fully Comfortable" is the number of participants who did not choose 5 for their comfort level. Each number was instantiated with a text explaining what frequency/comfort/still share when uncomfortable at that level meant. lr = least restrictive, mr = most restrictive.}
\label{tab:app_comfort}
\small
\begin{tabular}{l r r r r r}
\toprule
\textbf{Credential} & 
\textbf{Own Credential} & 
\textbf{Avg Frequency} & 
\textbf{Avg Comfort} & 
\textbf{Not Fully Comfortable} & 
\textbf{Avg Still Share} \\
& \textbf{\# Participants} & \textbf{(1=low, 5=high)} & \textbf{(1=high, 6=low)} & \textbf{\# Participants} & \textbf{(1=lr, 5=mr)} \\
\midrule
Official ID & 1012 & 2.46 & 3.58 & 923 & 2.92 \\
Driver's License & 866 & 3.42 & 3.38 & 730 & 3.04 \\
Birth Certificate & 876 & 4.42 & 3.92 & 789 & 3.15 \\
Marriage Certificate & 262 & 4.33 & 3.82 & 232 & 3.05 \\
Visa & 331 & 3.44 & 3.61 & 291 & 3.09 \\
Bank Acc. Validation & 752 & 3.44 & 4.61 & 736 & 3.54 \\
University Transcript & 628 & 3.86 & 2.98 & 453 & 2.4 \\
Diploma & 720 & 4.05 & 2.77 & 472 & 2.38 \\
Employment History & 494 & 3.83 & 3.09 & 374 & 2.18 \\
Professional Licenses & 214 & 3.7 & 3.03 & 152 & 2.47 \\
Health Insurance & 921 & 2.43 & 3.3 & 771 & 2.38 \\
Prescriptions & 733 & 2.63 & 3.45 & 650 & 2.42 \\
Medical Records & 566 & 3.82 & 4.1 & 533 & 2.94 \\
Disability Status & 51 & 3.2 & 3.59 & 46 & 2.5 \\
\bottomrule
\end{tabular}
\end{table*}

\subsection{Request Response for Justified Scenarios}
\label[appendix]{subsec:app_justified_request}

\begin{figure}[htb]
    \centering
    \includegraphics[width=\linewidth]{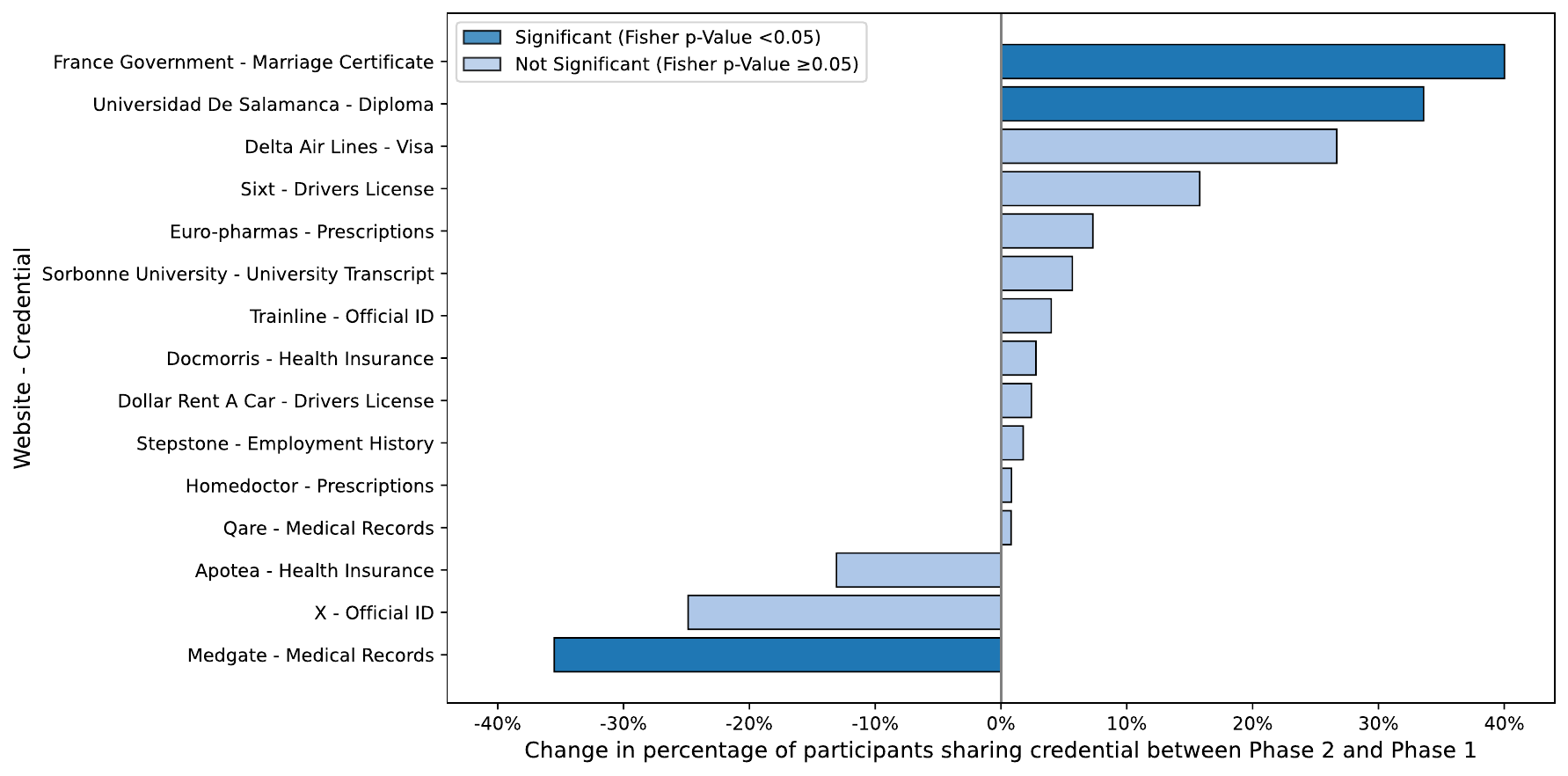}
    \caption{Comparison of how many participants said they would disclose a credential for a website in the user survey, as compared to how many control group participants in the user study disclosed the credential to the website. Positive percentages indicate increased disclosure in the user study. For all website-credential pairs, disclosure is justified.}
    \label{fig:chart_overshare_yes}
\end{figure}

For all user study scenarios in which disclosure is justified, we compared the responses of the control group with the number of users who said that they would disclose the same credential to the same website in our survey. The results are presented in \cref{fig:chart_overshare_yes}. Compared to the unjustified scenarios, we see a lower percentage of significant increases in oversharing ($\sim$13\% for justified vs. $\sim$28\% for unjustified). Furthermore, in the justified scenarios, we see a scenario where users disclosed significantly less when seeing a request, which never happened in the unjustified scenarios. This indicates that while increased disclosure when seeing a request occurs both when the request is justified and unjustified, the effect is stronger in the unjustified cases, which lead to oversharing. Thus, requests lead more to oversharing when disclosure is unjustified than to the correct disclosure behavior when disclosing is justified.

\subsection{Expert Types}

\begin{figure}[tb]
     \centering
     \begin{subfigure}[b]{0.45\linewidth}
         \centering
        \includegraphics[width=\linewidth]{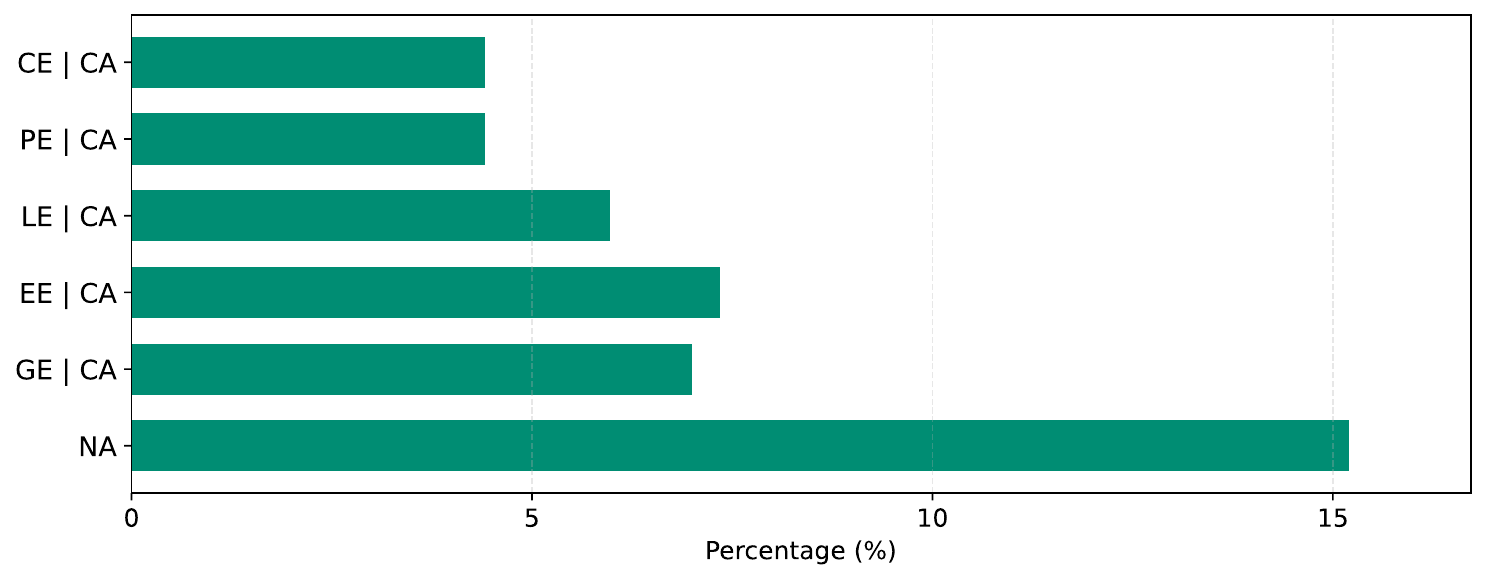}
        \caption{\name{} only shows expert opinion with 5-10\% yes. Unjustified credential request.}
        \label{fig:expert_type_S3}
     \end{subfigure}
     \hfill
     \begin{subfigure}[b]{0.45\linewidth}
         \centering
        \includegraphics[width=\linewidth]{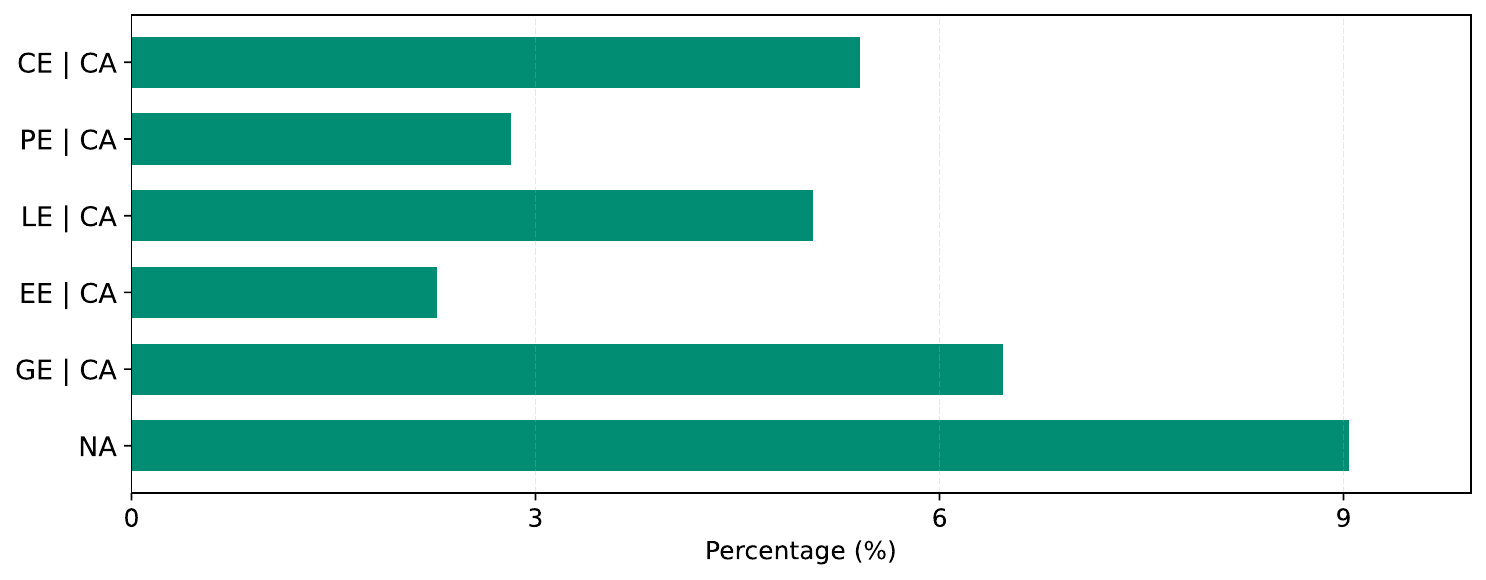}
        \caption{\name{} shows expert and user opinion with 5-10\% yes. Unjustified credential request.}
        \label{fig:expert_type_S5}
     \end{subfigure}
     \hfill
     \begin{subfigure}[b]{0.45\linewidth}
         \centering
        \includegraphics[width=\linewidth]{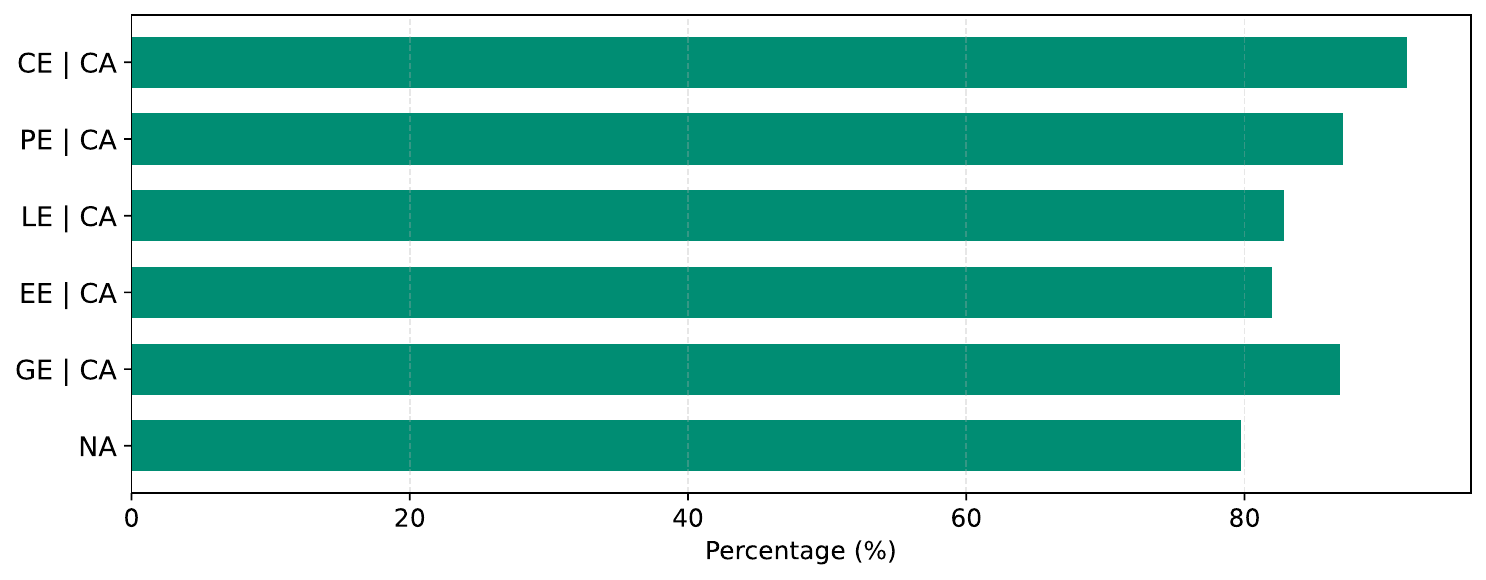}
        \caption{\name{} shows expert and user opinion with 85-90\% yes. Justified credential request.}
        \label{fig:expert_type_S6}
     \end{subfigure}
     \hfill
    \caption{User study result showing the percentage of users who disclosed their credentials in the group with no \name{} and the group with a \name{} displaying the correct information with different expert types.}
    \label{fig:expert_types}
\end{figure}

\cref{fig:expert_types} presents the user study results showing how often participants disclosed their credentials in the group without a \name{}, in the group with a \name{}, showing the recommendation of a generic expert, and in the groups with a \name{}, showing the recommendation of a specific expert type. The results are split into the scenarios: unjustified request with the \name{} only showing the correct expert recommendation, unjustified request with the \name{} showing the correct expert recommendation and user opinion, and justified request with the \name{} showing the correct expert recommendation and user opinion. None of the differences between the responses for different expert types is significant, and there is no clear pattern for how the participants reacted to the expert types in the different scenarios. This indicates that the type of expert in the expert recommendation does not matter to users for EUDI credential disclosure decisions.

\subsection{Demographic Disclosure Differences}

\begin{figure}[tb]
    \centering
    \includegraphics[width=0.8\linewidth]{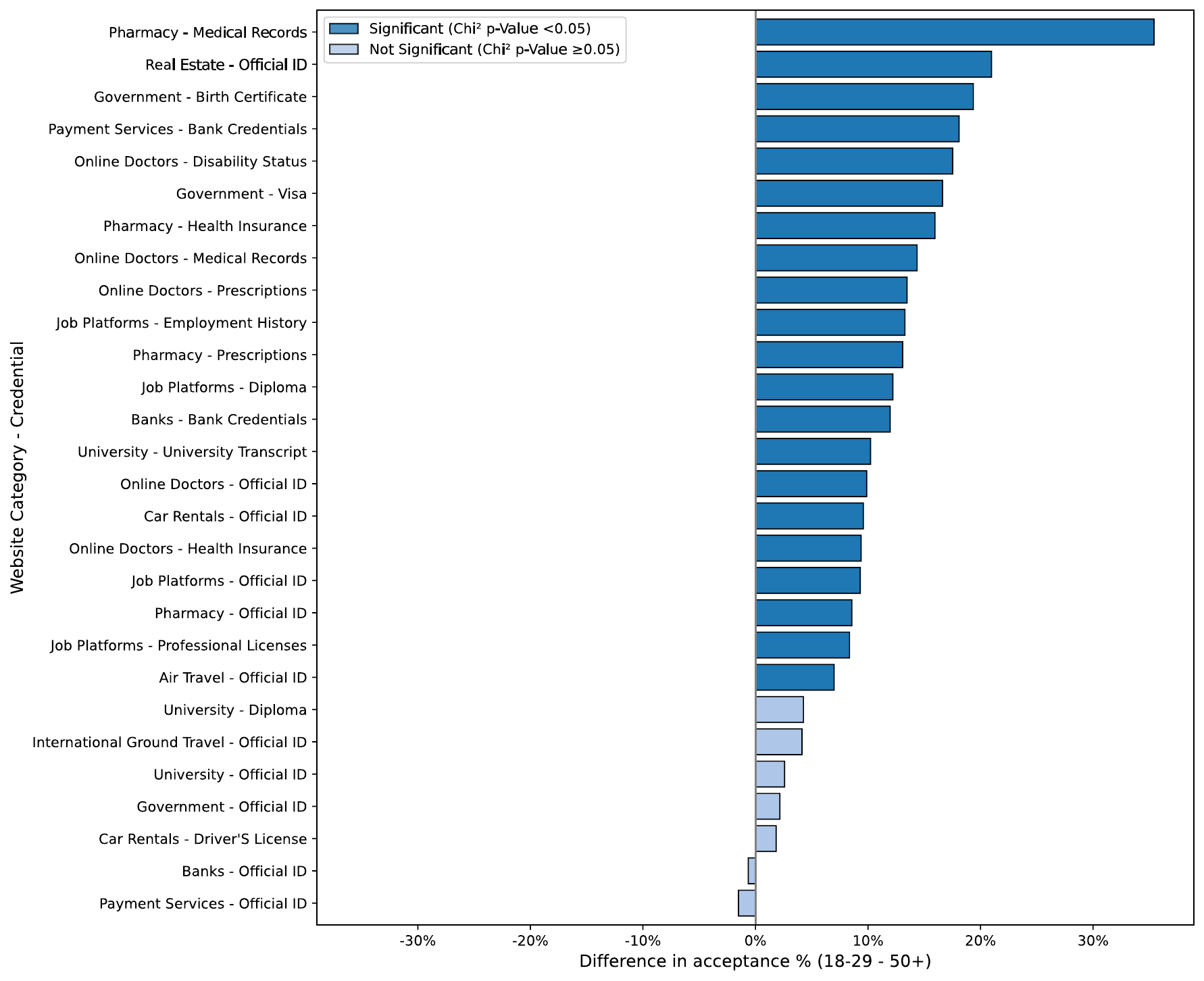}
    \caption{Results of the user survey comparing the percentage of users who stated they would disclose a credential to a website in the 18-29 and the 50+ age groups. Limited to the website-credential pairs where at least 50\% of participants in either age group would disclose the credential. Positive percentages indicate increased disclosure in the 18-29 age group.}
    \label{fig:disclosure_age_comparison}
\end{figure}

\Cref{fig:disclosure_age_comparison} presents the user survey results showing how often participants in the 18-29 age group stated that they would disclose a credential for a website compared to the 50+ age group. The results indicate that users in the 50+ age group are significantly less likely to disclose credentials.

\section{Additional Experiment Screenshots}
\label[appendix]{sec:detailed_survey_images}

In this section, we show additional screenshots from our survey and user study as described in \cref{sec:setup}.

\begin{figure}[tb]
     \centering
     \begin{subfigure}[b]{0.45\linewidth}
         \centering
        \includegraphics[width=\linewidth]{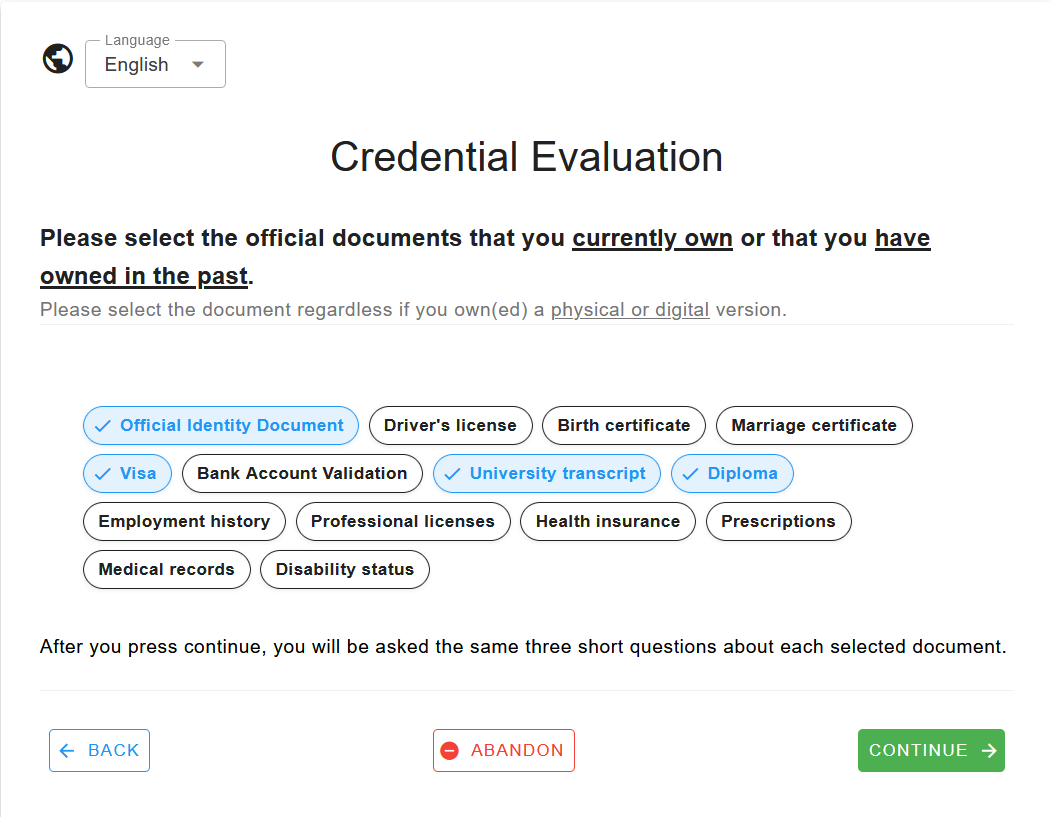}
        \caption{Credential selector.}
        \label{fig:survey_image_credential_choice}
     \end{subfigure}
     \hfill
     \begin{subfigure}[b]{0.45\linewidth}
        \centering
        \includegraphics[width=\linewidth]{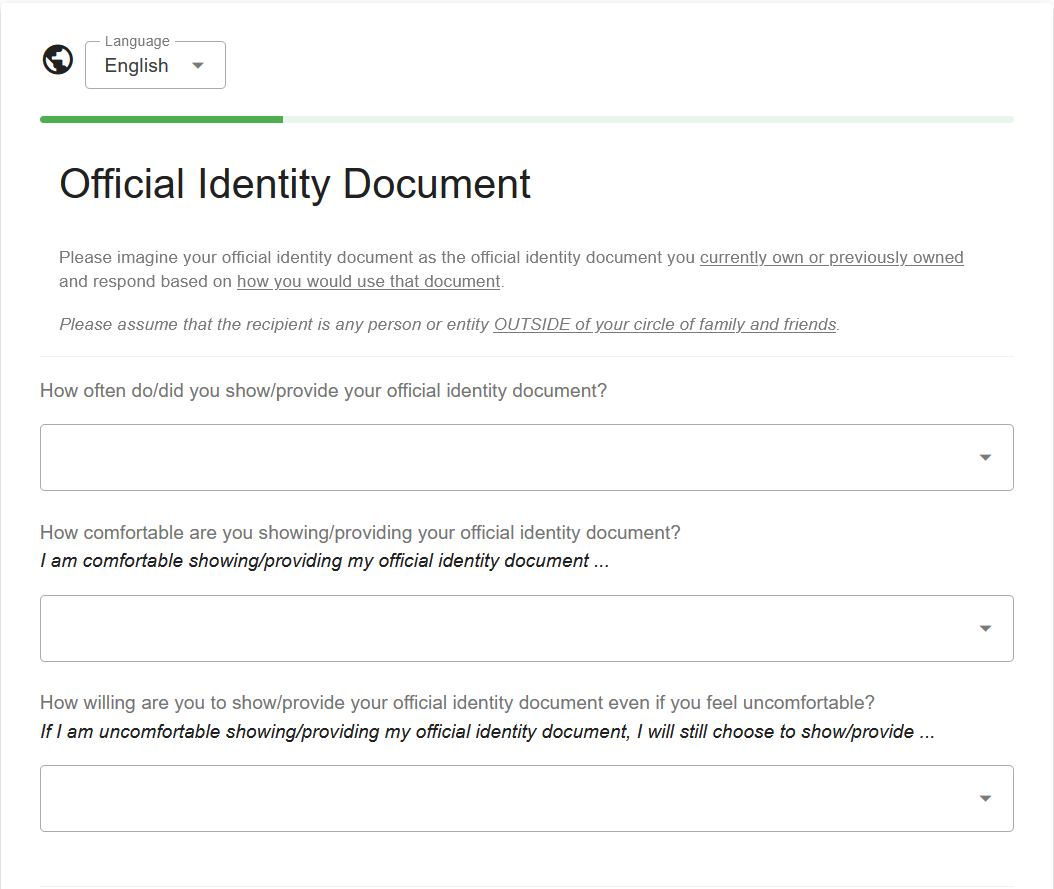}
        \caption{Credential task questions.}
        \label{fig:survey_image_credential_question}
     \end{subfigure}
    \caption{User survey screens for the credential task. Participants first choose the credentials they own and then answer the questions for each credential.}
    \label{fig:survey_image_credential}
\end{figure}

\begin{figure}[tb]
     \centering
    \includegraphics[width=0.8\linewidth]{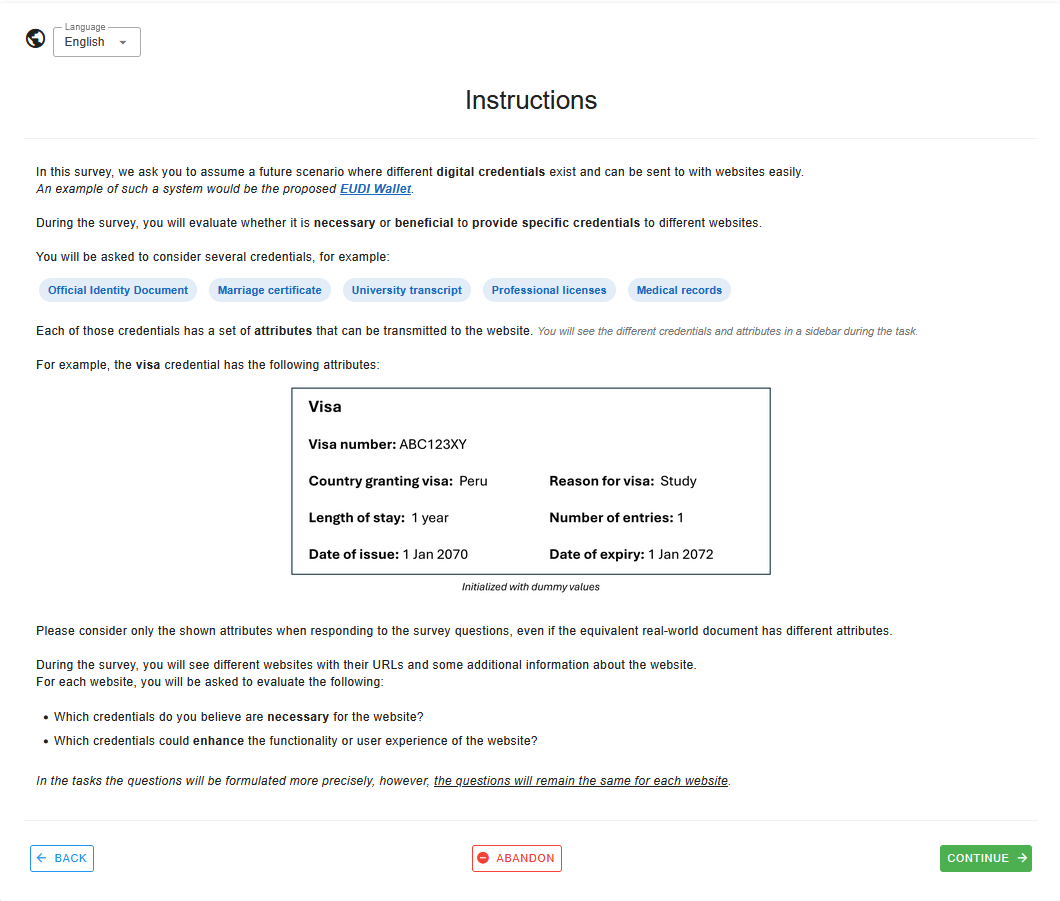}
    \caption{Survey screens showing the instruction page for the website task. The version shown is for users (experts saw different evaluation criteria at the end of the instructions).}
    \label{fig:survey_image_website_instruction}
\end{figure}

\cref{fig:survey_image_credential} shows the credential selector and the credential task questions in the user survey. \cref{fig:survey_image_website_instruction} shows the instruction page for the website task in the survey.

\begin{figure}[tb]
     \centering
     \begin{subfigure}[b]{0.45\linewidth}
         \centering
        \includegraphics[width=\linewidth]{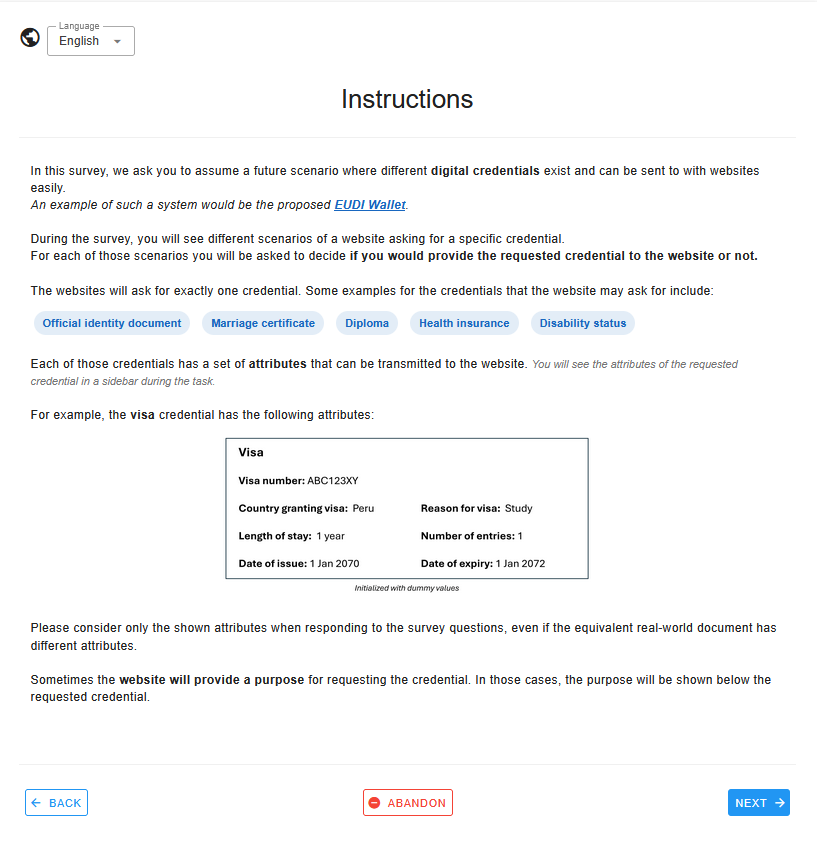}
        \caption{Instruction page describing EUDI.}
        \label{fig:user_study_instruction_1}
     \end{subfigure}
     \hfill
     \begin{subfigure}[b]{0.45\linewidth}
        \centering
        \includegraphics[width=\linewidth]{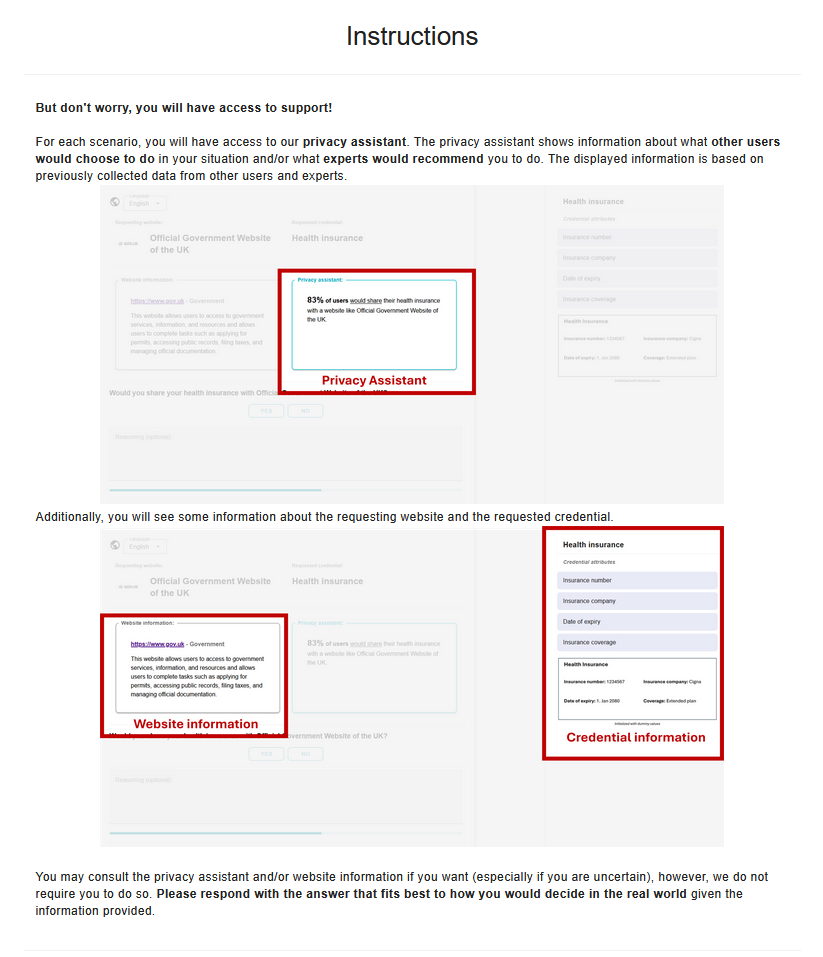}
        \caption{Instruction page introducing \name{}.}
        \label{fig:user_study_instruction_2}
     \end{subfigure}
    \caption{User study screens showing the instruction pages for the scenario task.}
    \label{fig:user_study_instruction_combined}
\end{figure}

\begin{figure}[tb]
     \centering
    \begin{subfigure}[b]{0.45\linewidth}
         \centering
        \includegraphics[width=\linewidth]{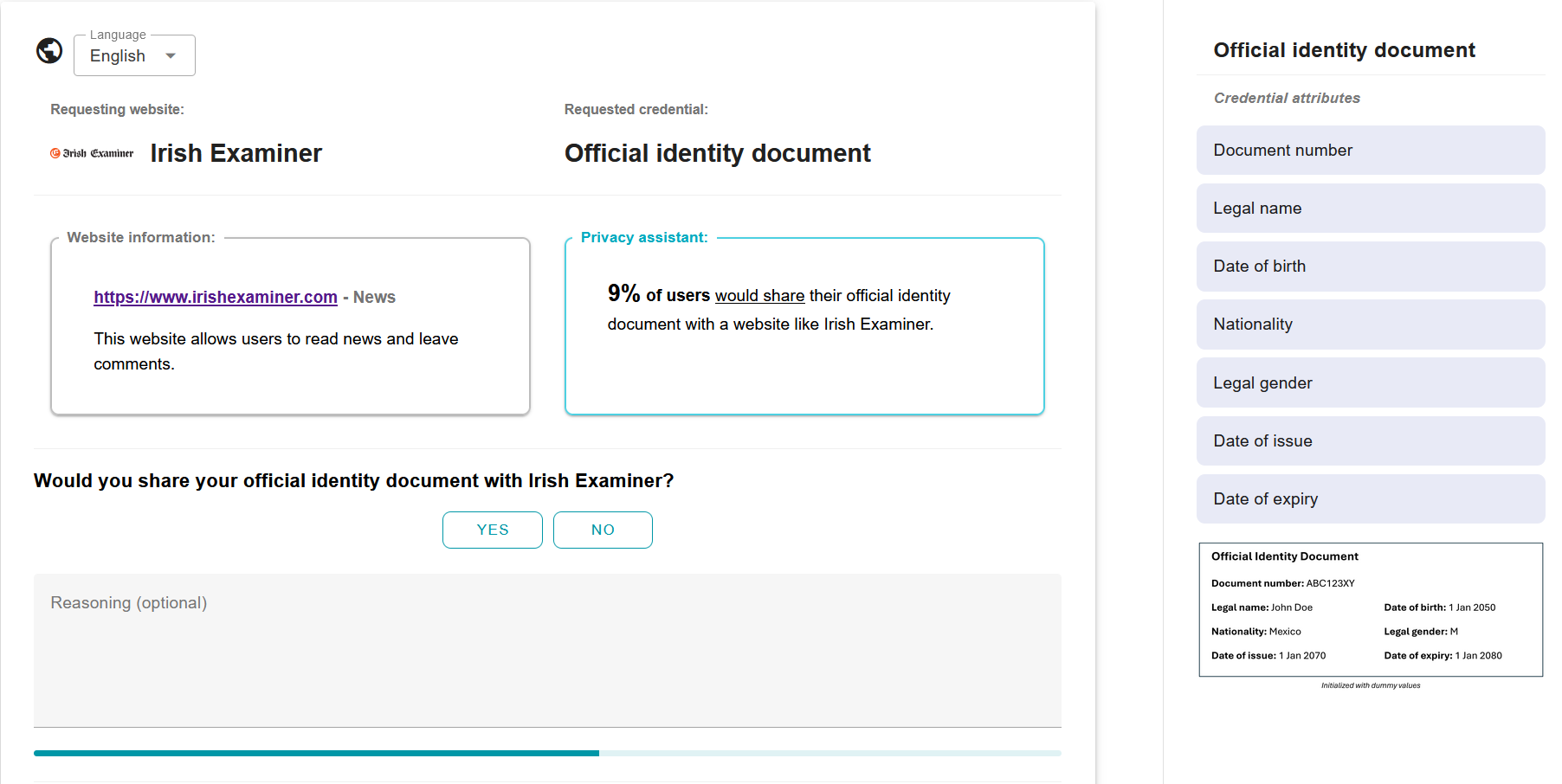}
        \caption{Scenario with \name{} with a single statement.}
        \label{fig:survey_image_scenario_single_support}
     \end{subfigure}
     \hfill
     \begin{subfigure}[b]{0.45\linewidth}
        \centering
        \includegraphics[width=\linewidth]{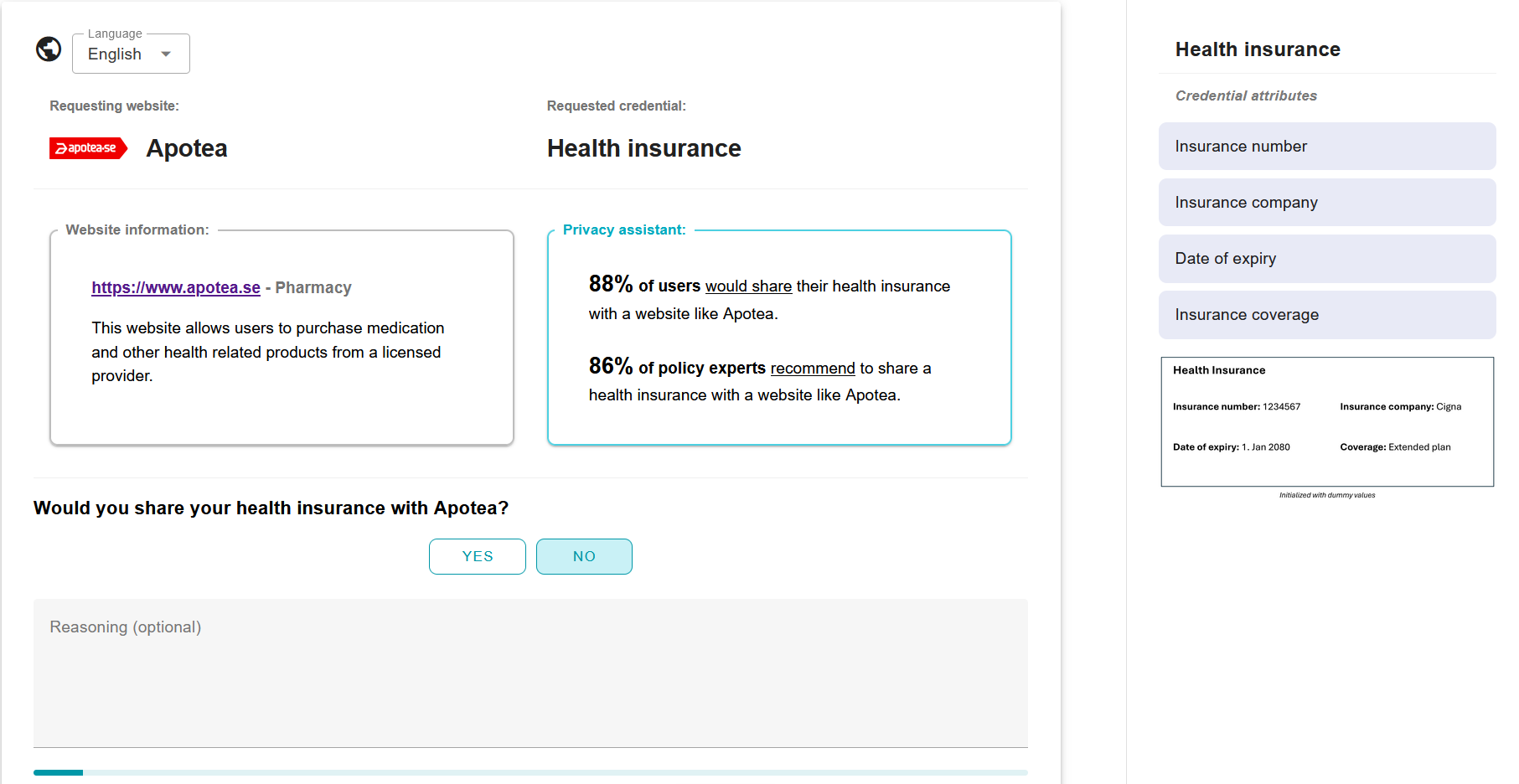}
        \caption{Scenario with \name{} with a double statement.}
        \label{fig:survey_image_scenario_double_support}
     \end{subfigure}
     \hfill
    \begin{subfigure}[b]{0.45\linewidth}
         \centering
        \includegraphics[width=\linewidth]{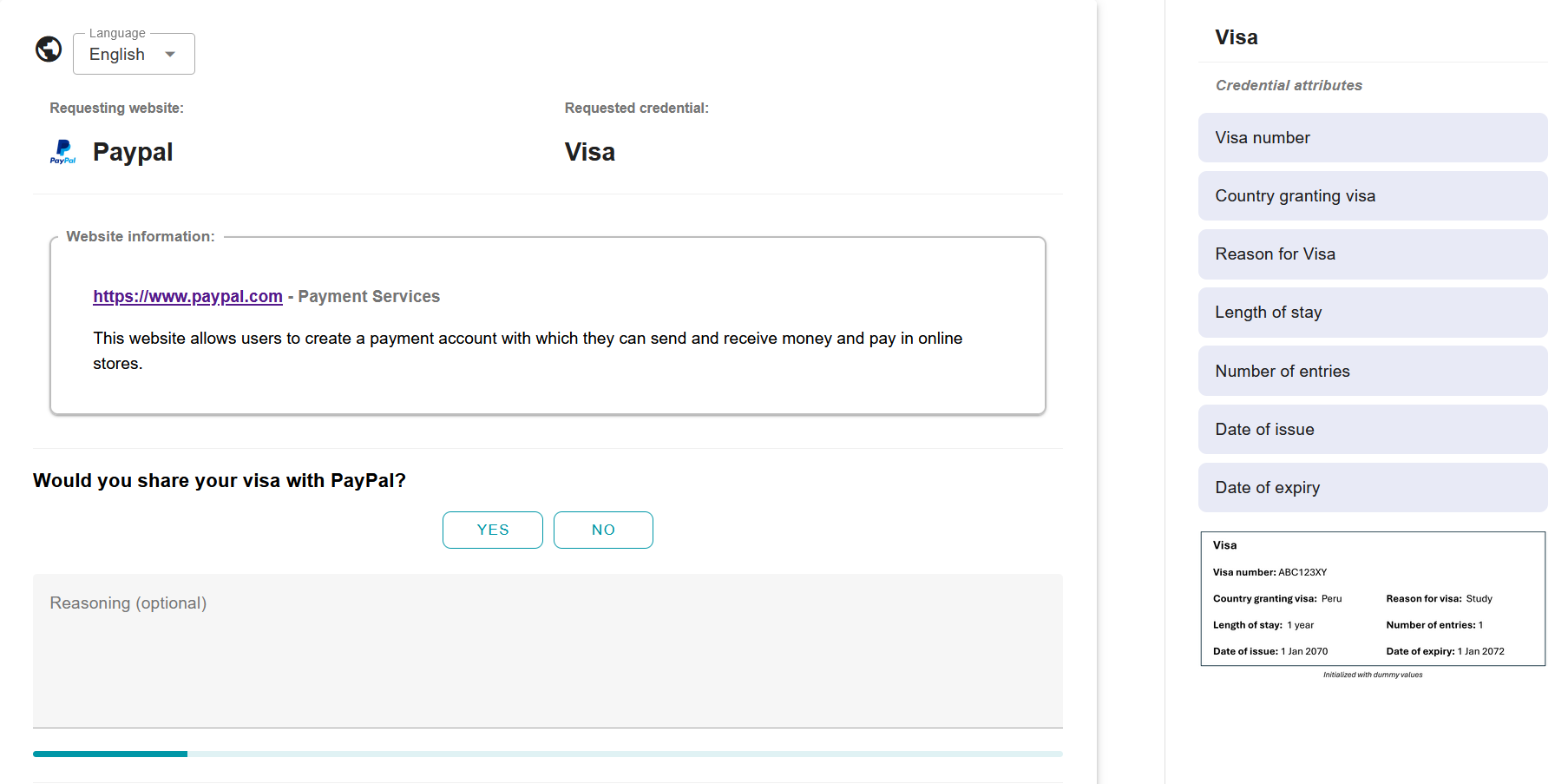}
        \caption{Scenario without \name{}.}
        \label{fig:survey_image_scenario_control}
     \end{subfigure}
     \hfill
     \begin{subfigure}[b]{0.45\linewidth}
        \centering
        \includegraphics[width=\linewidth]{images_for_paper/scenario_single_with_purpose.png}
        \caption{Scenario with website-provided purpose.}
        \label{fig:survey_image_scenario_purpose}
     \end{subfigure}
    \caption{Survey screens for the scenario task with different versions of \name{} and with and without a website-provided purpose. The sidebar always shows all attributes of the requested credential.}
    \label{fig:survey_image_scenario_full}
\end{figure}

\cref{fig:user_study_instruction_combined} shows the instruction pages of the credential task in the user study. The first page describes the EUDI and EUDI credentials, and the second page (not shown to the control group) introduces the \name{}. \cref{fig:survey_image_scenario_full} shows the screens of the scenario task for different options for \name{} (i.e., no \name{}, single statement, double statement), and with and without a website-provided purpose.

\section{Participant Demographics}
\label[appendix]{sec:app_demographics}

In this section, we show the demographic breakdown of the participants of our user survey and user study. 

\begin{figure}[tbp!]
     \centering
    \begin{subfigure}[b]{0.45\linewidth}
         \centering
        \includegraphics[width=\linewidth]{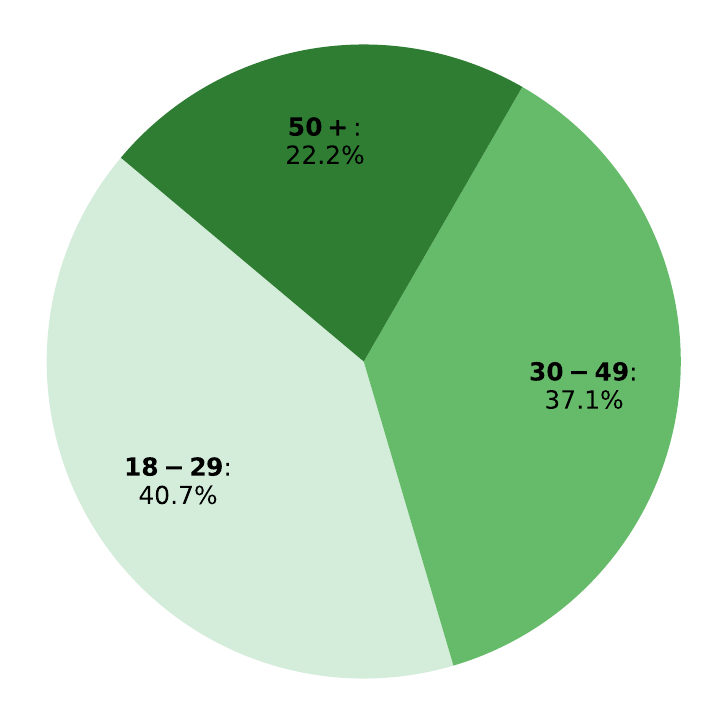}
        \caption{User survey participants age group breakdown.}
        \label{fig:survey_participant_data_age}
     \end{subfigure}
     \hfill
     \begin{subfigure}[b]{0.45\linewidth}
        \centering
        \includegraphics[width=\linewidth]{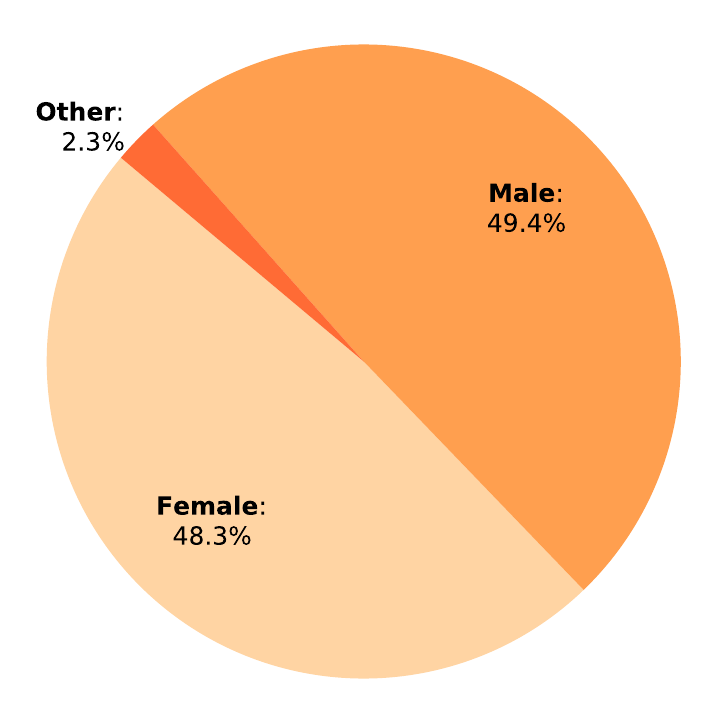}
        \caption{User survey participants gender breakdown.}
        \label{fig:survey_participant_data_gender}
     \end{subfigure}
     \hfill
    \begin{subfigure}[b]{0.45\linewidth}
         \centering
        \includegraphics[width=\linewidth]{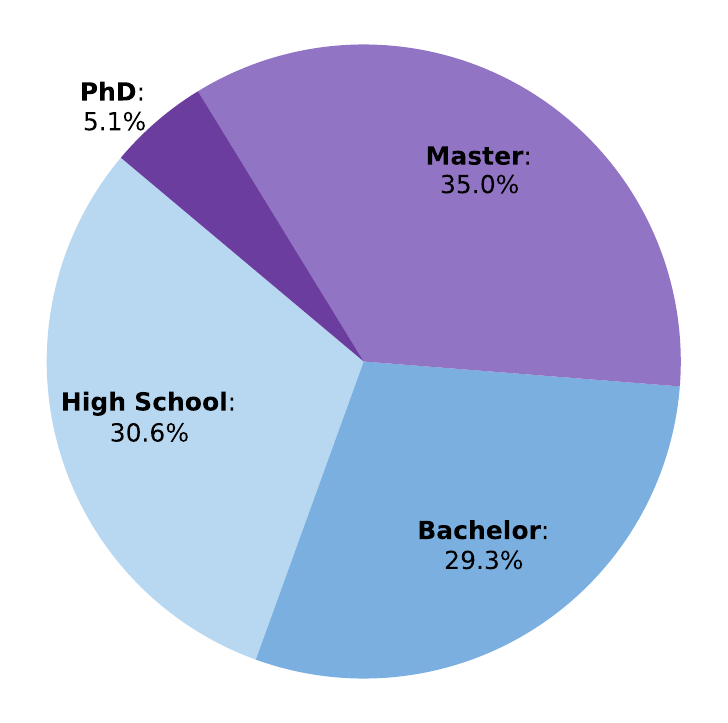}
        \caption{User survey participants highest level of education breakdown.}
        \label{fig:survey_participant_data_education}
     \end{subfigure}
     \hfill
     \begin{subfigure}[b]{0.45\linewidth}
        \centering
        \includegraphics[width=\linewidth]{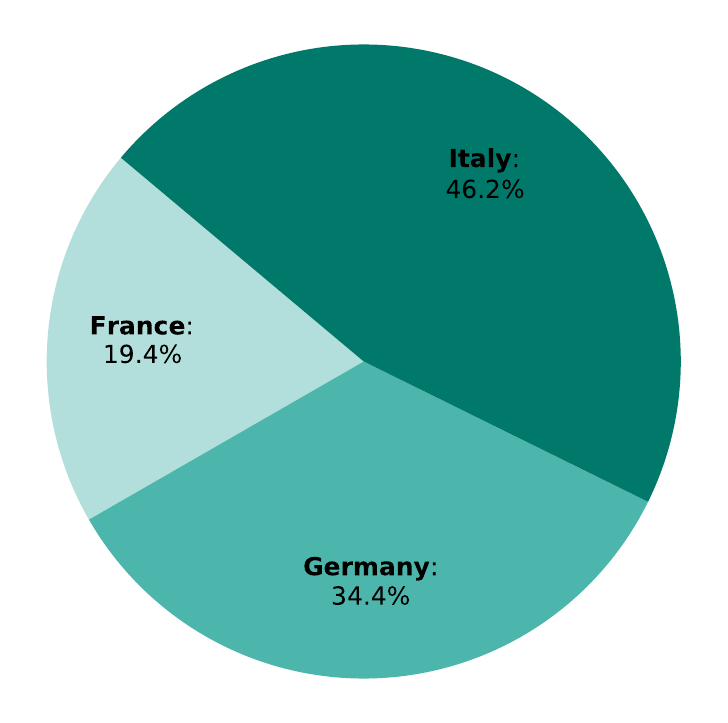}
        \caption{User survey participants place of residency breakdown.}
        \label{fig:survey_participant_data_residency}
     \end{subfigure}
    \caption{Demographic breakdown of the user survey participants.}
    \label{fig:survey_participant_data}
\end{figure}

\cref{fig:survey_participant_data} shows the demographic breakdown of the participants of the user survey. Experts are not shown, as they were not asked to provide demographic data to protect their anonymity. Residency options were Germany, France, and Italy, as participants were required to reside in one of those countries.

\begin{figure}[tbp!]
     \centering
    \begin{subfigure}[b]{0.45\linewidth}
         \centering
        \includegraphics[width=\linewidth]{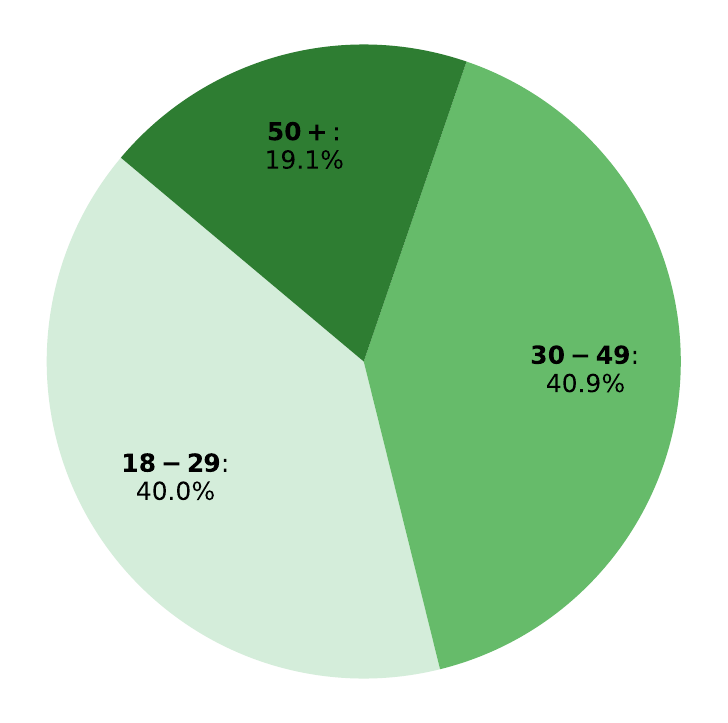}
        \caption{User study participants age group breakdown.}
        \label{fig:study_participant_data_age}
     \end{subfigure}
     \hfill
     \begin{subfigure}[b]{0.45\linewidth}
        \centering
        \includegraphics[width=\linewidth]{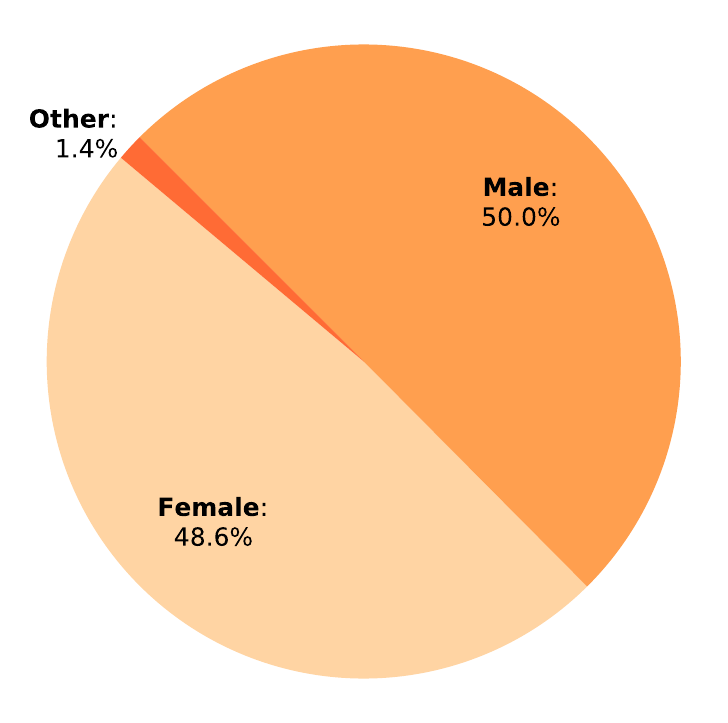}
        \caption{User study participants gender breakdown.}
        \label{fig:study_participant_data_gender}
     \end{subfigure}
     \hfill
    \begin{subfigure}[b]{0.45\linewidth}
         \centering
        \includegraphics[width=\linewidth]{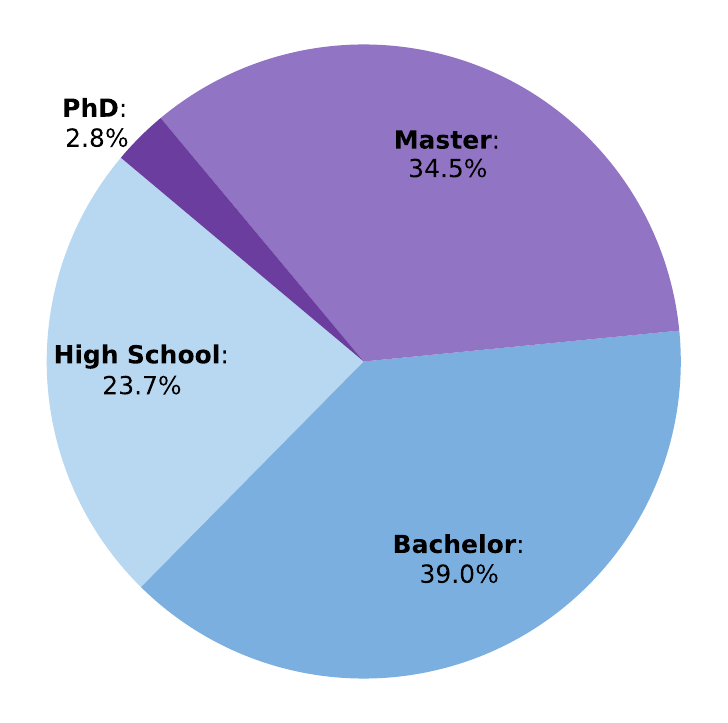}
        \caption{User study participants highest level of education breakdown.}
        \label{fig:study_participant_data_education}
     \end{subfigure}
    \caption{Demographic breakdown of the user study participants.}
    \label{fig:study_participant_data}
\end{figure}

\cref{fig:study_participant_data} shows the demographic breakdown of the participants of the user study.
\section{Details on Experiment Groups}
\label[appendix]{sec:app_deteils_groups}

An overview of the control, baseline, and test groups is shown in \cref{subsec:setup_study}. More details are shown in \cref{tab:experimental_groups}. The control group participants were not divided into classes and could see any of the control scenarios. The baseline participants were divided into a class that saw expert types and a class that did not. Both classes could see any of the baseline scenarios, but the class with expert types would see the "(E types)" version when an expert recommendation was shown. The test participants were divided into 16 classes, T1.1-T6.3, as shown in the table. Participants in normal frequency classes saw 2 test scenarios, and participants in high frequency classes saw 4 test scenarios. Otherwise, the participants saw baseline scenarios that did not have expert types (i.e., "(E types)" scenarios excluded) and that did not come from the same set. This is because the baseline and test scenarios in a set were the same website-credential pairs with different \name{} versions or different website-provided purposes, and we did not want participants to get confused by seeing different data for the same pair. For the same reason, a participant who saw any version for a credential-website pair would not be shown another version for the same pair. For example, a participant in group T3.3 would see two test scenarios from S3 and 8 baseline scenarios from any of S1, S2, S4, S5, S6. Another example, a participant in the baseline group, who saw the B3.1 version of the first scenario of S3, will not see the B3.2 version of the first scenario of S3 (i.e., they are blocked from seeing the first scenario of S3 again).

\begin{table*}[tb]
\centering
\caption{All user study scenarios in the control, baseline, and test groups. In each set, all scenarios are the same with only the purpose or \name{} changing. CA = Credential Assistant, U = User, E = Expert.}
\label{tab:experimental_groups}
\small
\setlength{\tabcolsep}{4pt}
\begin{tabularx}{\textwidth}{c c c c c X X X}
\toprule
Set &
\# &
Correct &
With &
Opinion &
Control &
Baseline &
Test \\
& scen. & answer & purpose & & & & \\
\midrule

\multirow{3}{*}{S1} &
\multirow{3}{*}{10} &
\multirow{3}{*}{N/A} &
\multirow{3}{*}{No} &
\multirow{3}{*}{U} &
C1.1: No CA &
B1.1: CA 81-85\% &
T1.1: CA 51-55\% \\
& & & & &
& &
T1.2 (high freq.): CA 51-55\% \\
& & & & &
& &
T1.3: CA 91-95\% \\

\midrule

\multirow{3}{*}{S2} &
\multirow{3}{*}{5} &
\multirow{3}{*}{No} &
\multirow{3}{*}{Yes} &
\multirow{3}{*}{U} &
C2.1: No purpose, no CA &
B2.1: No purpose, U = no &
T2.1: Vague purpose, U = no \\
& & & & &
C2.2: Vague purpose, no CA &
& T2.2: Ext. purpose, U = no \\
& & & & &
C2.3: Ext. purpose, no CA &
& \\

\midrule

\multirow{3}{*}{S3} &
\multirow{3}{*}{10} &
\multirow{3}{*}{No} &
\multirow{3}{*}{No} &
\multirow{3}{*}{U or E} &
C3.1: No CA &
B3.1: U = no &
T3.1: U = yes \\
& & & & &
& B3.2: E = no &
T3.2 (high freq.): U = yes \\
& & & & &
& B3.3 (E types): E = no &
T3.3: E = yes \\

\midrule

\multirow{3}{*}{S4} &
\multirow{3}{*}{5} &
\multirow{3}{*}{Yes} &
\multirow{3}{*}{Yes} &
\multirow{3}{*}{U} &
C4.1: No purpose, no CA &
B4.1: No purpose, U = yes &
T4.1: Vague purpose, U = no \\
& & & & &
C4.2: Vague purpose, no CA &
B4.2: Vague purpose, U = yes &
T4.2: Ext. purpose, U = no \\
& & & & &
C4.3: Ext. purpose, no CA &
B4.3: Ext. purpose, U = yes &
\\

\midrule

\multirow{3}{*}{S5} &
\multirow{3}{*}{10} &
\multirow{3}{*}{No} &
\multirow{3}{*}{No} &
\multirow{3}{*}{U and E} &
C5.1: No CA &
B5.1: U and E = no &
T5.1: U and E = yes \\
& & & & &
& B5.2 (E types): U and E = no &
T5.2: U = yes, E = no \\
& & & & &
& &
T5.3: E = yes, U = no \\

\midrule

\multirow{3}{*}{S6} &
\multirow{3}{*}{10} &
\multirow{3}{*}{Yes} &
\multirow{3}{*}{No} &
\multirow{3}{*}{U and E} &
C6.1: No CA &
B6.1: U and E = yes &
T6.1: U and E = no \\
& & & & &
& B6.2 (E types): U and E = yes &
T6.2: U = no, E = yes \\
& & & & &
& &
T6.3: E = no, U = yes \\

\bottomrule
\end{tabularx}
\end{table*}

\end{document}